\documentclass[aps,twocolumn,secnumarabic,amsmath,amssymb,
nofootinbib,superscriptaddress]{revtex4-1}

\usepackage{graphics}      
\usepackage{graphicx}      
\usepackage{longtable}     
\usepackage{url}           
\usepackage{bm}            
\usepackage{siunitx}
\usepackage{textcomp}       
\usepackage{gensymb}        
\usepackage{color}
\usepackage{chngcntr}

\usepackage{simplewick}

\usepackage{feynmp-auto}

\usepackage[titletoc,toc,title]{appendix}

\begin{document}
\title{Three-wave scattering in magnetized plasmas: \\from cold fluid to quantized Lagrangian}

\author{Yuan Shi}
\email{yshi@pppl.gov}
\affiliation{Department of Astrophysical Sciences, Princeton University, Princeton, NJ 08544 USA}
\affiliation{Princeton Plasma Physics Laboratory, Princeton University, Princeton, NJ 08543 USA}

\author{Hong Qin}
\affiliation{Department of Astrophysical Sciences, Princeton University, Princeton, NJ 08544 USA}
\affiliation{Princeton Plasma Physics Laboratory, Princeton University, Princeton, NJ 08543 USA}
\affiliation{School of Nuclear Science and Technology, University of Science and Technology of China, Hefei, Anhui 230026, China}

\author{Nathaniel J. Fisch}
\affiliation{Department of Astrophysical Sciences, Princeton University, Princeton, NJ 08544 USA}
\affiliation{Princeton Plasma Physics Laboratory, Princeton University, Princeton, NJ 08543 USA}

\date           {\today}

\setlength{\parskip}{0pt}

\begin{abstract}
Large amplitude waves in magnetized plasmas, generated either by external pumps or internal instabilities, can scatter via three-waves interactions. While three-wave scatterings in either forward or backward geometry are well-known, what happens when waves propagate at angles with one another in magnetized plasmas remains largely unknown, mainly due to the analytical difficulty of this problem. In this paper, we overcome this analytical difficulty and find a convenient formula for three-wave coupling coefficients in cold, uniform, magnetized plasmas in the most general geometry. This is achieved by systematically solving the fluid-Maxwell model to second order using a multiscale perturbative expansion. The general formula for the coupling coefficient becomes transparent when we reformulate it as the \textit{S} matrix element of a quantized Lagrangian. Using the quantized Lagrangian, it is possible to bypass the perturbative solution and directly obtain the nonlinear coupling coefficient from the linear response of plasmas. To illustrate how to evaluate the cold coupling coefficient, we give a set of examples where the participating waves are either quasi-transverse or quasi-longitudinal. In these examples, we determine the angular dependence of three-wave scattering, and demonstrate that backscattering is not necessarily the strongest scattering channel in magnetized plasmas, in contrast to what happens in unmgnetized plasmas. Our approach gives a more complete picture, beyond the simple collimated geometry, of how injected waves can decay in magnetic confinement devices, as well as how lasers can be scattered in magnetized plasma targets.
\end{abstract}

\maketitle

\section{Introduction}\label{sec:intro}
Coherent three-wave scattering is perhaps the simplest and the most common type of nonlinear interaction in plasmas. It happens, for example, in magnetic confinement devices, where waves injected by antenna arrays decay to other waves \cite{Chang74,Liu86}. In the case where the wave is injected to drive current in a tokamak \cite{Fisch78,Fisch87}, there is a possibility that the lower hybrid current drive is affected by unwanted decays near the tokamak periphery \cite{Cesario06}. Even more importantly, three-wave scattering also happens, for example, in laser implosion experiments \cite{Myatt13}, where high intensity lasers interact with plasmas. During magnetized implosions, where the magnetic field is imposed to enhance particle confinement \cite{Hohenberger12,Slutz12,Gotchev09}, multiple laser beams may scatter and reflect one another via magnetic resonances. In fact, the magnetic resonances can be utilized to mediate energy transfer between laser beams to achieve pulse amplification \cite{Shi17}, where three-wave scattering plays an essential role.

Despite of its importance, coherent three-wave scattering, well-studied in unmagnetized plasma \cite{Davidson12,Weiland77}, remains poorly understood when plasmas become magnetized, except in the simple forward or backward geometry, where the participating waves are collimated. This situation is mostly due to the analytical difficulty when external magnetic field is present. 
Such difficulty deserves to be overcome in the midst of recent developments in strong magnetic field technologies \cite{Wagner04,Santos15,Fujioka13}. Using these technologies, magnetic fields on the order of mega-Gauss or even giga-Gauss can be produced. Such strong magnetic field makes electron gyrofrequency comparable to the plasma frequency in laser implosion experiments, in which the anisotropy introduced by the magnetic field can play a prominent role. Since multiple laser beams usually propagate at angles to one another and with the magnetic field during laser-driven implosions, understanding the angular dependence of three-wave scattering in magnetized plasma becomes indispensable for making a knowledgeable choice of the experimental setups to optimize laser-plasma coupling.

By far, most theoretical work on laser scattering in magnetized plasmas is focused on the simple collimated geometry. In this simple geometry, three kinds of theories have been developed. 
The first kind is coupled mode theory, which searches for normal modes of the nonlinear equations \cite{Sjölund67,Shivamoggi82}. The normal modes are typically linear combinations of fluctuating quantities, and the equations satisfied by normal modes are formally simple. However, these equations hide the complexity of the nonlinear problem inside their complicated coupling coefficients, from which little physical meaning has been extracted.
The second kind is nonlinear current theory, which describes three-wave parametric interaction by adding a nonlinear source term into the Maxwell's equation. The nonlinear current can be expressed in terms of a coupling tensor, which is combined with the dielectric tensor to give a nonlinear dispersion relation of the system.
Using fluid models, parametric growth rates have been obtained for extraordinary wave pump \cite{Grebogi80,Barr84,Vyas16}, lower hybrid wave pump \cite{Sanuki77}, as well as the right- and left-circularly polarized wave pumps \cite{Laham98}. To capture thermal effects, a simple treatment retains only thermal corrections to the dielectric tensor \cite{Platzman68}. A more complete treatment also include thermal corrections to the coupling tensor \cite{Ram82,Boyd85}. However, beyond the simple collimated geometry, such treatment becomes so cumbersome that decades of efforts have been spent on just simplifying the expressions \cite{Stenflo70,Stenflo94,Brodin12}, with very little extractable physical results \cite{Larsson76,Stenflo04}.
Beside the coupled mode theory and the nonlinear current theory, the third kind of theory uses Lagrangian formulation. In this more systematic approach, the interaction Lagrangian is obtained either from the Low's Lagrangian \cite{Galloway71,Boyd78}, or the oscillation-center Lagrangian \cite{Dodin17} by expanding plasma response to the third order. Although transparent in formalism, three-wave interactions in magnetized plasma, where the waves are not collimated, remains to be analyzed systematically, in generality, and in detail.

In this paper, we overcome the analytical difficulty in fluid theory and obtain angular dependence of three-wave scattering in cold, uniform, magnetized plasmas in the most general geometry. This is achieved by systematically solving the fluid-Maxwell system to second order in the perturbation series, where secular terms are removed using a multiscale expansion. Using this technique, we manage to obtain an expression for the coupling coefficient that is not only explicit, but also convenient, from which illuminating physical results can be extracted. Moreover, we show that the formula for the coupling coefficient, which contains six permutations of the same structure, naturally arises as the scattering matrix (\textit{S} matrix) element of a quantized Lagrangian. This refreshing perspective, emerging from detailed cold fluid calculations, offers a high-level methodology, through which three-wave coupling can be easily computed.
The cold fluid results are applicable when the wave lengths of participating waves are much longer than both the Debye length and the typical gyroradius. Within the applicable range of the fluid model, our non-relativistic perturbative treatment is valid when the amplitudes of waves are small enough, so that the linear eigenmode structures are preserved, and spectrum broadening is limited.

This paper is organized as follows. In Sec.~\ref{sec:fluid}, we solve the fluid-Maxwell system to second order using a multiscale expansion, in the case where the fluctuation contains a discrete spectrum of waves. In Sec.~\ref{sec:3waves}, we simplify the general equation in the simple case where there are only three linear waves participating in the interaction. In Sec.~\ref{sec:Lagrangian}, we distill the classical theory into a quantized Lagrangian, where the formula for three-wave coupling becomes obvious. In Sec.~\ref{sec:quasi}, we illustrate the general cold fluid results using a set of examples, where the participating waves are either purely electrostatic or purely electromagnetic. The conclusion and discussion are given in Sec.~\ref{sec:discussion}, and supplementary materials are provided in the Appendixes.   

\section{Perturbative solution of fluid-Maxwell system}\label{sec:fluid}
In the fluid regime, where both the Debye length and the typical gyroradius are much smaller than the shortest wavelength, charged particles in the plasma respond collectively to perturbations. In this situation, the plasma system is well described by the fluid-Maxwell equations
\begin{eqnarray}
\label{eq:continuity}
\partial_t n_{s}&=&-\nabla\cdot(n_{s}\mathbf{v}_{s}),\\
\label{eq:momentum}
\partial_t\mathbf{v}_{s}&=&-\mathbf{v}_{s}\cdot\nabla\mathbf{v}_{s}+\frac{e_s}{m_s}(\mathbf{E}+\mathbf{v}_{s}\times\mathbf{B}),\\
\label{eq:Faraday}
\partial_t\mathbf{B}&=&-\nabla\times\mathbf{E},\\
\label{eq:Ampere}
\partial_t\mathbf{E}&=& c^2\nabla\times\mathbf{B}-\frac{1}{\epsilon_0}\sum_{s}e_sn_{s}\mathbf{v}_{s}.
\end{eqnarray}
The continuity equation [Eq.~(\ref{eq:continuity})] describes the conservation of particles of species \textit{s}, whose density is $n_s$ and average velocity is $\mathbf{v}_{s}$. The momentum equation [Eq.~(\ref{eq:momentum})] governs how the velocity field $\mathbf{v}_{s}$ change due to both the advection and the Lorentz force, where $e_s$ and $m_s$ are the charge and mass of individual particles of species \textit{s}. Finally, the magnetic field $\mathbf{B}$ evolves according to the Faraday's law [Eq.~(\ref{eq:Faraday})], and the electric field $\mathbf{E}$ evolves according to the Maxwell-Amp$\grave{\text{e}}$re's law [Eq.~(\ref{eq:Ampere})], where the current density is contributed by all charged species in the system.

The fluid-Maxwell equations [Eqs.~(\ref{eq:continuity})-(\ref{eq:Ampere})] are a system of nonlinear hyperbolic partial differential equations. Such a system of equations are in general difficult to solve. Nevertheless, when fluctuation near equilibrium is small, nonlinearities may be regarded as perturbations, and the equations may be solved perturbatively. To see when nonlinearities may be regarded as perturbations, we can normalize equations such that all quantities become dimensionless numbers. For example, we may normalize time to the plasma frequency $\omega_p$ and distance to the skin depth $c/\omega_p$. We may further normalize mass to electron mass $m_e$, charge to elementary charge $e$, density to unperturbed density $n_{s0}$, and velocity to the speed of light $c$. Finally, we can normalize electric field to $m_ec\omega_p/e$ and normalize magnetic field to $m_e\omega_p/e$.  With the above normalizations, the fluid-Maxwell equation can be written in dimensionless form. In this form, nonlinearities are products of small numbers and are therefore even smaller, provided that the perturbations are small.

In the absence of nonlinearities, the general solution to the fluid-Maxwell system is a spectrum of linear waves with constant amplitudes. Now imaging turning on nonlinearities adiabatically, then waves start to scatter one another, whose amplitudes start to evolve slowly in space and time. This physical picture may be translated into a formal mathematical procedure. Formally, to solve the fluid-Maxwell equations peturbatively, it is helpful to keep track of terms by inserting an auxilliary small parameter $\lambda\ll1$ in the perturbation series, and let the adiabatic parameter $\lambda\rightarrow 1$ in the end, mimicking the adiabatic ramping up of nonlinearities. The electric field, the magnetic field, the density, and the velocity can be expanded in asymptotic series
\begin{eqnarray}
\label{eq:expandE}
\mathbf{E}&=&\mathbf{E}_0+\lambda\mathbf{E}_1+\lambda^2\mathbf{E}_2+\dots,\\
\mathbf{B}&=&\mathbf{B}_0+\lambda\mathbf{B}_1+\lambda^2\mathbf{B}_2+\dots,\\
n_s&=&n_{s0}+\lambda n_{s1}+\lambda^2n_{s2}+\dots,\\
\label{eq:expandV}
\mathbf{v}_s&=&\mathbf{v}_{s0}+\lambda\mathbf{v}_{s1}+\lambda^2\mathbf{v}_{s2}+\dots,
\end{eqnarray}
where a self-consistent equilibrium is given by $\mathbf{E}_0=\mathbf{0}$ and $\mathbf{v}_{s0}=\mathbf{0}$, while the background magnetic field $\mathbf{B}_0$ and density $n_{s0}$ are some constants. It is well-known that if we only expand field amplitudes, the naive asymptotic solution will contain secular terms for nonlinear problems. To remove the secular terms, we also need to do a multiscale expansion  \cite{Debnath11} in both space and time
\begin{eqnarray}
x^i&=&x^i_{(0)}+\frac{1}{\lambda} x^i_{(1)}+\frac{1}{\lambda^2}x^i_{(2)}+\dots,\\
t&=&t_{(0)}+\frac{1}{\lambda} t_{(1)}+\frac{1}{\lambda^2}t_{(2)}+\dots,
\end{eqnarray}
where $x^i$ is the $i$-th components of vector $\mathbf{x}$. In the above expansion, $x^i_{(0)}$ is the shortest spatial scale. In comparison, one unit of $x^i_{(1)}$ is $1/\lambda$ times longer that one unit of $x^i_{(0)}$, and so on. Similarly, $t_{(0)}$ is the fastest time scale, and one unit of $t_{(n)}$ is $1/\lambda^n$ times longer that one unit of $t_{(0)}$. In the above multiscale expansion, different spatial and temporal scales are regarded as independent
\begin{eqnarray}
\label{eq:multispace}
\partial_i^{(a)}x^j_{(b)}=\delta_i^j\delta^{(a)}_{(b)},\\
\label{eq:multitime}
\partial_t^{(a)}t_{(b)}=\delta^{(a)}_{(b)},
\end{eqnarray}
and by chain rule, the total spatial and temporal derivatives are
\begin{eqnarray}
\partial_i&=&\partial_i^{(0)}+\lambda\partial_i^{(1)}+\lambda^2\partial_i^{(2)}+\dots,\\
\partial_t&=&\partial_{t(0)}+\lambda\partial_{t(1)}+\lambda^2\partial_{t(2)}+\dots.
\end{eqnarray}
Using the multiscale expansion (\ref{eq:multispace})-(\ref{eq:multitime}), together with expansion in field amplitudes (\ref{eq:expandE})-(\ref{eq:expandV}), secular terms can be removed and the perturbative solution is well behaved. In Appendix \ref{app:Multiscale}, we demonstrate how the multiscale expansion can be successively applied to a hyperbolic system of ordinary differential equations.

\subsection{First order equations}\label{sec:first}
Although the first order equations and their solutions are well-known \cite{Stix92}, here let us briefly review some important results, in order to introduce some new notations that will be used in the next subsection. To obtain first order equations, we expand fields, space, and time in fluid-Maxwell equations, and collect all the $O(\lambda)$ terms
\begin{eqnarray}
\label{eq:B1}
\partial_{t(0)}\mathbf{B}_1&=&-\nabla_{(0)}\times\mathbf{E}_1,\\
\label{eq:v1}
\partial_{t(0)}\mathbf{v}_{s1}&=&\frac{e_s}{m_s}(\mathbf{E}_1+\mathbf{v}_{s1}\times\mathbf{B}_0),\\
\label{eq:n1}
\partial_{t(0)} n_{s1}&=&-n_{s0}\nabla_{(0)}\cdot\mathbf{v}_{s1},\\
\label{eq:E1}
\Box^{(0)}_{ij}E_1^j&=&-\frac{1}{\epsilon_0}\sum_{s}e_sn_{s0}\partial_{t(0)}v^i_{s1}.
\end{eqnarray} 
Here, we have written the equations in the order that we are going to use them. The electric field equation (\ref{eq:E1}) is obtained by substituting the Faraday's law (\ref{eq:Faraday}) into the Maxwell-Amp$\grave{\text{e}}$re's equation (\ref{eq:Ampere}), and then making the multiscale expansion. This procedure introduces the zeroth order differential operator
\begin{equation}
\Box^{(0)}_{ij}:=(\partial_{t(0)}^2-c^2\nabla_{(0)}^2)\delta_{ij}+c^2\partial_i^{(0)}\partial_j^{(0)}.
\end{equation}
This operator is the d'Alembert wave operator projected in the transverse direction. 

Since the first order equations are linear, the general solution is a superposition of plane waves. Let us write the electric field in the form
\begin{equation}
\label{eq:E1k}
\mathbf{E}_1=\frac{1}{2}\sum_{\mathbf{k}\in\mathbb{K}_1}\mathbf{\mathcal{E}}_{\mathbf{k}}^{(1)} e^{i\theta_{\mathbf{k}}},
\end{equation}
where $\mathbf{\mathcal{E}}_{\mathbf{k}}^{(1)}(t_{(1)},\mathbf{x}_{(1)};t_{(2)},\mathbf{x}_{(2)};\dots)$ is the slowly varying complex wave amplitude, and $\theta_{\mathbf{k}}=i\mathbf{k}\cdot\mathbf{x}_{(0)}-i\omega_{\mathbf{k}}t_{(0)}$ is the fast varying wave phase. The summation of wave vector $\mathbf{k}$ is over a discrete spectrum $\mathbb{K}_1$. In order for $\mathbf{E}_1\in\mathbb{R}^3$ to be a real vector, whenever $\mathbf{k}\in\mathbb{K}_1$ is in the spectrum, then $-\mathbf{k}$ must also be in the spectrum. Moreover, the amplitude $\mathbf{\mathcal{E}}_{\mathbf{k}}^{(1)}$ must satisfy the reality condition $\mathbf{\mathcal{E}}_{-\mathbf{k}}^{(1)}=\mathbf{\mathcal{E}}_{\mathbf{k}}^{(1)*}$. Therefore, it is natural to introduce notations
\begin{eqnarray}
\label{eq:notationz}
\mathbf{z}_{-\mathbf{k}}&=&\mathbf{z}_{\mathbf{k}}^{*},\\
\label{eq:notationa}
\alpha_{-\mathbf{k}}&=&-\alpha_{\mathbf{k}},
\end{eqnarray}
for any complex vector $\mathbf{z}\in\mathbb{C}^3$ and real scalar $\alpha\in\mathbb{R}$ that are labeled by subscript $\mathbf{k}$. For example, the complex vector $\mathbf{\mathcal{E}}_{-\mathbf{k}}=\mathbf{\mathcal{E}}_{\mathbf{k}}^{*}$, and the real scalar $\theta_{-\mathbf{k}}=-\theta_{\mathbf{k}}$. Using the above notations, the reality condition is conveniently built into the symbols.
In spectral expansion Eq.~(\ref{eq:E1k}), it is tempting to write the summation over discrete wave vector $\mathbf{k}$ as an integral over some continuous spectrum. However, such a treatment will be very cumbersome due to double counting, because wave amplitude $\mathbf{\mathcal{E}}_{\mathbf{k}}$, which can vary on slow spatial and temporal scales, already has an spectral width. 

The first order magnetic field $\mathbf{B}_1$, velocity field $\mathbf{v}_{s1}$, and density field $n_{s1}$ can be expressed in terms of the first order electric field $\mathbf{E}_1$. Substituting expression (\ref{eq:E1k}) for the electric field into the first order fluid-Maxwell equations (\ref{eq:B1})-(\ref{eq:n1}), we immediately find
\begin{eqnarray}
\label{eq:Bwave}
\mathbf{B}_1&=&\frac{1}{2}\sum_{\mathbf{k}\in\mathbb{K}_1}\frac{\mathbf{k}\times\mathbf{\mathcal{E}}^{(1)}_\mathbf{k}}{\omega_{\mathbf{k}}}e^{i\theta_{\mathbf{k}}},\\
\label{eq:vwave}
\mathbf{v}_{s1}&=&\frac{ie_s}{2m_s}\sum_{\mathbf{k}\in\mathbb{K}_1}\frac{\mathbb{F}_{s,\mathbf{k}}\mathbf{\mathcal{E}}^{(1)}_\mathbf{k}}{\omega_{\mathbf{k}}}e^{i\theta_{\mathbf{k}}},\\
\label{eq:nwave}
n_{s1}&=&\frac{ie_sn_{s0}}{2m_s}\sum_{\mathbf{k}\in\mathbb{K}_1}\frac{\mathbf{k}\cdot\mathbb{F}_{s,\mathbf{k}}\mathbf{\mathcal{E}}^{(1)}_\mathbf{k}}{\omega_{\mathbf{k}}^2}e^{i\theta_{\mathbf{k}}}.
\end{eqnarray}
Here,  we introduce the forcing operator $\mathbb{F}_{s,\mathbf{k}}:\mathbb{C}^{3}\rightarrow\mathbb{C}^{3}$, acting on any complex vector $\mathbf{z}\in\mathbb{C}^{3}$ by 
\begin{equation}\label{eq:F}
\mathbb{F}_{s,\mathbf{k}}\mathbf{z}:=\gamma_{s,\mathbf{k}}^2[\mathbf{z}+i\beta_{s,\mathbf{k}}\mathbf{z}\times\mathbf{b}-\beta_{s,\mathbf{k}}^2(\mathbf{z}\cdot\mathbf{b})\mathbf{b}].
\end{equation}
In the above definition, $\mathbf{b}$ is the unit vector in the $\mathbf{B}_0$ direction
, $\gamma_{s,\mathbf{k}}^2:=1/(1-\beta_{s,\mathbf{k}}^2)$ is the magnetization factor, $\beta_{s,\mathbf{k}}:=\Omega_{s}/\omega_{\mathbf{k}}$ is the magnetization ratio, and $\Omega_{s}=e_sB_0/m_s$ is the gyrofrequency of species $s$. It is clear from Eq.~(\ref{eq:vwave}) that the forcing operator $\mathbb{F}_{s,\mathbf{k}}$ is related to the linear electric susceptibility $\chi_{s,\mathbf{k}}$ by
\begin{equation}
\label{eq:chi}
\chi_{s,\mathbf{k}}=-\frac{\omega_{ps}^2}{\omega^2_{\mathbf{k}}}\mathbb{F}_{s,\mathbf{k}},
\end{equation}
where $\omega^2_{ps}=e_s^2n_{s0}/\epsilon_0m_s$ is the plasma frequency of species $s$. While the susceptibility $\chi_{s,\mathbf{k}}$ is typically used in linear theories, the forcing operator $\mathbb{F}_{s,\mathbf{k}}$ will be much more convenient when we discuss nonlinear effects. Note that in the limit $B_0\rightarrow 0$, the forcing operator $\mathbb{F}_{s,\mathbf{k}}\rightarrow\mathbf{I}$ becomes the identity operator, and $\chi_{s}$ becomes the cold unmagnetized susceptibility.


The forcing operator $\mathbb{F}_{s,\mathbf{k}}$ will be extremely useful later on when we solve the second order equations. Therefore, let us observe a number of important properties of this operator. For brevity, we will suppress the subscript $s,\mathbf{k}$, with the implied understanding that all quantities have the same subscript. First, the operator satisfies the vector identity 
\begin{equation}\label{eq:Fvector}
\mathbb{F}\mathbf{z}=\mathbf{z}+i\beta(\mathbb{F}\mathbf{z})\times\mathbf{b}.
\end{equation}
This identity guarantees that the velocity field $\mathbf{v}_{s1}$, given by Eq.~(\ref{eq:vwave}), satisfies the first order momentum equation (\ref{eq:v1}). Second, $\mathbb{F}$ is a self-adjoint operator with respect to the inner product $\langle \mathbf{w},\mathbf{z}\rangle:=\mathbf{w}^\dagger\mathbf{z}$,
\begin{equation}
\label{eq:Fadj}
\mathbf{w}^\dagger\mathbb{F}\mathbf{z}=(\mathbb{F}\mathbf{w})^\dagger\mathbf{z},
\end{equation} 
for all complex vectors $\mathbf{z}, \mathbf{w}\in\mathbb{C}^{3}$. 
Using this property, we can move $\mathbb{F}$ from acting on one vector to acting on the other vector in an inner product pair. Third, it is a straightforward calculation to show that
\begin{equation}
\label{eq:F2}
\mathbb{F}^2=\mathbb{F}-\omega\frac{\partial\mathbb{F}}{\partial\omega},
\end{equation}
where the dependence of $\mathbb{F}$ on $\omega$ comes from $\beta$ and $\gamma$ in definition (\ref{eq:F}). Indeed, using its definition, $\mathbb{F}$ satisfies an obvious identity
\begin{equation}\label{eq:-F}
\mathbb{F}(-\omega)=\mathbb{F}^*(\omega),
\end{equation}
which can also be written as $\mathbb{F}_{-\mathbf{k}}=\mathbb{F}^*_{\mathbf{k}}$.
Lastly, when two frequencies $\omega_1$ and $\omega_2$ are involved, we have an nontrivial quadratic identity
\begin{equation}\label{eq:F12}
(\beta_1-\beta_2)\mathbb{F}_1\mathbb{F}_2=\beta_1\mathbb{F}_1-\beta_2\mathbb{F}_2,
\end{equation}
which can be shown by straight forward calculation. Using this identity, we can reduce higher powers of the forcing operators to their linear combinations. Combining with property Eq.~(\ref{eq:-F}), the above identity can generate a number of other similar identities. Properties (\ref{eq:Fvector})-(\ref{eq:F12}) will enable important simplifications when we solve the second order equations.

Having expressed other first order perturbations in terms of $\mathbf{E}_1$, the electric field equation  (\ref{eq:E1}) constrains the relations between the wave amplitude $\mathbf{\mathcal{E}}^{(1)}_{\mathbf{k}}$, the wave frequency $\omega_{\mathbf{k}}$, and the wave vector $\mathbf{k}$. Substituting the expression (\ref{eq:v1}) for $\mathbf{v}_{s1}$ into the electric field equation, we obtain the first order electric field equation in the momentum space
\begin{equation}
\label{eq:E1Fourier}
\omega_{\mathbf{k}}^2\mathbf{\mathcal{E}}^{(1)}_\mathbf{k}+c^2\mathbf{k}\times(\mathbf{k}\times\mathbf{\mathcal{E}}^{(1)}_\mathbf{k})=\sum_s\omega_{ps}^2\mathbb{F}_{s,\mathbf{k}}\mathbf{\mathcal{E}}^{(1)}_\mathbf{k},
\end{equation}
which must be satisfied for individual wave vector $\mathbf{k}$ in the spectrum. The above equation can be written in a matrix form $\mathbb{D}_{\mathbf{k}}\mathbf{\mathcal{E}}^{(1)}_\mathbf{k}=0$, where the dispersion tensor
\begin{equation}
\label{eq:Dk}
\mathbb{D}_{\mathbf{k}}^{ij}:=(\omega_{\mathbf{k}}^2-c^2\mathbf{k}^2)\delta^{ij} +c^2k^ik^j-\sum_s\omega_{ps}^2\mathbb{F}_{s,\mathbf{k}}^{ij}.
\end{equation}
The matrix equation has nontrivial solutions when the wave vector $\mathbf{k}$ and wave frequency $\omega_{\mathbf{k}}$ are such that the linear dispersion relation $\det \mathbb{D}(\mathbf{k},\omega_{\mathbf{k}})=0$ is satisfied. When the dispersion relation is indeed satisfied, solving the matrix equation gives wave polarizations. It is well-known that in magnetized plasmas, the eignemodes are two mostly electromagnetic waves and a number of mostly electrostatic hybrid waves. In Appendix \ref{app:Linear}, we review the dispersion relations and wave polarizations when waves propagate at arbitrary angles with respect to the background magnetic field. 

Finally, to introduce one more operator that will be useful for solving the second order equations, let us calculate the wave energy. The average energy carried by linear waves can be found by summing up average energy carried by fields and particles. For a single linear wave with wave vector $\mathbf{k}$, after averaging on $t_{(0)}$ and $\mathbf{x}_{(0)}$ scale, the wave energy
\begin{eqnarray}
\label{eq:U}
\nonumber
U_\mathbf{k}&=&\frac{\epsilon_0}{2}\langle\mathbf{E}_1^2\rangle_{(0)} +\frac{1}{2\mu_0}\langle\mathbf{B}_1^2\rangle_{(0)}+\frac{1}{2}\sum_sn_{s0}m_s\langle\mathbf{v}_{s1}^2\rangle_{(0)}\\
&=&\frac{\epsilon_0}{4}\mathbf{\mathcal{E}}^{(1)*}_\mathbf{k}\cdot\mathbb{H}_{\mathbf{k}}\mathbf{\mathcal{E}}^{(1)}_\mathbf{k},
\end{eqnarray}
where we introduce the normalized wave energy operator
\begin{eqnarray}
\label{eq:Hk}
\mathbb{H}_{\mathbf{k}}&:=&2\mathbb{I}-\sum_s\frac{\omega_{ps}^2}{\omega_\mathbf{k}} \frac{\partial\mathbb{F}_{s,\mathbf{k}}}{\partial\omega_\mathbf{k}}\\
\nonumber
&=&\frac{1}{\omega_\mathbf{k}}\frac{\partial(\omega_\mathbf{k}^2\epsilon_\mathbf{k})}{\partial\omega_\mathbf{k}}.
\end{eqnarray}
Here, $\epsilon_\mathbf{k}=\mathbb{I}+\sum_s\chi_{s,\mathbf{k}}$ is the dielectric tensor, and we have used the Eq.~(\ref{eq:chi}), which relates the forcing operator to the susceptibility. When evaluating $\langle\mathbf{B}_1^2\rangle$, we have used expression (\ref{eq:Bwave}) for $\mathbf{B}_1$, followed by simplification using the momentum space electric field equation (\ref{eq:E1Fourier}). This term is then combined with $\langle\mathbf{v}_{s1}^2\rangle$, calculated using Eq.~(\ref{eq:vwave}) for $\mathbf{v}_{s1}$. The final result is simplified using identity (\ref{eq:F2}) for the forcing operator $\mathbb{F}_{s,\mathbf{k}}$. Now that we have introduced the wave energy operator $\mathbb{H}_{\mathbf{k}}$, the momentum space electric field equation (\ref{eq:E1Fourier}) can be converted into a form that is closely related to the wave energy
\begin{equation}
\label{eq:dE1}
\frac{\partial\omega_{\mathbf{k}}}{\partial k_l}\omega_\mathbf{k}\mathbb{H}_{\mathbf{k}}^{ij}\mathcal{E}^{(1)j}_\mathbf{k}
=c^2(2k_l\delta_{ij}-k_i\delta_{jl}-k_j\delta_{il})\mathcal{E}^{(1)j}_\mathbf{k}.
\end{equation}
This form of the first order electric field equation is obtained by taking $\partial/\partial k_l$ derivative on both side of Eq.~(\ref{eq:E1Fourier}). Notice that although $\mathbf{\mathcal{E}}^{(1)}_\mathbf{k}$ is labeled by $\mathbf{k}$, it does not explicitly depend on $\mathbf{k}$. This alternative form of the first order electric field equation will be useful when we solve the second order equations. 

\subsection{Second order equations}\label{sec:second}
To obtain the second order equations, we collect all the $O(\lambda^2)$ terms in the asymptotic expansions. The resultant second order equations are 
\begin{eqnarray}
\label{eq:B2}
\partial_{t(0)}\mathbf{B}_2&=&-\partial_{t(1)}\mathbf{B}_1-\nabla_{(1)}\times\mathbf{E}_1-\nabla_{(0)}\times\mathbf{E}_2,\\
\label{eq:v2}
\nonumber
\partial_{t(0)}\mathbf{v}_{s2}&=&-\partial_{t(1)}\mathbf{v}_{s1}-\mathbf{v}_{s1}\cdot\nabla_{(0)}\mathbf{v}_{s1}\\
&&+\frac{e_s}{m_s}\Big(\mathbf{v}_{s1}\times\mathbf{B}_1+\mathbf{E}_2+\mathbf{v}_{s2}\times\mathbf{B}_0\Big),\\
\label{eq:n2}
\nonumber
\partial_{t(0)} n_{s2}&=&-\partial_{t(1)} n_{s1}-\nabla_{(0)}\cdot(n_{s1}\mathbf{v}_{s1})\\
&&-n_{s0}\big(\nabla_{(1)}\cdot\mathbf{v}_{s1}+\nabla_{(0)}\cdot\mathbf{v}_{s2}\big),\\
\label{eq:E2}
\nonumber
\Box^{(0)}_{ij}E_2^j&=&-\Box^{(1)}_{ij}E_1^j-\frac{1}{\epsilon_0}\sum_{s}e_s\Big[n_{s0}\partial_{t(1)}v^i_{s1}\\
&&+\partial_{t(0)}(n_{s1}v^i_{s1})+n_{s0}\partial_{t(0)}v^i_{s2}\Big].
\end{eqnarray}
Again, the electric field equation (\ref{eq:E2}) is obtained by substituting Faraday's law into the Maxwell-Amp$\grave{\text{e}}$re's equation. In doing so, we introduce the first order differential operator
\begin{eqnarray}
\nonumber
\Box^{(1)}_{ij}:&=&2\big(\partial_{t(0)}\partial_{t(1)}-c^2\partial_l^{(0)}\partial_l^{(1)}\big)\delta_{ij}\\
&+&c^2\big(\partial_i^{(0)}\partial_j^{(1)}+\partial_i^{(1)}\partial_j^{(0)}\big).
\end{eqnarray}
This operator mixes fast and slow scales, and will govern how wave amplitudes vary on the slow scales due to interactions that happen on the fast scale.

To solve the second order equations, notice that although the second order equations are nonlinear in $\mathbf{B}_1$, $\mathbf{v}_{s1}$, and $n_{s1}$, they are nevertheless linear in  $\mathbf{E}_2$, $\mathbf{B}_2$, $\mathbf{v}_{s2}$, and $n_{s2}$. Therefore, we may solve for the second order perturbations from the linear equations, regarding nonlinearities in first order perturbations as source terms. The general solution to such a system of linear equations is again a superposition of plane waves. Let us write the second order electric field
\begin{equation}
\label{eq:E2k}
\mathbf{E}_2=\frac{1}{2}\sum_{\mathbf{k}\in\mathbb{K}_2}\mathbf{\mathcal{E}}_{\mathbf{k}}^{(2)} e^{i\theta_{\mathbf{k}}}.
\end{equation}
Similar to the first order expansion (\ref{eq:E1k}), in the above expression, $\mathbf{\mathcal{E}}_{\mathbf{k}}^{(2)}(t_{(1)},\mathbf{x}_{(1)};t_{(2)},\mathbf{x}_{(2)};\dots)$ is the second order slowly varying complex wave amplitude, $\theta_{\mathbf{k}}$ is the fast wave phase, and $\mathbb{K}_2$ is the spectrum of second order fluctuations, which contains $-\mathbf{k}$ whenever $\mathbf{k}\in\mathbb{K}_2$. 
The second order spectrum $\mathbb{K}_2$ is highly constrained and will need to be determined from the second order electric field equation, once the first order spectrum $\mathbb{K}_1$ is given.

Before we can 
determine $\mathbb{K}_2$ and $\mathbf{\mathcal{E}}_{\mathbf{k}}^{(2)}$, we need to express $\mathbf{B}_2$ in terms of $\mathbf{E}_2$. Plugging in expressions for the first order fluctuations Eqs.~(\ref{eq:E1k}) and (\ref{eq:Bwave}) into the second order Faraday's law Eq.~(\ref{eq:B2}), the second order magnetic field can be expressed as
\begin{eqnarray}
\label{eq:Bwave2}
\mathbf{B}_2&=&\frac{1}{2}\sum_{\mathbf{k}\in\mathbb{K}_2}\frac{\mathbf{k}\times\mathbf{\mathcal{E}}^{(2)}_\mathbf{k}}{\omega_{\mathbf{k}}}e^{i\theta_{\mathbf{k}}}\\
\nonumber
&+&\frac{1}{2}\sum_{\mathbf{k}\in\mathbb{K}_1}\Big(\frac{\nabla_{(1)} \times\mathbf{\mathcal{E}}_{\mathbf{k}}^{(1)}}{i\omega_\mathbf{k}} +\frac{\mathbf{k}\times\partial_{t(1)}\mathbf{\mathcal{E}}_{\mathbf{k}}^{(1)}}{i\omega_\mathbf{k}^2}\Big)e^{i\theta_{\mathbf{k}}}.
\end{eqnarray}
The first line has the same structure as $\mathbf{B}_1$, except now the summation is over the second order spectrum $\mathbb{K}_2$. The second line involves slow derivatives of the first order amplitude $\mathbf{\mathcal{E}}^{(1)}_\mathbf{k}$. These derivatives, still unknown at this step, will be determined later from the second order electric field equation. 

Similarly, the second order velocity $\mathbf{v}_{s2}$ can be solved from Eq.~(\ref{eq:v2}). One way of solving this equation is by first taking the Fourier transform on $t_{(0)}$ and $\mathbf{x}_{(0)}$ scale. Then in the Fourier space, the resultant algebraic equation can be readily solved using the property (\ref{eq:Fvector}) of the forcing operator. After taking the inverse Fourier transform, the second order velocity can be expressed as
\begin{eqnarray}
\label{eq:vwave2}
\nonumber
\mathbf{v}_{s2}&=&\frac{ie_s}{2m_s}\sum_{\mathbf{k}\in\mathbb{K}_2} \frac{\mathbb{F}_{s,\mathbf{k}}\mathbf{\mathcal{E}}_{\mathbf{k}}^{(2)}}{\omega_\mathbf{k}}e^{i\theta_{\mathbf{k}}} \\
&+&\frac{e_s}{2m_s}\sum_{\mathbf{k}\in\mathbb{K}_1}\frac{\mathbb{F}^2_{s,\mathbf{k}}\partial_{t(1)}\mathbf{\mathcal{E}}_{\mathbf{k}}^{(1)}}{\omega_\mathbf{k}^2}e^{i\theta_{\mathbf{k}}}\\
\nonumber
&-&\frac{e_s^2}{4m_s^2}\!\sum_{\mathbf{q},\mathbf{q}'\in\mathbb{K}_1} \frac{\mathbb{F}_{s,\mathbf{q}+\mathbf{q}'}(\mathbf{L}^{s}_{\mathbf{q},\mathbf{q}'}\!+\!\mathbf{T}^{s}_{\mathbf{q},\mathbf{q}'})}{\omega_\mathbf{q}+\omega_\mathbf{q}'}e^{i\theta_{\mathbf{q}}+i\theta_{\mathbf{q}'}}.
\end{eqnarray}
The first two lines of the above expression is in analogy to the expression (\ref{eq:Bwave2}) for $\mathbf{B}_2$. The third line comes from beating of nonlinearities. In particular, the $\mathbf{v}_{s1}\times\mathbf{B}_1$ nonlinearity introduce a longitudinal beating 
\begin{equation}
\label{eq:L}
\mathbf{L}^{s}_{\mathbf{q},\mathbf{q}'}=\frac{(\mathbb{F}_{s,\mathbf{q}}\mathbf{\mathcal{E}}_{\mathbf{q}}^{(1)})\times(\mathbf{q}'\times\mathbf{\mathcal{E}}_{\mathbf{q}'}^{(1)})}{\omega_\mathbf{q}\omega_{\mathbf{q}'}}.
\end{equation}
In addition, the Euler derivative $\mathbf{v}_{s1}\cdot\nabla_{(0)}\mathbf{v}_{s1}$, which is responsible for generating turbulence in neutral fluids, gives rise to a turbulent beating
\begin{equation}
\label{eq:T}
\mathbf{T}^{s}_{\mathbf{q},\mathbf{q}'}=\frac{(\mathbb{F}_{s,\mathbf{q}}\mathbf{\mathcal{E}}_{\mathbf{q}}^{(1)})(\mathbf{q}\cdot\mathbb{F}_{s,\mathbf{q}'}\mathbf{\mathcal{E}}_{\mathbf{q}'}^{(1)})}{\omega_\mathbf{q}\omega_{\mathbf{q}'}}.
\end{equation}
The third line in Eq.~(\ref{eq:vwave2}) may be simplified using the quadratic property (\ref{eq:F12}) of the forcing operator. This simplification will be done later when we discuss interaction of three waves in the next section.

Using similar method, we can find the expression for the second order density $n_{s2}$. Although the expression for $n_{s2}$ is not indispensable for studying three-wave scattering, we present it here because it will become useful when one studies four-wave or even higher order interactions. The second order density can be expressed as
\begin{eqnarray}
\label{eq:nwave2}
\nonumber
n_{s2}&=&\frac{e_sn_{s0}}{2m_s}\Bigg[\sum_{\mathbf{k}\in\mathbb{K}_2}\frac{i\mathbf{k}\cdot\mathbb{F}_{s,\mathbf{k}}\mathbf{\mathcal{E}}^{(2)}_\mathbf{k}}{\omega_{\mathbf{k}}^2}e^{i\theta_{\mathbf{k}}}\\
\nonumber
&+&\!\sum_{\mathbf{k}\in\mathbb{K}_1}\!\bigg(\!\frac{\mathbf{k}\!\cdot\!(\mathbb{F}_{s,\mathbf{k}}\!+\!\mathbb{F}^2_{s,\mathbf{k}})\partial_{t(1)}\mathbf{\mathcal{E}}_{\mathbf{k}}^{(1)}}{\omega_\mathbf{k}^3}\!+\!\frac{\nabla_{(1)}\!\cdot\!\mathbb{F}_{s,\mathbf{k}}\mathbf{\mathcal{E}}_{\mathbf{k}}^{(1)}}{\omega_\mathbf{k}^2}\!\bigg)\!e^{i\theta_{\mathbf{k}}}\Bigg]\\
&-&\frac{e_s^2n_{s0}}{4m_s^2}\!\sum_{\mathbf{q},\mathbf{q}'\in\mathbb{K}_1} \frac{(\mathbf{q}+\mathbf{q}')\cdot\mathbf{R}^{s}_{\mathbf{q},\mathbf{q}'}}{(\omega_\mathbf{q}+\omega_\mathbf{q}')^2}e^{i\theta_{\mathbf{q}}+i\theta_{\mathbf{q}'}}.
\end{eqnarray}
The above three lines are in analogy to those for $\mathbf{v}_{s2}$ in Eq.~(\ref{eq:vwave2}). In the third line, the quadratic response 
\begin{equation}
\label{eq:R}
\mathbf{R}^{s}_{\mathbf{q},\mathbf{q}'}=\mathbb{F}_{s,\mathbf{q}+\mathbf{q}'} (\mathbf{L}^{s}_{\mathbf{q},\mathbf{q}'}+\mathbf{T}^{s}_{\mathbf{q},\mathbf{q}'}) +(1+\frac{\omega_\mathbf{q}}{\omega_\mathbf{q}'})\mathbf{C}^{s}_{\mathbf{q},\mathbf{q}'},
\end{equation}
where the longitudinal beating $\mathbf{L}^{s}_{\mathbf{q},\mathbf{q}'}$ and the turbulent beating $\mathbf{T}^{s}_{\mathbf{q},\mathbf{q}'}$ are given by Eqs.~(\ref{eq:L}) and (\ref{eq:T}). The third term, proportional to $\mathbf{C}^{s}_{\mathbf{q},\mathbf{q}'}$, comes from the divergence of the nonlinear current $\nabla_{(0)}\cdot(n_{s1}\mathbf{v}_{s1})$, which introduces the current beating 
\begin{equation}
\label{eq:C}
\mathbf{C}^{s}_{\mathbf{q},\mathbf{q}'}=\frac{(\mathbb{F}_{s,\mathbf{q}}\mathbf{\mathcal{E}}_{\mathbf{q}}^{(1)})(\mathbf{q}'\cdot\mathbb{F}_{s,\mathbf{q}'}\mathbf{\mathcal{E}}_{\mathbf{q}'}^{(1)})}{\omega_\mathbf{q}\omega_{\mathbf{q}'}}.
\end{equation}
Although the form of $\mathbf{C}^{s}_{\mathbf{q},\mathbf{q}'}$ is similar to that of  $\mathbf{T}^{s}_{\mathbf{q},\mathbf{q}'}$
, the physics of these two types of beating are nevertheless very different.

Having expressed second order fluctuations 
in terms of $\mathbf{E}_2$, we can obtain an equation that only involves electric perturbations. Substituting expressions (\ref{eq:vwave}), (\ref{eq:nwave}), and (\ref{eq:vwave2}) into the second order electric field equation (\ref{eq:E2}), we can eliminate $\mathbf{v}_{s1}$, $n_{s1}$, and $\mathbf{v}_{s2}$. The resultant equation can be simplified using the first order electric field equation (\ref{eq:dE1}), as well as property (\ref{eq:F2}) of the forcing operator. The second order electric field equation can then be put into a rather simple and intuitive form
\begin{eqnarray}
\label{eq:E2s}
\nonumber
&&\sum_{\mathbf{k}\in\mathbb{K}_2}\mathbb{D}_{\mathbf{k}}\mathbf{\mathcal{E}}^{(2)}_\mathbf{k}e^{i\theta_{\mathbf{k}}} +i\sum_{\mathbf{k}\in\mathbb{K}_1}\omega_{\mathbf{k}}\mathbb{H}_{\mathbf{k}}d_{t(1)}^{\mathbf{k}}\mathbf{\mathcal{E}}^{(1)}_\mathbf{k} e^{i\theta_{\mathbf{k}}}\\
&=&\frac{i}{2}\sum_{s,\mathbf{q},\mathbf{q}'\in\mathbb{K}_1}\mathbf{S}^{s}_{\mathbf{q},\mathbf{q}'}e^{i\theta_{\mathbf{q}}+i\theta_{\mathbf{q}'}}.
\end{eqnarray}
The left-hand-side are modifications of the first order spectrum, as consequences of three-wave scatterings on the right-hand-side. In the above equation, the dispersion tensor $\mathbb{D}_{\mathbf{k}}=\mathbb{D}^*_{-\mathbf{k}}$ is defined by Eq.~(\ref{eq:Dk}), the normalized wave energy operator $\mathbb{H}_{\mathbf{k}}=\mathbb{H}^*_{-\mathbf{k}}$ is defined by Eq.~(\ref{eq:Hk}), and $d_{t(1)}^{\mathbf{k}}=d_{t(1)}^{-\mathbf{k}}$ is the advective derivative
\begin{equation}
\label{eq:dt1}
d_{t(1)}^{\mathbf{k}}:=\partial_{t(1)}+\frac{\partial\omega_{\mathbf{k}}}{\partial\mathbf{k}}\cdot\nabla_{(1)},
\end{equation}
which advects the wave envelope at the wave group velocity $\mathbf{v}_g=\partial\omega_{\mathbf{k}}/\partial\mathbf{k}$ on the slow scale $t_{(1)}$ and $\mathbf{x}_{(1)}$. In Eq.~(\ref{eq:E2s}), the three-wave scattering strength 
\begin{equation}
\label{eq:S}
\mathbf{S}^{s}_{\mathbf{q},\mathbf{q}'}=\frac{e_s\omega_{ps}^2}{2m_s}\Big(\mathbf{R}^{s}_{\mathbf{q},\mathbf{q}'}+\mathbf{R}^{s}_{\mathbf{q}',\mathbf{q}}\Big),
\end{equation}
where the quadratic response $\mathbf{R}^{s}_{\mathbf{q},\mathbf{q}'}$ is given by Eq.~(\ref{eq:R}). Notice that the scattering strength $\mathbf{S}^{s}_{\mathbf{q},\mathbf{q}'}$ is proportional to the density $n_{s0}$. This is intuitive because three-wave scattering cannot happen in the vacuum. Hence, all three-wave scatterings come from charged particle response, which is additive and therefore proportional to the density. Also notice that $\mathbf{S}^{s}_{\mathbf{q},\mathbf{q}'}$ is proportional to the charge-to-mass ratio. This is also intuitive because $e_s/m_s$ is the coefficient by which charged particles respond to the electric field. 

Let us observe a number of properties of the scattering strength $\mathbf{S}^{s}_{\mathbf{q},\mathbf{q}'}$. First, by construction, the scattering strength is symmetric with respect to $\mathbf{q},\mathbf{q}'$, namely,
\begin{equation}
\mathbf{S}^{s}_{\mathbf{q},\mathbf{q}'}=\mathbf{S}^{s}_{\mathbf{q}',\mathbf{q}}.
\end{equation}
In addition, using notation (\ref{eq:notationz}) and (\ref{eq:notationa}), it is easy to see that reality condition for $\mathbf{S}_{\mathbf{q},\mathbf{q}'}$ is
\begin{equation}
\mathbf{S}_{\mathbf{q},\mathbf{q}'}^{s*}=-\mathbf{S}^{s}_{-\mathbf{q},-\mathbf{q}'}.
\end{equation}
Moreover, it turns out that the scattering strength $\mathbf{S}^{s}_{\mathbf{q},\mathbf{q}'}$ satisfies the important identity
\begin{equation}
\label{eq:SDC}
\mathbf{S}^{s}_{\mathbf{q},-\mathbf{q}}=\mathbf{0}.
\end{equation}
This identity can be shown by straight forward calculation using the limiting form $\mathbb{F}(\omega)\rightarrow\mathbf{b}\mathbf{b}$ when $\omega\rightarrow 0$. Identity (\ref{eq:SDC}) guarantees that no zero-frequency mode with $\omega_{\mathbf{k}}=0$ will arise in the second order electric field equation. 
Without this important identity, any change in the wave amplitude would be faster then the zero-frequency mode, a situation that would violate the multiscale assumption. Fortunately, due to identity (\ref{eq:SDC}), the multiscale perturbative solution is well justified.

Now that we have obtained the second order electric field equation (\ref{eq:E2s}), we can use it to constrain the spectrum $\mathbb{K}_2$ and the amplitude $\mathbf{\mathcal{E}}_{\mathbf{k}}^{(2)}$.  In order to satisfy  (\ref{eq:E2s}), the coefficient of each Fourier exponent $e^{i\theta_{\mathbf{k}}}$ must be matched on both sides of the equation. 
To match the spectrum on the right-hand-side of Eq.~(\ref{eq:E2s}), which is generated by beating of first order perturbations, we can take the second order spectrum to be
\begin{equation}
\label{eq:K2}
\mathbb{K}_2=(\mathbb{K}_1^0\bigoplus\mathbb{K}_1^0)\setminus\mathbb{K}_1^0,
\end{equation}
where the set $\mathbb{K}_1^0:=\mathbb{K}_1\bigcup\{\mathbf{0}\}$. We define the direct sum of two sets $G_1,G_2\subseteq G$, where $G$ is an additive group, by $G_1\bigoplus G_2:=\{g_1+g_2|g_1\in G_1, g_2\in G_2\}$. We can exclude the zero vector $\mathbf{0}$ from the second order spectrum $\mathbb{K}_2$ using property (\ref{eq:SDC}) of the scattering strength. We also excluded vectors that are already contained in the first order spectrum $\mathbb{K}_1$, such that the matrix $\mathbb{D}_{\mathbf{k}}$ is invertible for all $\mathbf{k}\in\mathbb{K}_2$. Since the matrix is invertible, the second order amplitude $\mathbf{\mathcal{E}}_{\mathbf{k}}^{(2)}$ is determined by
\begin{equation}
\mathbf{\mathcal{E}}_{\mathbf{k}}^{(2)}=i\mathbb{D}^{-1}_{\mathbf{k}}\sum_{s}\mathbf{S}^{s}_{\mathbf{q},\mathbf{q}'},
\end{equation}
where $\mathbf{q}, \mathbf{q}'\in\mathbb{K}_1$ are such that $\mathbf{k}=\mathbf{q}+\mathbf{q}'\in\mathbb{K}_2$. Here, the factor $1/2$ has been removed using the symmetry property $2\mathbf{S}^{s}_{\mathbf{q},\mathbf{q}'}=\mathbf{S}^{s}_{\mathbf{q},\mathbf{q}'}+\mathbf{S}^{s}_{\mathbf{q}',\mathbf{q}}$. We can put the above abstract notations in more intuitive language as follows. The first order spectrum contains all the ``on-shell" waves, which satisfy the dispersion relation $\det\mathbb{D}(\mathbf{k},\omega_\mathbf{k})=0$ for all $\mathbf{k}\in\mathbb{K}_1$. While the second order spectrum $\mathbb{K}_2$ contains all the ``off-shell" waves generated by beating. These ``off-shell" waves do not satisfy the linear dispersion relation, and their amplitude is driven by the beating of two ``on-shell" waves.

To illustrate the abstract notations introduced above, let us consider the simplest example where the spectrum $\mathbb{K}_1$ contains only one ``on-shell" wave, namely, $\mathbb{K}_1=\{\mathbf{k},-\mathbf{k}\}$. In this case, the second order spectrum $\mathbb{K}_2=\{2\mathbf{k},-2\mathbf{k}\}$ contains the second harmonic. Matching the Fourier exponents, the ``on-shell" equation is 
\begin{equation}
\omega_{\mathbf{k}}\mathbb{H}_{\mathbf{k}}d_{t(1)}^{\mathbf{k}}\mathbf{\mathcal{E}}^{(1)}_\mathbf{k}=\mathbf{0}.
\end{equation}
The other ``on-shell" equation is the complex conjugate of the above equation. 
Since $\mathbb{H}_{\mathbf{k}}$ enters the wave energy (\ref{eq:U}), this matrix is positive definite and therefore nondegenrate. Hence, the above equation can be written as $d_{t(1)}^{\mathbf{k}}\mathbf{\mathcal{E}}^{(1)}_\mathbf{k}=\mathbf{0}$, which says that the wave amplitude is a constant of advection. Next, matching coefficients of the other Fourier exponent, we obtain the ``off-shell" equation for the second harmonic is
\begin{equation}
\mathbb{D}_{2\mathbf{k}}\mathbf{\mathcal{E}}^{(2)}_{2\mathbf{k}}=i\sum_{s}\mathbf{S}^{s}_{\mathbf{k},\mathbf{k}}.
\end{equation}
After inverting the matrix $\mathbb{D}_{2\mathbf{k}}$, this equation gives the amplitude of the second harmonic in terms of the amplitude of the first harmonic. Moreover, since the complex amplitude $\mathbf{\mathcal{E}}^{(2)}_{2\mathbf{k}}$ also encodes the phase information, the above equation also tells how the second harmonic is phase-locked with the fundamental.

\section{Scattering of three resonant on-shell waves}\label{sec:3waves}
In this section, we illustrate the general theory developed in Sec.~\ref{sec:fluid} with the
simplest nontrivial example where the spectrum contains exactly three resonant ``on-shell" waves. 
Without loss of generality, suppose the three waves satisfies the resonance conditions
\begin{eqnarray}
\label{eq:resonantK}
\mathbf{k}_{1}&=&\mathbf{k}_{2}+\mathbf{k}_{3},\\
\label{eq:resonantW}
\omega_{\mathbf{k}_1}&=&\omega_{\mathbf{k}_2}+\omega_{\mathbf{k}_3},
\end{eqnarray}
where all $\omega$'s are positive. The above resonance condition can also be written more compactly as $\theta_{\mathbf{k}_1}=\theta_{\mathbf{k}_2}+\theta_{\mathbf{k}_3}$. In this case, the spectrum $\mathbb{K}_1=\{\mathbf{k}_{1}, \mathbf{k}_{2}, \mathbf{k}_{3}, (\mathbf{k}\rightarrow-\mathbf{k})\}$. Using Eq.~(\ref{eq:K2}), we find the second order spectrum $\mathbb{K}_2=\{2\mathbf{k}_{1}, 2\mathbf{k}_{2}, 2\mathbf{k}_{3}, \mathbf{k}_{1}+\mathbf{k}_{2}, \mathbf{k}_{2}-\mathbf{k}_{3}, \mathbf{k}_{3}+\mathbf{k}_{1}, (\mathbf{k}\rightarrow-\mathbf{k})\}$. Notice that resonant waves, such as $\mathbf{k}_{1}=\mathbf{k}_{2}+\mathbf{k}_{3}$, are not contained in the second order spectrum $\mathbb{K}_2$. In this way, we avoid the ambiguous partition between $\mathbf{\mathcal{E}}^{(2)}_{\mathbf{k}}$, and $d_{t(1)}^{\mathbf{k}}\mathbf{\mathcal{E}}^{(1)}_\mathbf{k}$. In another word, all perturbative corrections to the first order amplitude $\mathbf{\mathcal{E}}^{(1)}_{\mathbf{k}}$ are accounted for by its slow derivatives.

Using the electric field equation (\ref{eq:E2s}), we can extract the ``off-shell" equations by matching coefficients of Fourier exponents. There are twelve ``off-shell" equations, six of which are complex conjugations of the following six ``off-shell" equations
\begin{eqnarray}
\mathbb{D}_{2\mathbf{k}_1}\mathbf{\mathcal{E}}^{(2)}_{2\mathbf{k}_1}&=&i\sum_{s}\mathbf{S}^{s}_{\mathbf{k}_1,\mathbf{k}_1},\\
\mathbb{D}_{2\mathbf{k}_2}\mathbf{\mathcal{E}}^{(2)}_{2\mathbf{k}_2}&=&i\sum_{s}\mathbf{S}^{s}_{\mathbf{k}_2,\mathbf{k}_2},\\
\mathbb{D}_{2\mathbf{k}_3}\mathbf{\mathcal{E}}^{(2)}_{2\mathbf{k}_3}&=&i\sum_{s}\mathbf{S}^{s}_{\mathbf{k}_3,\mathbf{k}_3},\\
\mathbb{D}_{\mathbf{k}_{1}+\mathbf{k}_{2}}\mathbf{\mathcal{E}}^{(2)}_{\mathbf{k}_{1}+\mathbf{k}_{2}} &=&i\sum_{s}\mathbf{S}^{s}_{\mathbf{k}_1,\mathbf{k}_2},\\
\mathbb{D}_{\mathbf{k}_{2}-\mathbf{k}_{3}}\mathbf{\mathcal{E}}^{(2)}_{\mathbf{k}_{2}-\mathbf{k}_{3}} &=&i\sum_{s}\mathbf{S}^{s}_{\mathbf{k}_2,-\mathbf{k}_3},\\
\mathbb{D}_{\mathbf{k}_{3}+\mathbf{k}_{1}}\mathbf{\mathcal{E}}^{(2)}_{\mathbf{k}_{3}+\mathbf{k}_{1}} &=&i\sum_{s}\mathbf{S}^{s}_{\mathbf{k}_3,\mathbf{k}_1}.
\end{eqnarray}
Since the dispersion tensor $\mathbb{D}_\mathbf{q}$ for ``off-shell" waves are non-degenerate, the second order amplitudes $\mathbf{\mathcal{E}}^{(2)}_{\mathbf{k}}$ can be found by simply inverting the above matrix equations, which gives the second order amplitudes in terms of the first order amplitudes.

Similarly, we can extract the ``on-shell" equations from the second order electric field equation (\ref{eq:E2s}). There are six ``on-shell" equations, three of which are complex conjugation of the following three ``on-shell" equations 
\begin{eqnarray}
\label{eq:onE1}
\omega_{\mathbf{k}_1}\mathbb{H}_{\mathbf{k}_1}d_{t(1)}^{\mathbf{k}_1}\mathbf{\mathcal{E}}^{(1)}_{\mathbf{k}_1} &=&\sum_{s}\mathbf{S}^{s}_{\mathbf{k}_2,\mathbf{k}_3},\\
\label{eq:onE2}
\omega_{\mathbf{k}_2}\mathbb{H}_{\mathbf{k}_2}d_{t(1)}^{\mathbf{k}_2}\mathbf{\mathcal{E}}^{(1)}_{\mathbf{k}_2} &=&\sum_{s}\mathbf{S}^{s}_{\mathbf{k}_1,-\mathbf{k}_3},\\
\label{eq:onE3}
\omega_{\mathbf{k}_3}\mathbb{H}_{\mathbf{k}_3}d_{t(1)}^{\mathbf{k}_3}\mathbf{\mathcal{E}}^{(1)}_{\mathbf{k}_3} &=&\sum_{s}\mathbf{S}^{s}_{\mathbf{k}_1,-\mathbf{k}_2}.
\end{eqnarray}
These ``on-shell" equations govern how the first order amplitudes $\mathbf{\mathcal{E}}^{(1)}_{\mathbf{k}}$ evolve on the slow scales due to scattering of the three waves. The left-hand-side of these equations is basically the passive advection of wave envelopes at the wave group velocities. The right-hand-side of these equations is redistribution of wave action and energy due to three-wave scattering.

\subsection{Action conservation of on-shell equations}\label{sec:action}
By the conservative nature of the redistribution process, the ``on-shell" equations (\ref{eq:onE1})-(\ref{eq:onE3}) conserve the total wave action $U/\omega$, as well as the total wave energy $U$. As will be proven in the next paragraph, the local conservation laws of wave actions are 
\begin{eqnarray}
\label{eq:action12}
d_{t(1)}^{\mathbf{k}_1}\frac{U_{\mathbf{k}_1}}{\omega_{\mathbf{k}_1}} +d_{t(1)}^{\mathbf{k}_2}\frac{U_{\mathbf{k}_2}}{\omega_{\mathbf{k}_2}}&=&0,\\
\label{eq:action23}
d_{t(1)}^{\mathbf{k}_3}\frac{U_{\mathbf{k}_3}}{\omega_{\mathbf{k}_3}}-d_{t(1)}^{\mathbf{k}_2}\frac{U_{\mathbf{k}_2}}{\omega_{\mathbf{k}_2}}&=&0,
\end{eqnarray}
where $U_{\mathbf{k}}$, given by Eq.~(\ref{eq:U}), is the energy of the linear wave with wave vector $\mathbf{k}$. The first conservation law (\ref{eq:action12}) implies that the total number of wave quanta in the incident wave and the scattered wave is a constant. This is intuitive because, in the absence of damping, whenever a quanta of the $\mathbf{k}_1$ mode is annihilated, it is consumed to create a quanta of the $\mathbf{k}_2$ mode. Analogously, the second conservation law (\ref{eq:action23}) says that whenever a quanta of the $\mathbf{k}_2$ mode is created, a quanta of the $\mathbf{k}_3$ mode must also be created by the three-wave process (\ref{eq:resonantK}). As a consequence of wave action conservation, the total wave energy is also conserved during resonant three-wave interaction
\begin{equation}
\label{eq:action123}
d_{t(1)}^{\mathbf{k}_1}U_{\mathbf{k}_1}+d_{t(1)}^{\mathbf{k}_2}U_{\mathbf{k}_2}+d_{t(1)}^{\mathbf{k}_3}U_{\mathbf{k}_3}=0.
\end{equation}
This local energy conservation law can be obtained by linearly combining Eqs.~(\ref{eq:action12}) and (\ref{eq:action23}), and use the frequency resonance condition (\ref{eq:resonantW}). The conservation of wave energy is also intuitive, because in the absence of damping and other waves, three-wave scattering can only redistribute energy among the three waves.

The above conservation laws 
can be proven by noting the following properties of the scattering strength $\mathbf{S}^{s}_{\mathbf{q},\mathbf{q}'}$. First, using formula (\ref{eq:S}) for the scattering strength, together with the quadratic identity (\ref{eq:F12}) of the forcing operator $\mathbb{F}$, we can obtain a simple expression for $\mathbf{S}^{s}_{\mathbf{k}_2,\mathbf{k}_3}$ 
\begin{eqnarray}
\label{eq:S23}
\nonumber
\mathbf{S}_{2,3}\!&=&\!\frac{e\omega_p^2\omega_1}{2m\omega_2\omega_3}\Big[\frac{(\mathbf{\mathcal{E}}_3\!\cdot\!\mathbb{F}_2\mathbf{\mathcal{E}}_2)(\mathbb{F}_1\mathbf{k}_3) \!+\!(\mathbf{\mathcal{E}}_2\!\cdot\!\mathbb{F}_3\mathbf{\mathcal{E}}_3)(\mathbb{F}_1\mathbf{k}_2)}{\omega_1}\\
&&\hspace{20pt}+\frac{(\mathbb{F}_3\mathbf{\mathcal{E}}_3)(\mathbf{k}_1\!\cdot\!\mathbb{F}_2\mathbf{\mathcal{E}}_2) \!-\!(\mathbb{F}_1\mathbf{\mathcal{E}}_3)(\mathbf{k}_3\!\cdot\!\mathbb{F}_2\mathbf{\mathcal{E}}_2)}{\omega_2}\\
\nonumber
&&\hspace{20pt}+\frac{(\mathbb{F}_2\mathbf{\mathcal{E}}_2)(\mathbf{k}_1\!\cdot\!\mathbb{F}_3\mathbf{\mathcal{E}}_3) \!-\!(\mathbb{F}_1\mathbf{\mathcal{E}}_2)(\mathbf{k}_2\!\cdot\!\mathbb{F}_3\mathbf{\mathcal{E}}_3)}{\omega_3}\Big],
\end{eqnarray}
where we have abbreviated $\omega_j:=\omega_{\mathbf{k}_j}$, $\mathcal{E}_{j}:=\mathcal{E}_{\mathbf{k}_j}^{(1)}$, $\mathbb{F}_j:=\mathbb{F}_{s,\mathbf{k}_j}$, and suppressed other species label $s$ for simplicity. The expression for $\mathbf{S}_{1,-3}$ can be obtained easily from Eq.~(\ref{eq:S23}) using the replacement rule $1\rightarrow 2, 2\rightarrow 1, 3\rightarrow-3$, where the minus sign is interpreted using notations (\ref{eq:notationz}) and (\ref{eq:notationa}). Similarly, to obtain the expression for $\mathbf{S}_{1,-2}$, we can replace $1\rightarrow 3, 2\rightarrow 1, 3\rightarrow-2$ in Eq.~(\ref{eq:S23}). Having obtained expressions for $\mathbf{S}_{2,3}$, $\mathbf{S}_{1,-3}$, and $\mathbf{S}_{1,-2}$, we can use the self-adjoint property (\ref{eq:Fadj}) of the forcing operator to show, by straight forward calculations, that the scattering strength for three resonant waves satisfies the following identities
\begin{eqnarray}
\label{eq:actionS12}
\frac{\mathbf{\mathcal{E}}_{1}\cdot\mathbf{S}^{*}_{2,3}}{\omega_{1}^2}+\frac{\mathbf{\mathcal{E}}^{*}_{2}\cdot\mathbf{S}_{1,-3}}{\omega_{2}^2}&=&0,\\
\label{eq:actionS23}
\frac{\mathbf{\mathcal{E}}^{*}_{2}\cdot\mathbf{S}_{1,-3}}{\omega_{2}^2} -\frac{\mathbf{\mathcal{E}}^{*}_{3}\cdot\mathbf{S}_{1,-2}}{\omega_{3}^2}&=&0.
\end{eqnarray}
Then the action conservation Eqs.~(\ref{eq:action12}) and (\ref{eq:action23}), as well as the energy conservation Eq.~(\ref{eq:action123}), are immediate consequences of the above identities. 

One may be puzzled by the expression (\ref{eq:S23}) for $\mathbf{S}_{2,3}$. After all, why $\mathbf{S}_{2,3}$ is given by those six particular combinations of vectors $\mathbb{F}_{\mathbf{q}}\mathbf{\mathcal{E}}_{\mathbf{q}'}$ and $\mathbb{F}_{\mathbf{q}}\mathbf{\mathbf{q}'}$, weighted by inner products $\mathbf{\mathcal{E}}_\mathbf{q}\!\cdot\!\mathbb{F}_{\mathbf{q}'}\mathbf{\mathcal{E}}_{\mathbf{q}'}$ and $\mathbf{q}\!\cdot\!\mathbb{F}_{\mathbf{q}'}\mathbf{\mathcal{E}}_{\mathbf{q}'}$, as well as signed frequencies $\pm1/\omega$? At first glance, there seems to be no obvious pattern. However, action conservation laws, given by Eqs.~(\ref{eq:actionS12}) and (\ref{eq:actionS23}), clearly indicate that $\mathbf{S}_{2,3}$, $\mathbf{S}_{1,-3}$, and $\mathbf{S}_{1,-2}$ are originated from a single term from the variational principle. In Sec.~\ref{sec:Lagrangian}, we will write down the Lagrangian that generate the three ``on-shell" equations (\ref{eq:onE1})-(\ref{eq:onE3}). From the Lagrangian, it will become obvious why Eq.~(\ref{eq:S23}) looks the way it is.

\subsection{Three-wave equations}
Before unveiling the deeper reason leading to the expression of the scattering strength, let us first extract a number of observable consequences of three-wave interactions. When one is not concerned with the vector dependence of the complex wave amplitude $\mathbf{\mathcal{E}}_\mathbf{k}$, the ``on-shell" equations (\ref{eq:onE1})-(\ref{eq:onE3}) can be written as three scalar equations, called the three-wave equations. To remove the vector dependence, let us decompose $\mathbf{\mathcal{E}}^{(1)}_\mathbf{k}=\mathbf{e}_\mathbf{k}\varepsilon_\mathbf{k}$, where $\mathbf{e}_\mathbf{k}$ is the complex unit vector satisfying $\mathbf{e}^*_\mathbf{k}\cdot\mathbf{e}_\mathbf{k}=1$. This decomposition is not unique due to the U(1) symmetry $\mathbf{e}_\mathbf{k}\rightarrow e^{i\alpha}\mathbf{e}_\mathbf{k}$ and $\varepsilon_\mathbf{k}\rightarrow e^{-i\alpha}\varepsilon_\mathbf{k}$. By requiring the scalar amplitude $\varepsilon_\mathbf{k}\in\mathbb{R}$ to be real valued, the symmetry group of the above decomposition is reduced to the $\mathbb{Z}_2$ symmetry $\varepsilon\rightarrow-\varepsilon$. The convective derivative of the complex wave amplitude 
\begin{equation}
d_{t(1)}^\mathbf{k}\mathbf{\mathcal{E}}^{(1)}_\mathbf{k}=\mathbf{e}_\mathbf{k}d_{t(1)}^\mathbf{k}\varepsilon_\mathbf{k}+\varepsilon_\mathbf{k} d_{t(1)}^\mathbf{k}\mathbf{e}_\mathbf{k},
\end{equation}
can be decomposed into change due to the scalar amplitude and the change due to the rotation of the complex unit vector.

The left-hand-sides of the ``on-shell" equations are closely related to the energy of the linear waves. Denote the dimensionless wave energy coefficient
\begin{equation}
\label{eq:Ucoef}
u_\mathbf{k}:=\frac{1}{2}\mathbf{e}_\mathbf{k}^\dagger\mathbb{H}_\mathbf{k}\mathbf{e}_\mathbf{k}.
\end{equation} 
Then the wave energy Eq.~(\ref{eq:U}) can be written as $U_\mathbf{k}=\epsilon_0u_\mathbf{k}\varepsilon_\mathbf{k}^2/2$. Notice that the energy coefficient $u_\mathbf{k}>0$ is always real and positive, because the matrix $\mathbb{H}_\mathbf{k}$ is Hermitian and positive definite. 
Taking inner product with $\mathbf{e}_\mathbf{k}^*$ on both sides of the ``on-shell" equations and sum the result with its Hermitian conjugate, we obtain $u_\mathbf{k}d_{t(1)}^{\mathbf{k}}\varepsilon_\mathbf{k}+\frac{1}{2}\varepsilon_\mathbf{k} d_{t(1)}^{\mathbf{k}}u_\mathbf{k}=\sum_s[\mathbf{e}_\mathbf{k}^\dagger\mathbf{S}_{\mathbf{q},\mathbf{q}'}^s/\omega_\mathbf{k}+\text{h.c.}]/4$.
From this expression, we see the combination $\varepsilon_\mathbf{k} u_\mathbf{k}^{1/2}$ will be particularly convenient. Let us nondimensionalize  the electric field amplitude by electron mass
\begin{equation}
a_\mathbf{k}:=\frac{e\varepsilon_\mathbf{k}}{m_ec\omega_\mathbf{k}}u_\mathbf{k}^{1/2}.
\end{equation}
Then the ``on-shell" equations can then be written in terms of the normalized wave amplitude $d_{t(1)}^{\mathbf{k}}a_\mathbf{k}=e/(4m_ec\omega_\mathbf{k} u_\mathbf{k}^{1/2})\sum_s(\mathbf{e}_\mathbf{k}^\dagger\mathbf{S}^s_{\mathbf{q},\mathbf{q}'}/\omega_\mathbf{k}+\text{h.c.}).$
From this equation, we see only the real part of $\mathbf{e}^\dagger\mathbf{S}$ affects how the amplitude change, while the imaginary part affects how the direction $\mathbf{e}$ rotates on the complex unit sphere.

The right-hand-sides of the ``on-shell" equations are originated from a single scattering term. As can be seen from identities (\ref{eq:actionS12}) and (\ref{eq:actionS23}), there exist some dimensionless scattering strength $\Theta^s$, such that
\begin{equation}
\label{eq:aRHS}
\frac{e_s\omega_{ps}^2}{2m_sc}\frac{\varepsilon_1\varepsilon_2^*\varepsilon_3^*}{\omega_1\omega_2\omega_3}\Theta^s:=-\frac{\mathbf{\mathcal{E}}_{1}\cdot\mathbf{S}^{*}_{2,3}}{\omega_{1}^2} =\frac{\mathbf{\mathcal{E}}^{*}_{2}\cdot\mathbf{S}_{1,\bar{3}}}{\omega_{2}^2} =\frac{\mathbf{\mathcal{E}}^{*}_{3}\cdot\mathbf{S}_{1,\bar{2}}}{\omega_{3}^2},
\end{equation}
where we have abbreviate $\varepsilon_j:=\varepsilon_{\mathbf{k}_j}$, 
and used the notation $\bar{j}=-j$. Using formula (\ref{eq:S23}) for $\mathbf{S}_{2,3}$, we see that the normalized scattering strength can be written as the summation of strengths of 
six scattering channels
\begin{eqnarray}
\label{eq:Theta3}
\nonumber
\Theta^s&=&\Theta_{1,\bar{2}\bar{3}}^s+\Theta_{\bar{2},\bar{3}1}^s+\Theta_{\bar{3},1\bar{2}}^s\\
&+&\Theta_{1,\bar{3}\bar{2}}^s+\Theta_{\bar{2},1\bar{3}}^s+\Theta_{\bar{3},\bar{2}1}^s,
\end{eqnarray}
where the normalized scattering strength due to each channel is given by the simple formula
\begin{equation}
\label{eq:Thetaijl}
\Theta_{i,jl}^s=\frac{1}{\omega_j}(c\mathbf{k}_i\cdot\mathbf{f}_{s,j})(\mathbf{e}_i\cdot\mathbf{f}_{s,l}) .
\end{equation}
In the above formula, the vector $\mathbf{f}_{s,j}$ is defined by $\mathbf{f}_{s,j}:=\mathbb{F}_{s,\mathbf{k}_j}\mathbf{e}_j$, and we have abbreviated $\mathbf{e}_j:=\mathbf{e}_{\mathbf{k}_j}$. In general, the normalized scattering strength $\Theta^s=\Theta^s_r+i\Theta^s_i$ contains both real and imaginary parts. In Sec.~\ref{sec:Lagrangian}, we will show that the normalized scattering strength $\Theta^s$ is related to the reduced \textit{S} matrix element of the quantized theory, and the six scattering channels correspond to the six ways of contracting a single interaction vertex. 

Having expressed both the left- and the right-hand-side of the ``on-shell" equations as scalars, we can now write down the three-wave equations
\begin{eqnarray}
\label{eq:3waves1}
d_{t(1)}^{\mathbf{k}_1}a_1&=&-\frac{\Gamma}{\omega_1}a_2a_3,\\
\label{eq:3waves2}
d_{t(1)}^{\mathbf{k}_2}a_2&=&\phantom{+}\frac{\Gamma}{\omega_2}a_3a_1,\\
\label{eq:3waves3}
d_{t(1)}^{\mathbf{k}_3}a_3&=&\phantom{+}\frac{\Gamma}{\omega_3}a_1a_2,
\end{eqnarray}
where $a_j:=a_{\mathbf{k}_j}$ are the real-valued normalized wave amplitudes, and $\Gamma$ is the coupling coefficient. Notice that due to the residual $\mathbb{Z}_2$ symmetry  $a_j\rightarrow-a_j$, the sign of $\Gamma$ is insignificant, as long as Eq.~(\ref{eq:3waves1}) has the opposite sign as Eqs.~(\ref{eq:3waves2}) and (\ref{eq:3waves3}). Combining Eqs.~(\ref{eq:aRHS})-(\ref{eq:Thetaijl}), the coupling coefficient is given by 
\begin{equation}
\label{eq:coupling}
\Gamma=\sum_s\frac{Z_s\omega_{ps}^2\Theta^s_r}{4M_s(u_1u_2u_3)^{1/2}},
\end{equation}
where $Z_s:=e_s/e$ is the normalized charge, $M_s:=m_s/m_e$ is the normalized mass of species $s$, and $u_j:=u_{\mathbf{k}_j}$ is the wave energy coefficient. As expected, only the real part $\Theta^s_r$ of the normalized scattering strength affects the wave amplitude. Also notice when density $n_{s0}\rightarrow 0$, coupling due to species $s$ vanishes as expected. The numerator of the coupling coefficient measures how strong the three waves are coupled by the scattering strength, 
and the denominator measures how energetically expensive to excite the linear waves, as measured by the wave energy coefficients.

It is instructive to count how many degrees of freedom does the three-wave coupling coefficient $\Gamma$ contains. For each wave, its 4-momentum is constrained by one dispersion relation. Once the 4-momentum is fixed, the wave polarization is determined by the dispersion tensor up to the wave amplitude, which $\Gamma$ does not dependent. Therefore, for each wave, there are three degrees of freedom. Now that the resonant conditions give another four constrains, there are in total $3\times3-4=5$ independent variables. Hence, in the absence of additional symmetry, the three-wave coupling coefficient $\Gamma$ is a function of five independent variables in a given plasma.

Once the coupling coefficient is obtained in a given situation, the nonlinear three-wave equations Eqs.~(\ref{eq:3waves1})-(\ref{eq:3waves3}) may be solved using a number of techniques. For the homogeneous problem, where the spatial derivatives are zero, the equations become a system of nonlinear ordinary differential equations, and the general solution are given by the Jacobi elliptic functions \cite{Jurkus60,Armstrong62}. Similarly, in one dimension, the steady state problem, where the time derivatives are zero, can also be solved in terms of the Jacobi elliptic functions \cite{Harvey75}. As a trivial extension, traveling wave solutions in one spatial dimension can also be found \cite{Armstrong70,Nozaki73,Ohsawa74}, using the coordinate transform $\xi=x-vt$. In addition to these periodic solutions, the nonlinear three-wave equations also has compact solutions, such as the N-soliton solutions \cite{Zakharov75,Turner88}. More general solutions may also be constructed using the inverse scattering method \cite{Ablowitz74,Kaup79}. In this paper, we will not be concerned with solving the three-wave equations, and only focus on calculating the coupling coefficient.

Without solving the three-wave equations, a number of experimental observables can already be extracted from the coupling coefficient. For example, $\Gamma$ can be related to the growth rate of parametric instabilities. Consider the parametric decay instability where a pump wave with frequency $\omega_1$ decays into two waves with frequencies $\omega_2$ and $\omega_3$. Suppose the pump has constant amplitude $a_1$, and the decay waves have no spatial variation. Then solving the linearized three-wave equations, we find $a_2$ and $a_3$ grow exponentially with rate
\begin{equation}
\label{eq:GrowthRate}
\gamma_0=\frac{|\Gamma a_1|}{\sqrt{\omega_2\omega_3}}.
\end{equation}
The experimentally observed linear growth rate will be somewhat different than $\gamma_0$ due to wave damping. Wave damping, both collisional and collisionless, can be taken into account by inserting a phenomenological damping term $\nu a$ into the left-hand-side of the three-wave equations. Solving the linearized equations, the growth rate, modified by wave damping, is
\begin{equation}
\label{eq:GrowthRateDamped}
\gamma=\sqrt{\gamma_0^2+\Big(\frac{\nu_2-\nu_3}{2}\Big)^2}-\frac{\nu_2+\nu_3}{2},
\end{equation} 
where $\nu_2$ and $\nu_3$ are the phenomenological damping rates of the two decay waves. In addition to wave damping, the experimentally observed growth rate can also be modified by frequency mismatch $\delta\omega=\omega_1-\omega_2-\omega_3$. When the frequency mismatch is much smaller than the spectral width of waves, the three waves can still couple almost resonantly. To find the growth rate in the presence of small $\delta\omega$, promote amplitude $a$ to be complex and change variable $\alpha_j:=a_je^{-it\delta\omega/2}$ for $j=2$ and $3$. This change of variable is equivalent to modifying the damping rates to $\nu'_2:=\nu_2+i\delta\omega/2$ and $\nu_3^{'*}:=\nu_3-i\delta\omega/2$. Therefore, the growth rate of parametric decay instability, modified by both weak damping and small frequency mismatch is
\begin{equation}
\gamma'=\sqrt{\gamma_0^2+\Big(\frac{\nu_2-\nu_3+i\delta\omega}{2}\Big)^2}-\frac{\nu_2+\nu_3}{2}.
\end{equation} 
The frequency mismatch $\delta\omega$ not only introduces amplitude modification, but also results in phase modification. In the following discussions, we shall only be concerned with the growth rate $\gamma_0$ as observable, ignoring wave damping and frequency mismatch.

\section{Lagrangian of three-wave interaction}\label{sec:Lagrangian}
Now that we know how the coupling coefficient can be related to experimental observables, let us unveil why its formula looks the ways it is. Recall in the previous section, we show that the three-wave scattering strengths $\mathbf{S}_{\mathbf{q},\mathbf{q}'}$ satisfies the action conservation laws. 
Motivated by these conservation laws, here in this section, we show that the three ``on-shell" equations (\ref{eq:onE1})-(\ref{eq:onE3}) can be derived from a classical three-wave Lagrangian. More importantly, we will show that all terms in the classical interaction Lagrangian arise from essentially one term after quantizing the Lagrangian. 

To write down the Lagrangian, it is more convenient to use the gauge field $A^{\mu}$ instead of the electric or magnetic fields. Since we will later quantize the Lagrangian, it is convenient to use the temporal gauge $A^{0}=0$. In temporals gauge, the electric field is related to the vector potential by 
\begin{eqnarray}
\mathbf{A}_{\mathbf{k}}=\frac{\mathcal{E}_{\mathbf{k}}}{\omega_{\mathbf{k}}},
\end{eqnarray} 
which, in the natural units $\hbar=c=1$, has the dimension of energy $M$. Similarly, we can dimensionalize the wave energy operator $\mathbb{H}$ by
\begin{equation}
\Lambda_{\mathbf{k}}:=\omega_{\mathbf{k}}\mathbb{H}_{\mathbf{k}},
\end{equation}
which then has the dimension of energy $M$ as it should. 

Having defined the necessary operators, we can now write down the classical three-wave action for the three ``on-shell" equations
\begin{equation}
\label{eq:Sc}
S_{c}=\int d^4x_{(1)}(\mathcal{L}_{c0}+\mathcal{L}_{cI}),
\end{equation}
where the integrations over space and time are on the slow scales $x_{(1)}$ and $t_{(1)}$. Abbreviating the subscripts $\mathbf{k}_j$ as $j$, the Lagrangian of freely advecting wave envelopes
\begin{equation}
\mathcal{L}_{c0}=\sum_{j=1}^3\mathbf{A}_j^*\cdot i\Lambda_jd^j_{t(1)}\mathbf{A}_j,
\end{equation}
where the complex amplitude $\mathbf{A}_j(x_{(1)},t_{(1)})$ is a function of the slow spatial and temporal scales, and the advective derivative $d^j_{t(1)}$ is defined by Eq.~(\ref{eq:dt1}). It is easy to show that $\mathcal{L}_{c0}$ gives rise to a real-valued action $S_{c0}$ after integrating by part. The second term in the classical action [Eq.~(\ref{eq:Sc})] is the three-wave interaction Lagrangian 
\begin{equation}
\label{eq:LcI}
\mathcal{L}_{cI}=-i(\Xi-\Xi^*),
\end{equation}
which is obviously real-valued. Using Eq.~(\ref{eq:aRHS}), the three waves interact through the coupling 
\begin{equation}
\label{eq:Xi}
\Xi=A_1 A_2^* A_3^*\sum_s\frac{e_s\omega_{ps}^2}{2m_sc}\Theta^s,
\end{equation}
where $\Theta^s$ is the normalized scattering strength [Eq.~(\ref{eq:Theta3})], and the $A$'s are the scalar amplitudes of the three waves. Clearly, the coupling $\Xi$ has mass dimension $M^4$, and hence the action $S_{cI}$ is dimensionless in the natural unit as expected. 
Now that we have written down the Lagrangian, we can find the classical equations of motion by taking variations with respect to $\mathbf{A}_1$, $\mathbf{A}_2$, and $\mathbf{A}_3$, or equivalently, their independent complex conjugates. Using the self-adjointness [Eq.~(\ref{eq:Fadj})] of the forcing operator, it is straight forward to verify that the three ``on-shell" equations (\ref{eq:onE1})-(\ref{eq:onE3}) are the resultant equations.

The classical three-wave Lagrangian $\mathcal{L}_c=\mathcal{L}_{c0}+\mathcal{L}_{cI}$ has U(1) symmetries, which lead to the action conservation laws. For example, the Lagrangian is invariant under the following global U(1) transformation
\begin{eqnarray}
\mathbf{A}_1&\rightarrow& e^{i\alpha}\mathbf{A}_1,\\
\mathbf{A}_2&\rightarrow& e^{i\alpha}\mathbf{A}_2,\\
\mathbf{A}_3&\rightarrow& \mathbf{A}_3,
\end{eqnarray}
where $\alpha$ is an arbitrary real constant. Under the above transformation, the infinitesimal variation of the Lagrangian is zero $\delta\mathcal{L}_{c}=0$, while the infinitesimal variation $\delta\mathbf{A}_1=i\alpha\mathbf{A}_1$, $\delta\mathbf{A}_2=i\alpha\mathbf{A}_2$, and $\delta\mathbf{A}_3=0$, giving rise to a Noether's current. In fact, we have an even stronger symmetry $\delta\Xi=0$ for any $\alpha$. Therefore this U(1) symmetry leads to the identity
\begin{equation}
\mathbf{A}_1\cdot\frac{\delta\Xi}{\delta\mathbf{A}_1}-\mathbf{A}_2^*\cdot\frac{\delta\Xi}{\delta\mathbf{A}_2^*}=0,
\end{equation}
which is exactly the action conservation law Eq.~(\ref{eq:actionS12}). Using similar arguments, other action conservation laws can be derived from other global U(1) symmetries.

The large number of terms contained in the classical Lagrangian can be reduced to essentially two terms when we quantized the Lagrangian, in which the gauge field becomes real valued. Before introducing the quantized Lagrangian, it is helpful to review the second quantization notations. For simplicity, we will omit the subscripts for the slow spatial and temporal variables $x_{(1)}$ and $t_{(1)}$, with the implied understanding that all spatial and temporal dependences are on the full scales. Let us promote the gauge field $\mathbf{A}$ to quantized operator
\begin{equation}
\label{eq:Ahat}
\hat{\mathbf{A}}:=\int\frac{d^3\mathbf{k}}{(2\pi)^3}\frac{1}{\sqrt{2\omega_{\mathbf{k}}}}\Big(\mathbf{e}_{\mathbf{k}}\hat{a}_{\mathbf{k}}e^{-ikx}+\mathbf{e}_{\mathbf{k}}^*\hat{a}^\dagger_{\mathbf{k}}e^{ikx}\Big),
\end{equation}
where $kx:=\omega_{\mathbf{k}}t-\mathbf{k}\cdot\mathbf{x}$ is the Minkowski inner product, $\mathbf{e}_\mathbf{k}$ is the unit polarization vector, and the summation over branches of the dispersion relation is implied. The annihilation operator $\hat{a}_{\mathbf{k}}$ and the creation operator $\hat{a}^\dagger_{\mathbf{k}}$ satisfies the canonical commutation relations for bosons, where the nontrivial commutator is
\begin{equation}
[\hat{a}_{\mathbf{p}},\hat{a}^\dagger_{\mathbf{k}}]=(2\pi)^3\delta^{(3)}(\mathbf{p}-\mathbf{k}).
\end{equation} 
Using the standard normalization, the single boson state 
\begin{equation}
|\mathbf{k}\rangle:=\sqrt{2\omega_{\mathbf{k}}}\hat{a}^\dagger_{\mathbf{k}}|0\rangle,
\end{equation}
where $|0\rangle$ is the vacuum state. Then we have the following Wick contractions
\begin{eqnarray}
\contraction{}{\hat{\mathbf{A}}}{|}{\mathbf{k}}
\hat{\mathbf{A}}|\mathbf{k}\rangle&=&\mathbf{e}_\mathbf{k}e^{-ikx},\\
\contraction{\langle}{\mathbf{k}}{|}{\hat{\mathbf{A}}}
\langle\mathbf{k}|\hat{\mathbf{A}}&=&\mathbf{e}_\mathbf{k}^*e^{ikx}.
\end{eqnarray}
Let us also promote the displacement operator for species $s$ to act on the operator $\hat{\mathbf{A}}$ by
\begin{equation}
\hat{\Pi}_{s}\hat{\mathbf{A}}\!:=\!i\!\int\!\frac{d^3\mathbf{k}}{(2\pi)^3}\frac{1}{\sqrt{2\omega_{\mathbf{k}}}}\Big(\!\frac{\mathbb{F}_{s,\mathbf{k}}\mathbf{e}_{\mathbf{k}}}{\omega_{\mathbf{k}}}\hat{a}_{\mathbf{k}}e^{-ikx}\!-\!\frac{\mathbb{F}_{s,\mathbf{k}}^*\mathbf{e}_{\mathbf{k}}^*}{\omega_{\mathbf{k}}}\hat{a}^\dagger_{\mathbf{k}}e^{ikx}\!\Big),
\end{equation}
where the minus sign in front of the second term comes from notation Eq.~(\ref{eq:notationa}). Taking time derivative of the displacement operator, $\partial_t(\hat{\Pi}_{s}\hat{\mathbf{A}})$ is the velocity operator for species $s$, which is proportional to the current operator. 

Now we are ready to write down the quantized Lagrangian, which contains a kinetic term and a single three-wave coupling term
\begin{equation}
\label{eq:Lagrangian}
\mathcal{L}=\hat{\mathbf{A}}^{\dagger}i\Lambda d_t \hat{\mathbf{A}} -\sum_s\frac{e_s\omega_{ps}^2}{2m_s}(\hat{\Pi}_{s}\hat{\mathbf{A}})_i(\partial_i\hat{\mathbf{A}}_j)\partial_t(\hat{\Pi}_{s}\hat{\mathbf{A}})_j.
\end{equation}
Here, the $i$ and $j$ indices in the second term are the spatial indices, and summation over repeated indices is assumed. The first term $\mathcal{L}_0$ closely resembles the kinetic term of quantum electrodynamics (QED), with the Dirac spinor replaced by the gauge field, and the Dirac gamma matrices replaced by the $\Lambda$ energy matrix. The second term $\mathcal{L}_I$ is the three-wave interaction Lagrangian, which is nonvanishing only if the background density of some species $s$ is nonzero. 
Notice that the three-wave interaction is nonrenormalizable, which is not unexpected in an effective field theory.

To make sense of the quantized Lagrangian, we recognize that the displacement $\hat{\Pi}_{s}\hat{\mathbf{A}}$ is proportional to the polarization density $\mathbf{P}$, and the velocity $\partial_t(\hat{\Pi}_{s}\hat{\mathbf{A}})$ is proportional to the current density $\mathbf{J}$. Therefore, the three-wave interaction Lagrangian is of the form $\mathcal{L}_I\propto P^i(\partial_i\mathbf{A}_j)J^j$, where the polarization and current density are determined by linear response. Although one may not have guessed this form of the interaction Lagrangian, it makes the following intuitive sense: in the absence of the third wave, the electromagnetic field interacts with the particle fields through $\mathbf{A}_jJ^j$ in the temporal gauge; now when the third wave is present, it modulates the medium through which the electromagnetic field advects, giving rise to the $P^i(\partial_i\mathbf{A}_j)J^j$ interaction. In this interaction term, there is no reason why a particular wave should only be responsible for $\mathbf{P}$, $\mathbf{A}$, or $\mathbf{J}$. Therefore, the three waves can switch their roles, and the total interaction is given by linear superpositions of all possible permutations.

To see how the quantized Lagrangian, with the linear superposition principle built in, gives rise to the classical Lagrangian, let us compute the \textit{S} matrix element of three-wave decay $\mathbf{k}_1\rightarrow\mathbf{k}_2+\mathbf{k}_3$. The \textit{S} matrix element 
\begin{equation}
\langle\mathbf{k}_2,\mathbf{k}_3|i\mathcal{L}_I|\mathbf{k}_1\rangle=i\mathcal{M} e^{i(k_2+k_3-k_1)x},
\end{equation}
where the reduced matrix element $i\mathcal{M}$ can be represented using Feynman diagrams
\begin{fmffile}{w3}
	\begin{eqnarray}
	i\mathcal{M}=
	\begin{gathered}
	\begin{fmfgraph*}(40,40)
	\fmfkeep{w3}
	\fmfleft{i1}
	\fmfright{o2,o3}
	\fmf{photon}{i1,v1}
	\fmf{photon}{v2,o2}
	\fmf{photon}{v3,o3}
	\fmf{fermion}{v1,v3}
	\fmf{plain}{v1,v2}
	\fmfdot{v2,v3}
	\fmfv{label=$1$,label.angle=-120,label.dist=6}{v1}
	\fmfv{label=$2$,label.angle=-120,label.dist=6}{v2}
	\fmfv{label=$3$,label.angle=120,label.dist=6}{v3}
	\end{fmfgraph*}
	\end{gathered}+\text{5 permutations}.
	\end{eqnarray}
\end{fmffile}Since there are three external boson lines, each connecting to one of the three vertices, there are in total $3!=6$ Feynman diagrams. In the above Feynman diagram, interaction vertex to which 1 is connected to is the usual QED vertex, whereas vertices 2 and 3 appear only when there are background particle fields \cite{Shi16}. The arrow between vertices 1 and 3 indicates the direction of momentum flow, and also labels which vertex does the $\partial_t$ derivative acts on. The above Feynman diagram corresponds to the particular Wick contraction
\begin{eqnarray}
\label{eq:Feynman}
\begin{gathered}
\nonumber
\fmfreuse{w3}
\end{gathered}\hspace{-10pt}&=&-
\contraction[1.5ex]{i\frac{e_s\omega_{ps}^2}{2m_s} \langle}{\mathbf{k}_2}{,\mathbf{k}_3|(\hat{\Pi}}{_{s}\hat{\mathbf{A}})}
\bcontraction{i\frac{e_s\omega_{ps}^2}{2m_s} \langle\mathbf{k}_2,}{\mathbf{k}_3}{|(\hat{\Pi}_{s}\hat{\mathbf{A}})_j(\partial_j\hat{\mathbf{A}}_l)\partial_t(\hat{\Pi}}{_{s}\hat{\mathbf{A}}}
\contraction{i\frac{e_s\omega_{ps}^2}{2m_s} \langle\mathbf{k}_2,\mathbf{k}_3|(\hat{\Pi}_{s}\hat{\mathbf{A}})_j(\partial}{_j\hat{\mathbf{A}}_l}{)\partial_t(\hat{\Pi}_{s}\hat{\mathbf{A}})_l|}{\mathbf{k}}
i\frac{e_s\omega_{ps}^2}{2m_s} \langle\mathbf{k}_2,\mathbf{k}_3|(\hat{\Pi}_{s}\hat{\mathbf{A}})_j(\partial_j\hat{\mathbf{A}}_l)\partial_t(\hat{\Pi}_{s}\hat{\mathbf{A}})_l|\mathbf{k}_1\rangle\\
&=&i\frac{e_s\omega_{ps}^2}{2m_sc}\Theta_{1,\bar{2}\bar{3}}^s.
\end{eqnarray} 
Summing with the other five Feynman diagrams, the reduced \textit{S} matrix element in the quantum theory is related to the normalized scattering strength in the classical theory by the simple relation
\begin{equation}
\label{eq:MTheta}
\mathcal{M}=\sum_s\frac{e_s\omega_{ps}^2}{2m_sc}\Theta^s.
\end{equation} 
From the Lagrangian perspective, the classical three-wave coupling is related to the quantized interaction through the \textit{S} matrix 
\begin{equation}
i\Xi=A_1A_2^*A_3^*\langle\mathbf{k}_2,\mathbf{k}_3|i\mathcal{L}_I|\mathbf{k}_1\rangle e^{i(k_1-k_2-k_3)x}.
\end{equation}
Using the above relation, 
we immediately recovers the classical three-wave coupling by computing the \textit{S} matrix element using the quantized Lagrangian. Alternatively, one may simply regard Lagrangian Eq.~(\ref{eq:Lagrangian}) as a classical Lagrangian, and substitute Eq.~(\ref{eq:Ahat})
as the spectral expansion of the gauge field. Then after integrating over spacetime, $\int d^4x\exp[i(k_1-k_2-k_3)x]=(2\pi)^4\delta^{(4)}(k_1-k_2-k_3)$ will select out the six resonate terms from the interaction Lagrangian.

Now that we understand how the classical theory and the quantized theory are connected, we may postulate that the three-wave coupling always arises from the $P^i(\partial_i\mathbf{A}_j)J^j$ term in the effective Lagrangian, regardless of the plasma model that is used to calculate the linear response. 
In the cold fluid model, the linear response is expressed in terms of the forcing operator $\mathbb{F}$. By modifying this operator to include thermal or even quantum effects, and plugging it into the formalism we have developed, the three-wave scattering strength may be evaluated immediately. Having obtained the normalized scattering strength, as well as the wave energy coefficients in that particular plasma model, one can then compute the three-wave coupling coefficient using Eq.~(\ref{eq:coupling}). We have thus conjectured a prescription for computing three-wave coupling, without the need for going through the perturbative solution of the equations. The coupling coefficient then enters the three-wave equation, which governs the evolution of the envelopes of the three waves.

\section{Scattering of quasi-transverse and quasi-longitudinal waves}\label{sec:quasi}
\begin{figure}[t]
	\includegraphics[angle=0,width=5cm]{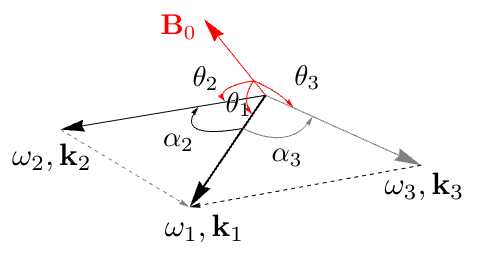}
	\caption{The most general geometry of three-wave scattering in an uniform plasma with a constant magnetic field. The three wave vectors $\mathbf{k}_1=\mathbf{k}_2+\mathbf{k}_3$ are in the same plane, and are at angles $\theta$'s with respect to the magnetic field.}
	\label{fig:ThreeWave}
\end{figure}

The three-wave coupling coefficient (\ref{eq:coupling}) can be readily evaluated in cold fluid model using wave energy coefficient Eq.~(\ref{eq:Ucoef}) and normalized scattering strength Eq.~(\ref{eq:Thetaijl}). In the most general geometry (Fig.~\ref{fig:ThreeWave}), we need to ensure that the resonant conditions Eqs.~(\ref{eq:resonantK}) and (\ref{eq:resonantW}) are satisfied by three otherwise arbitrary ``on-shell" waves. The evaluation becomes particularly easy when waves are either quasi-transverse (\textit{T}) or quasi-longitudinal (\textit{L}). In these situations, the wave dispersion relations are simplified, and hence matching resonance conditions becomes an easy task. Moreover, for both \textit{T} and \textit{L} waves, the wave polarization vector $\mathbf{e}$ are at special angles with the wave vector $\mathbf{k}$, so that the expressions for the wave energy and scattering strength can be further simplified. It is possible that \textit{T} and \textit{L} waves couple with other waves that have both electrostatic and electromagnetic components, but in this section, we will only give examples where all three waves are either \textit{T} or \textit{L} waves.

Although there are in general four different three-wave triplets: $\{T,T,T\}$, $\{T,T,L\}$, $\{T,L,L\}$, and $\{L,L,L\}$, only two of these triplets can couple resonantly. 
From Appendix~\ref{app:Linear}, we know the \textit{T} waves are electromagnetic waves with $\omega\gg\omega_p,|\Omega_e|$, and the \textit{L} waves are electrostatic waves with $\omega\rightarrow\omega_r$, for some resonance $\omega_r$. Since the frequency of \textit{T} waves are much higher than the frequency of \textit{L} waves, only the following types of interactions can match frequency resonance 
\begin{eqnarray}
&&T\rightleftharpoons T+L,\\
&&L\rightleftharpoons L+L.
\end{eqnarray}
A typical scenario for the \textit{TTL} interaction is the scattering of lasers. For example, an incident lasers is scattered inelastically by some plasma waves and thereafter propagates in some other direction with shifted frequency. Similarly, a typically scenario for the \textit{LLL} interaction is the scattering of antenna waves. For example, a plasma wave, launched by some antenna array, can decay into two other plasma waves propagating in some other directions. In what follows, we will consider these two scenarios in details.

\subsection{$T\rightleftharpoons T+L$ scattering}\label{sec:TTL}
Consider the decay of a pump laser ($\omega_1$) into a scattered laser ($\omega_2$) and a plasma wave ($\omega_3$). Since the frequency $\omega_{1,2}\gg\Omega_s$, the magnetization ratio $\beta_{1,2}\simeq 0$ and the magnetization factor $\gamma_{1,2}\simeq 1$ for any species. Consequently, the forcing operator $\mathbb{F}_{1,2}\simeq\mathbb{I}$ are approximately the identity operator, and the lasers are therefore transverse electromagnetic waves. As for the plasma wave, using the quasi-longitudinal approximation $\mathbf{e}_3\simeq\hat{\mathbf{k}}_3$, the inner products is purely real
\begin{equation}
\label{eq:TTL_inner}
\hat{\mathbf{k}}_{3}\cdot\hat{\mathbf{f}}^*_{s,3}\simeq\hat{\mathbf{k}}_{3}\cdot\mathbb{F}_{s,3}\hat{\mathbf{k}}_3=\gamma_{s,3}^2(1-\beta_{s,3}^2\cos^2\theta_3),
\end{equation}
where $\theta_3$ is the angle between $\mathbf{k}_3$ and $\mathbf{b}$ as shown in Fig.~\ref{fig:ThreeWave}, and $\hat{\mathbf{k}}_3$ is the unit vector along $\mathbf{k}_3$ direction. With these basic setup, we can readily evaluate Eq.~(\ref{eq:coupling}), the coupling coefficient.

Let us first calculate the wave energy coefficients Eq.~(\ref{eq:Ucoef}), which enters the denominator of the coupling coefficient. Since $\mathbb{F}_{1,2}\simeq\mathbb{I}$, the wave energy coefficients for the lasers are simply
\begin{equation}
\label{eq:TTL_U12}
u_1\simeq u_2\simeq 1.
\end{equation}
As for the plasma wave, after taking the frequency derivative in Eq.~(\ref{eq:Hk}), the wave energy coefficient for quasi-longitudinal wave is
\begin{equation}
\label{eq:TTL_U3}
u_3\simeq 1+\sum_s\frac{\omega_{ps}^2}{\omega_3^2}\gamma_{s,3}^4\beta_{s,3}^2\sin^2\theta_3.
\end{equation}
As expected, $u_3$ is always positive, although $\gamma_{s,3}^2$ can be either positive or negative, depending on whether $\beta_{s,3}$ is either smaller or larger than one.

To find the normalized scattering strength Eq.~(\ref{eq:Theta3}), which enters the numerator of $\Gamma$, we again use the fact $\omega_{1,2}\gg\omega_3$. Since the wave vectors are comparable in magnitudes, the dominant terms of the coupling strength are
the two terms proportional to $1/\omega_3$, 
if the inner product $\mathbf{e}_1\cdot\mathbf{f}_{2}^*\simeq\mathbf{f}_1\cdot\mathbf{e}_{2}^*\simeq\mathbf{e}_1\cdot\mathbf{e}_{2}^*$ is of oder unity. Using the resonance condition $\mathbf{k}_1-\mathbf{k}_2=\mathbf{k}_3$, the dominant term of the\textit{TTL} scattering strength
\begin{equation}
\label{eq:TTL_THETAr}
\Theta^s\simeq-\frac{ck_3}{\omega_3} (\hat{\mathbf{k}}_{3}\cdot\mathbb{F}_{s,3}\hat{\mathbf{k}}_3) (\mathbf{e}_1\cdot\mathbf{e}_{2}^*),
\end{equation}
where the inner product $\hat{\mathbf{k}}_{3}\cdot\mathbb{F}_{s,3}\hat{\mathbf{k}}_3$ is given explicitly by Eq.~(\ref{eq:TTL_inner}). Now that we have simplified both the denominator and the numerator of Eq.~(\ref{eq:coupling}), a simple formula for the three-wave coupling coefficient $\Gamma$ can be obtained.

Having obtained an explicit formula for the coupling coefficient, we can use it to obtain expressions for experimental observables. For example, the linear growth rate $\gamma_0$ [Eq.~(\ref{eq:GrowthRate})] can be decomposed as
\begin{equation}
\gamma_0=\gamma_R|\mathcal{M}_T|,
\end{equation}
where $\gamma_R$ is the backward Raman growth rate when the plasma is unmagnetized 
\begin{equation}
\gamma_R=\frac{\sqrt{\omega_1\omega_p}}{2}\Big|a_1\text{Re}(\mathbf{e}_1^*\cdot\mathbf{e}_2)\Big|,
\end{equation}
and $\mathcal{M}_T$ is the normalized growth rate of the \textit{TTL} scattering. The normalized growth rate is proportional to the  coupling coefficient $\Gamma=\omega_p^2\mu/4$ up to some kinematic factor
\begin{equation}
\label{eq:MT}
\mathcal{M}_T=\frac{1}{2}\Big(\frac{\omega_p^3}{\omega_1\omega_2\omega_3}\Big)^{1/2}\mu_{T},
\end{equation}
where the normalized coupling coefficient $\mu_T$ is given by
\begin{equation}
\label{eq:uT}
\mu_T\simeq \sum_s\frac{Z_s}{M_s}\frac{\omega_{ps}^2}{\omega_p^2}\frac{ck_3}{\omega_3}\frac{\hat{\mathbf{k}}_{3}\cdot\mathbb{F}_{s,3}\hat{\mathbf{k}}_3}{u_3^{1/2}},
\end{equation}
in the \textit{TTL} approximation. In the unmagnetized limit $B_0\rightarrow 0$, 
we have $\beta_3\rightarrow 0$ and $\gamma_3\rightarrow 1$. Since ion mass is much larger than electron mass, we have $\mu_T\rightarrow -ck_3/\omega_3$. Moreover, since the lasers can only couple through the Langmuir wave in cold unmagnetized plasma, we have $\omega_3\rightarrow\omega_p$. Then the normalized growth rate $\mathcal{M}_T\rightarrow ck_3/2\sqrt{\omega_1\omega_2}$. Finally, in backward scattering geometry $ck_3=ck_1+ck_2\simeq\omega_1+\omega_2\simeq2\omega_0$, where we have denoted $\omega_0:=\omega_1\simeq\omega_2$. We see $\mathcal{M}_T\rightarrow 1$ in the unmagnetized limit as expected. 

The normalized growth rate becomes particularly simple when waves propagate at special angles. For example, consider the situation where the three waves propagate along the magnetic field $\mathbf{B}_0$, and the plasma wave $\omega_3=\omega_p$ is the Langmuir wave. Since $\gamma_{s,3}^2$ remains finite as $\theta_3\rightarrow 0$, the normalized growth rate for collimated parallel wave propagation is
\begin{equation}
\label{eq:MTP}
\mathcal{M}_{T\parallel}^{P}\simeq-\frac{1}{2}\frac{ck_3}{\sqrt{\omega_1\omega_2}},
\end{equation}
where we have used $M_i\gg1$ to drop the summation over species. The above is exactly the same as the unmagnetized result, which is expected because the plasma wave is not affected by the parallel magnetic field. 

To give another simple example, consider the situation where the three waves are collimated and propagate perpendicular to the magnetic field $\mathbf{B}_0$. In cold electron-ion plasma, there are two \textit{L} waves in the perpendicular direction: the upper-hybrid (\textit{UH}) wave and the lower-hybrid (\textit{LH}) wave. Let us first consider scattering mediated by the \textit{UH} wave $\omega_3\simeq\omega_{UH}\simeq\sqrt{\omega_p^2+\Omega_e^2}$. In this situation, the magnetization factor $\gamma_{3,e}^2\simeq\omega_{UH}^2/\omega_p^2$ and $\gamma_{3,i}^2\simeq 1$. Since $M_i\gg1$, the dominant contribution for both the wave energy coefficient and the scattering strength comes from electrons. The wave energy coefficient $u_3\simeq\omega_{UH}^2/\omega_p^2$, and the normalized coupling coefficient $\mu_T\simeq-ck_3/\omega_p$. Therefore, the normalized growth rate for collimated perpendicular wave propagation mediated by the \textit{UH} wave is
\begin{equation}
\label{eq:MTUH}
\mathcal{M}_{T\perp}^{UH}\simeq-\frac{1}{2}\frac{ck_3}{\sqrt{\omega_1\omega_2}}\bigg(\frac{\omega_p}{\omega_{UH}}\bigg)^{1/2}.
\end{equation}
Similarly, let us consider scattering mediated by the \textit{LH} wave $\omega_3\simeq\omega_{LH}\simeq\sqrt{|\Omega_e|\Omega_i}\omega_p/\omega_{UH}$. Since the \textit{LH} frequency satisfies $\Omega_i\ll\omega_{LH}\ll|\Omega_e|$, the magnetization ratios $\beta_{3,e}\gg1$ and $\beta_{3,i}\ll1$. Consequently, the magnetization factor $\gamma_{3,e}\simeq-1/\beta_{3,e}^2$ and $\gamma_{3,i}\simeq 1$. When $\omega_p\sim|\Omega_e|$ are comparable, electron contributions again dominate. The wave energy coefficient $u_3\simeq\omega_{UH}^2/\Omega_e^2$, and the normalized coupling coefficient $\mu_T\simeq ck_3\omega_{LH}/\omega_{UH}|\Omega_e|$. Therefore, the normalized growth rate for \textit{LH} wave mediation in the collimated perpendicular geometry is
\begin{equation}
\label{eq:MTLH}
\mathcal{M}_{T\perp}^{LH}\simeq\frac{1}{2}\frac{ck_3}{\sqrt{\omega_1\omega_2}} \frac{\omega_p^{3/2}\omega_{LH}^{1/2}}{\omega_{UH}|\Omega_e|}.
\end{equation}
The above examples recover results known in \cite{Grebogi80}, who analyze the same problem in the restricted geometry where the the waves are collimated and propagate perpendicular to the magnetic field.

In more general geometry, where the waves are not collimated and propagate at angles with respect to the magnetic field, we can evaluate the normalized growth rate using the following procedure, mimicking what happens in an actual experiment where the plasma density and magnetic field strength are known. First, we shine a laser with frequency $\omega_1$ into the plasma at some angle $\theta_1$ with respect to the magnetic field. Then the wave vector $k_1$ is known from the dispersion relation. Second, we observe the scattered laser using some detector placed at angle $\theta_2$ with respect to the magnetic field, and point the detector at angle $\alpha_2$ with respect to the incoming laser. Suppose the detector can measure the frequency $\omega_2$ of the scattered laser, then from this frequency information, we immediately know $k_2$ from the dispersion relation, as well as $\omega_3=\omega_1-\omega_2$ from the resonance condition. Next, we can calculate $k_3=\sqrt{k_1^2+k_2^2-2k_1k_2\cos\alpha_2}$ using the resonance condition. Finally, we can determine $\theta_3$ by inverting $\omega_3=\omega_r(\theta_3)$, where $\omega_r$ is the angle-dependent resonance frequency. Using this procedure, the normalized growth rate can be readily evaluated numerically.

Conversely, we may diagnose the plasma density and magnetic field using information measured from laser scattering experiment. Using measured scattering intensities, which can be related to the growth rate, one may be able to fit plasma parameters such that the angular dependence of $\mathcal{M}_{T}$, measured from experiments, matches what is expected from the theory.

\subsubsection{Parallel pump}\label{sec:TTLPara}
\begin{figure}[t]
	\includegraphics[angle=0,width=8cm]{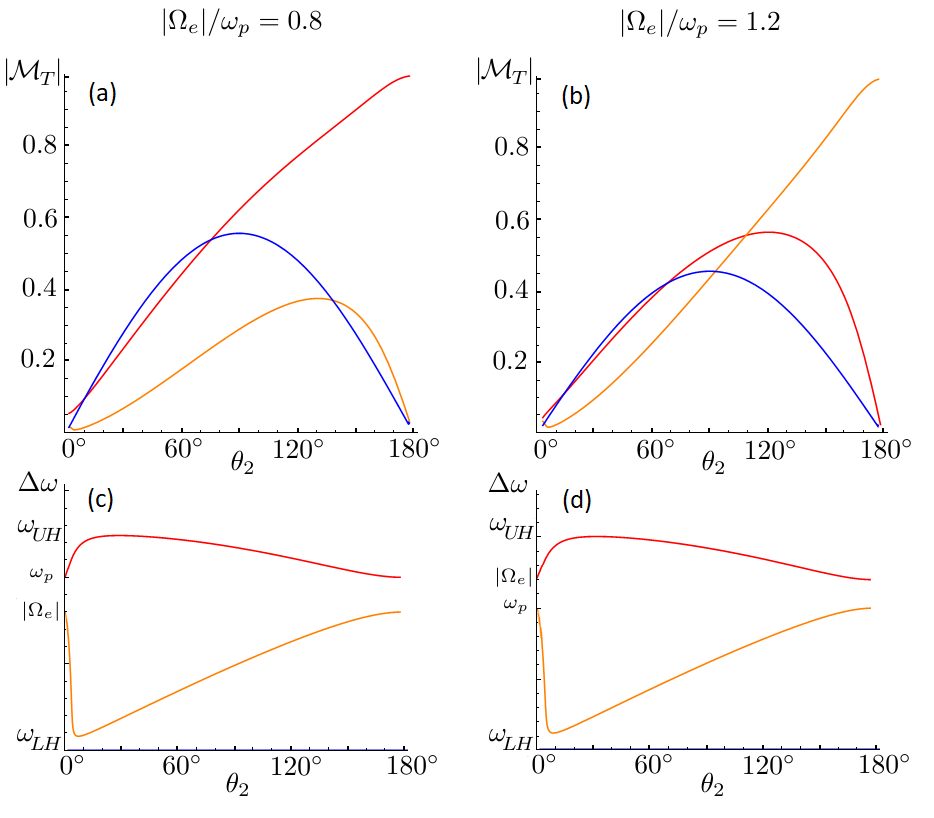}
	\caption{Scattering of a parallel pump laser in uniform hydrogen plasmas. 
	The pump laser has frequency $\omega_1/\omega_p\!=\!10$, and the scattered laser propagates at angle $\theta_2$ with respect to $\mathbf{k}_1\!\parallel\!\mathbf{B}_0$. The laser can scatter from the upper resonance (red), the lower resonance (orange), and the bottom resonance (blue). When the plasma is over-dense (a,c), the upper resonance is Langmuir-like; when the plasma is under-dense (b,d), the lower resonance is Langmuir-like. For Langmuir-like resonance, the frequency shift (c,d) $\Delta\omega\rightarrow\omega_p$, and the normalized growth rate (a,b) is monotonously increasing; while for cyclotron-like resonances, $\Delta\omega\rightarrow|\Omega_e|,\Omega_i$, and the normalized growth rate $|\mathcal{M}_{T}|$ peaks at intermediate $\theta_2$, while becoming zero for exact backscattering. See text for how maxima of $|\mathcal{M}_{T}|$ scales with plasma parameters.}
	\label{fig:TTL_Para}
\end{figure}

To demonstrate how to evaluate the normalized growth rate $\mathcal{M}_{T}$, consider the example where the incident laser propagate along the magnetic field, while the scattered laser propagate at some angle $\theta_2$. In this case $\alpha_2=\theta_2$, and by cylindrical symmetry, $\mathcal{M}_{T}$ depends on only one free parameter $\theta_2$, as plotted in Fig.~\ref{fig:TTL_Para} for hydrogen plasma with $\omega_1/\omega_p=10$. When there are only two charged species, as in the case of hydrogen plasma, there are three electrostatic resonances the lasers can scatter from (Fig.~\ref{fig:Resonance}). The first resonance is the upper resonance, whose frequency asymptotes to the upper-hybrid frequency $\omega_{UH}$ when $\theta_3\rightarrow\pi/2$. When scattered from the upper resonance (red curves), the scattered laser is frequency down-shifted ($\Delta\omega=\omega_1-\omega_2$) by the largest amount. The second resonance is the lower resonance, whose frequency asymptotes to the lower-hybrid frequency $\omega_{LH}$ when $\theta_3\rightarrow\pi/2$. When scattered from the lower resonance (orange curves), the scattered laser is frequency shifted by either $|\Omega_e|$ in over-dense plasma ($|\Omega_e|<\omega_p$), or by $\omega_p$ in under-dense plasma ($|\Omega_e|>\omega_p$), when $\theta_3\rightarrow0$. The third resonance is the bottom resonance, whose frequency asymptotes to $0$ when $\theta_3\rightarrow\pi/2$. When scattered from the bottom resonance (blue curves), the scattered laser is frequency shifted by at most $\Omega_i$ when $\theta_3\rightarrow0$. Since $\Omega_i$ is much smaller than other frequency scales, the frequency shift $\Delta\omega$ for scattering off the bottom resonance is not discernible in Fig.~\ref{fig:TTL_Para}c and \ref{fig:TTL_Para}d. In terms of the normalized growth rate (upper panels), we see $\mathcal{M}_{T}\rightarrow1$  when the laser is backscattered from the Langmuir resonance with $\Delta\omega\rightarrow\omega_p$, while $\mathcal{M}_{T}\rightarrow 0$ when the laser is scattered from the cyclotron resonances with $\Delta\omega\rightarrow|\Omega_e|,\Omega_i$. For Langmuir-like resonance, $\mathcal{M}_{T}$ increases monotonously with $\theta_2$. In contrast, for cyclotron-like resonances, $\mathcal{M}_{T}$ peaks near at intermediate $\theta_2$, and becomes zero for exact backscattering.

To better understand the angular dependence of the normalized growth rate $\mathcal{M}_{T}$, let us find its asymptotic expressions. In the limit $\omega_{1,2}\gg\omega_3$, the wave vector $k_2/k_1\simeq 1$ and $k_3/k_1\simeq 2\sin(\theta_2/2)$. At finite angle $\theta_2>0$, we can approximate $\theta_3\simeq(\pi-\theta_2)/2$
. For even larger $\theta_2$, we can also approximate the resonance frequency $\omega_3$ using Eqs.~(\ref{eq:wu_para})-(\ref{eq:wb_para}), because $\theta_3\sim 0$ is now small. These asymptotic geometric relations will be useful when we evaluate the coupling coefficient.

First, consider scattering off the Langmuir-like resonance $\omega_3\sim\omega_p$. Since $\gamma_{3,s}$ is finite, the lowest order angular dependence comes from $k_3$. Take the limit $\theta_3\rightarrow 0$, we get Eq.~(\ref{eq:MTP}). Now retain the angular dependence of $k_3$, we can grossly approximate
\begin{equation}
|\mathcal{M}_{T}^p|\simeq\sin\frac{\theta_2}{2}.
\end{equation}
This approximation is of course very crude, but it captures the monotonous increasing feature for scattering off the Langmuir-like resonance. In fact, the above result becomes a very good approximation when the magnetic field $B_0\rightarrow0$. In this unmagnetized limit, we recover the angular dependence of Raman scattering.

Second, consider scattering off the electron-cyclotron-like resonance $\omega_3\sim|\Omega_e|$. Notice that in this case, the magnetization factor $\gamma_{3,e}^2\gg 1$ for small $\theta_3$. Nevertheless, since both the numerator and the denominator contains this factor, $\mathcal{M}_{T}$ remains finite. For electrons, the magnetization ratio $\beta_{3,e}\simeq1$. Using Eq.~(\ref{eq:wl_para}), which is valid when $\omega_p\ne|\Omega_e|$, the magnetization factor $\gamma_{3,e}^2\simeq(\Omega_e^2-\omega_p^2)/(\omega_p^2\sin^2\theta_3)$. In comparison, $\beta_{3,i}\ll 1$ and $\gamma_{3,i}^2\simeq1$. Hence the dominant contribution comes from electrons. Substituting these into formula Eq.~(\ref{eq:MT}), we see to leading order
\begin{equation}
\label{eq:MTe}
|\mathcal{M}_{T}^e|\simeq\frac{1}{2}\bigg(\frac{\omega_p}{\omega_3}\bigg)^{1/2}\sin\theta_2,
\end{equation}
where $\omega_3$ as function of $\theta_2$ is given by Eq.~(\ref{eq:wl_para}), with $\theta_3\simeq(\pi-\theta_2)/2$. From Eq.~(\ref{eq:MTe}), we see $|\mathcal{M}_{T}^e|$ reaches maximum when the laser is scattered almost perpendicularly to the magnetic field. The maximum value scales roughly as $|\mathcal{M}_{T}^e|\sim\sqrt{\omega_p/|\Omega_e|}/2$, which can be very large in weakly magnetized plasmas, as long as the cold fluid approximation remains valid. Away from $\theta_2\sim\pi/2$, the normalized growth rate $|\mathcal{M}_{T}^e|$ falls off to zero. This falloff is expected, because exciting cyclotron resonance is energetically forbidden. 

In the end, consider scattering off ion-cyclotron-like resonance $\omega_3\sim\Omega_i$. In this case, the ion contribution to the wave energy coefficient is no longer negligible, because $\beta_{3,i}\simeq1$ and $\gamma_{3,i}^2\simeq\Omega_e/\Omega_i\tan^2(\theta_2/2)\gg1$, as can be seen from Eq.~(\ref{eq:wb_para}). The scattering strength is still dominated by electrons, for which $\beta_{3,e}\gg1$, and $\gamma_{3,e}^2\simeq-1/\beta_{3,e}^2\ll1$. Substituting these into Eq.~(\ref{eq:MT}), the normalized growth rate
\begin{equation}
\label{eq:MTi}
|\mathcal{M}_{T}^i|\simeq\frac{1}{2}\bigg(\frac{\phantom{.}\omega_p\phantom{.}\Omega_i}{|\Omega_e|\omega_3}\bigg)^{1/2}\sin\theta_2.
\end{equation}
We see the above result is rather similar to Eq.~(\ref{eq:MTe}), except that $\omega_3\sim\Omega_i$ has very weak angular dependence. Therefore, $|\mathcal{M}_{T}^i|$ is very well approximated by Eq.~(\ref{eq:MTi}). The normalized growth rate peaks almost at $\theta_2=\pi/2$, reaching a maximum $|\mathcal{M}_{T}^i|\sim\sqrt{\omega_p/|\Omega_e|}/2$, which can be very large in weakly magnetized plasmas. Similar to the electron cyclotron case, $|\mathcal{M}_{T}^i|$ falls off to zero for parallel scattering.

\subsubsection{Perpendicular pump}\label{sec:TTLPerp}

\begin{figure}[t]
	\includegraphics[angle=0,width=8cm]{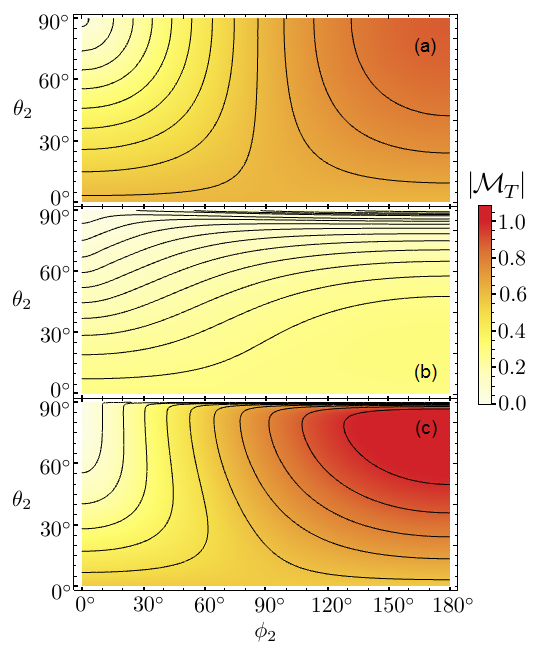}
	\caption{Normalized growth rate $|\mathcal{M}_{T}|$ for scattering of a perpendicular pump laser ($\mathbf{k}_1\!\perp\!\mathbf{B}_0$) in a uniform hydrogen plasma with $\omega_1/\omega_p=10$ and $|\Omega_e|/\omega_p=0.8$. In spherical coordinate, the scattered laser propagates at polar angle $\theta_2$ with respect to $\mathbf{B}_0$, and azimuthal angle $\phi_2$ measured from $\mathbf{k}_1$. The laser can scatter from the upper resonance (a), in which case backscattering is the strongest scattering mode. Alternatively, the laser can scatter off the lower resonance (b). In this case, one maximum of $|\mathcal{M}_{T}|$ is attained for backscattering, and another maximum is attained when the scattered laser propagate almost perpendicular to the incident laser along the magnetic field. Finally, the laser can scatter off the bottom resonance (c). In this case, exact backscattering is suppressed while nearly backward scattering is strong.}
	\label{fig:TTL_Perp}
\end{figure}

Consider the other special case where the pump laser propagates perpendicular to the magnetic field. In this geometry, it is natural to plot the normalized growth rate $|\mathcal{M}_{T}|$ in spherical coordinate (Fig.~\ref{fig:TTL_Perp}), where the polar angle $\theta_2$ is measured from the magnetic field $\mathbf{B}_0$, and the azimuthal angle $\phi_2$ is measured from the wave vector $\mathbf{k}_1$. By symmetry of this setup, it is obvious that $\mathcal{M}_{T}(\phi_2,\theta_2) =\mathcal{M}_{T}(\phi_2,\pi-\theta_2) =\mathcal{M}_{T}(-\phi_2, \theta_2)$. Therefore, it is sufficient to consider the range $\theta_2\in[0,\pi/2]$ and $\phi_2\in[0,\pi]$. By matching the $\mathbf{k}$ resonance, we can read $\theta_3$ from the spherical coordinates $(\phi_2,\theta_2)$, and thereafter read the frequency shift $\omega_3$ from Fig.~\ref{fig:Resonance}. As for the growth rate, in electron-ion plasma, 
when scattered from the upper resonance (Fig.~\ref{fig:TTL_Perp}a), backscattering has the largest growth rate. While for scattering off the lower resonance (Fig.~\ref{fig:TTL_Perp}b), $|\mathcal{M}_{T}|$ reaches maximum for both backscattering and nearly parallel scattering, where the scattered laser propagates almost parallel to the magnetic field. In comparison, when scattering off the bottom resonance  (Fig.~\ref{fig:TTL_Perp}c), the normalized growth rate peaks for nearly backward scattering, while falls to zero for exact backscattering.

To better understand the angular dependence of the normalized growth rate, let us consider its asymptotic expressions for two special cases. The first special case is when all waves lie in the plane perpendicular to the magnetic field, namely, when $\theta_2=90^\circ$. In this case, the angle $\theta_3$ is fixed to $90^\circ$, and the frequency of the plasma resonances are also fixed to $\omega_{UH}$, $\omega_{LH}$, or zero. Therefore, the angular dependence only comes from $k_3$. In the limit $\omega_{1,2}\gg\omega_3$, we have $k_3\simeq2k_1\sin(\phi_2/2)$. Using Eqs.~(\ref{eq:MTUH}) and (\ref{eq:MTLH}), it is easy to see, for scattering off \textit{UH} and \textit{LH} waves in the perpendicular plane
\begin{eqnarray}
|\mathcal{M}_{T\perp}^{UH}|&\simeq&\bigg(\frac{\omega_p}{\omega_{UH}}\bigg)^{1/2}\sin\frac{\phi_2}{2},\\
|\mathcal{M}_{T\perp}^{LH}|&\simeq&\frac{\omega_p^{3/2}\omega_{LH}^{1/2}}{\omega_{UH}|\Omega_e|}\sin\frac{\phi_2}{2}.
\end{eqnarray}
Now let us calculate $\mathcal{M}_{T\perp}^{b}$ for scattering off the bottom resonance. Using asymptotic expression Eq.~(\ref{eq:wb_perp}) for $\omega_3$, we see although the magnetization ratio $\beta_{3,s}\rightarrow\infty$, the product $\beta_{3,s}\cos\theta_3$ remains finite as $\theta_3\rightarrow\pi/2$. Since the magnetization factor $\gamma_{3,s}\simeq-1/\beta_{3,s}^2\ll1$, it is easy to see $\mathcal{M}_{T\perp}^{b}\propto\sqrt{\omega_3}$, which goes to zero when $\theta_3\rightarrow\pi/2$. Hence, for scattering off the bottom resonance in the perpendicular plane
\begin{equation}
|\mathcal{M}_{T\perp}^{b}|=0,
\end{equation}
is completely suppressed. Consequently, exact backscattering from the bottom resonance is also suppressed.

\begin{figure}[t]
	\includegraphics[angle=0,width=8cm]{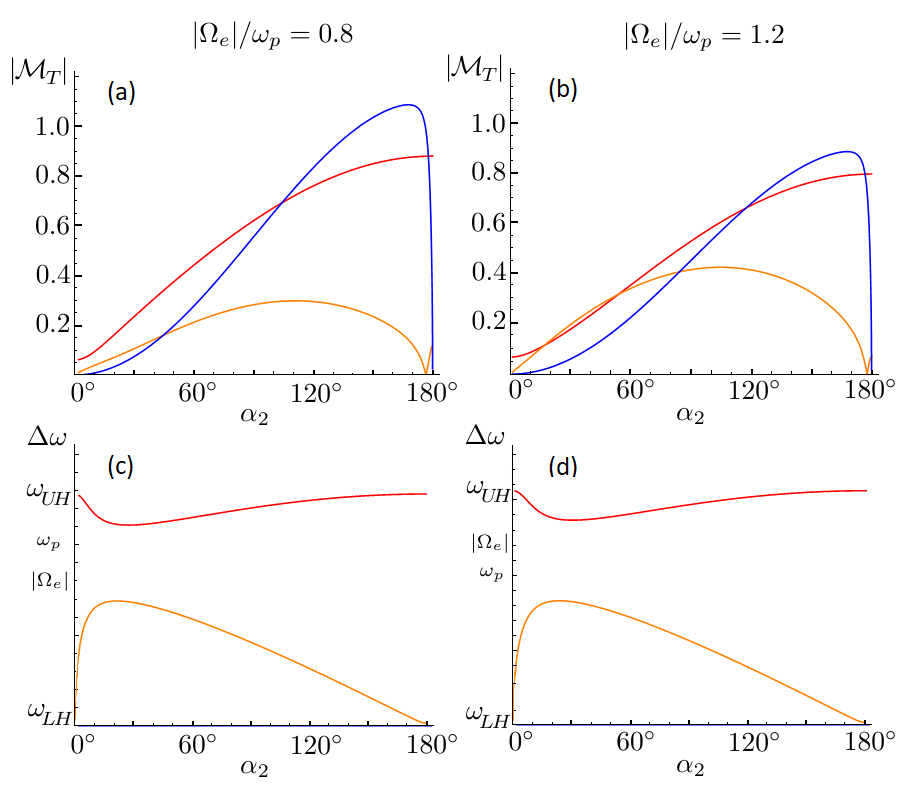}
	\caption{Scattering of a perpendicular pump laser with $\omega_1/\omega_p=10$ in a uniform hydrogen plasma. Figure (a) can be obtained from Fig.~\ref{fig:TTL_Perp} by taking a one dimensional cut along the unit sphere using the plane spanned by $\mathbf{k}_1\perp\mathbf{B}_0$. The scattered laser, propagating at angle $\alpha_2$ with respect to $\mathbf{k}_1$, can scatter from the upper resonance (red), the lower resonance (orange), and the bottom resonance (blue). Both the normalized growth rate $|\mathcal{M}_{T}|$ (a,b) and the frequency shifts $\Delta\omega$ (c,d) behave qualitatively the same in over-dense plasma (a,c) and under-dense plasma (b,d). As $\alpha_2$ increases from $0^\circ$ to $180^\circ$, $|\mathcal{M}_{T}|$ increases monotonously for scattering from the upper resonance. For scattering off the lower resonance, $|\mathcal{M}_{T}|$ hit zero near $\alpha_2\sim176^\circ$, where electron and ion contributions exactly cancel, and then increase to finite value at exact backscattering. In contrast, when the laser is scattered from the bottom resonance, $|\mathcal{M}_{T}|$ strongly peaks near $\alpha_2\sim170^\circ$, and becomes zero for exact backward scattering. See text for how $|\mathcal{M}_{T}|$ scales with plasma parameters.}
	\label{fig:TTL_Perp1D}
\end{figure}

To see how $\mathcal{M}_{T}^{b}$ climbs up from zero, consider another special case where $\mathbf{k}_2$ is in the plane spanned by $\mathbf{k}_1$ and $\mathbf{b}$. In this case, it is more natural to consider $\mathcal{M}_{T}$ as function of $\alpha_2$, the angle between $\mathbf{k}_1$ and $\mathbf{k}_2$, as plotted in Fig.~\ref{fig:TTL_Perp1D}. Let us find the asymptotic expression of $\mathcal{M}_{T}^{b}$ when $\alpha_2\sim\pi$. In this limit, we have $\theta_3\sim\pi/2$, and the resonance frequency $\omega_3$ can be approximated by Eq.~(\ref{eq:wb_perp}). Then the magnetization ratios $\beta_{3,e}^2\simeq\Omega_e^2/\Omega_i^2+|\Omega_e|/(\Omega_i\cos^2\theta_3)$ and $\beta_{3,i}^2\simeq1+\Omega_i/(|\Omega_e|\cos^2\theta_3)$. Consequently, the magnetization factors can be well approximated by $\gamma_{3,e}^2\simeq-1/\beta_{3,e}^2$ and $\gamma_{3,i}^2\simeq-|\Omega_e|\cos^2\theta_3/\Omega_i$. Moreover, since $\omega_{1,2}\gg\omega_3$, the angle $\theta_3\simeq\alpha_2/2$ and the wave vector $k_3\simeq2k_1\sin(\alpha_2/2)$. Substituting these into formula Eq.~(\ref{eq:MT}), we see when $\alpha_2\sim\pi$, the normalized growth rate
\begin{equation}
|\mathcal{M}_{T}^{b}|^2\simeq\frac{[\zeta(1+\zeta\cos^2\frac{\alpha_2}{2})]^{3/2}\sin^2\frac{\alpha_2}{2}\cos\frac{\alpha_2}{2}}{r^3+r[1+\zeta(1+\zeta\cos^2\frac{\alpha_2}{2})^2]\sin^2\frac{\alpha_2}{2}},
\end{equation}
where $r:=|\Omega_e|/\omega_p$ and $\zeta:=M_i/Z_i\gg1$. To see the lowest order angular dependence, we can use a cruder but simpler approximation $|\mathcal{M}_{T}^{b}|^2\simeq\zeta^{1/2}\cos(\alpha_2/2)/r$. We see $|\mathcal{M}_{T}^{b}|$ increases sharply from zero away from exact backscattering. Using result Eq.~(\ref{eq:MTi}), we find in the other limit $\alpha_2\sim0$, the normalized growth rate
\begin{equation}
|\mathcal{M}_{T}^{b}|\simeq\frac{\sin^2\frac{\alpha_2}{2}}{r^{1/2}}\Big(1-\frac{1}{\zeta}\tan^2\frac{\alpha_2}{2}\Big)^{-3/4}.
\end{equation} 
We see scattering from the bottom resonance can be strong when the plasma is weakly magnetized, as long as the scattering angle is away from exact forward or backward scattering.

In summary, the \textit{TTL} scattering in magnetized plasma is mostly due to density beating Eq.~(\ref{eq:TTL_THETAr}), and the modification due to the magnetic field can be represented by the normalized growth rate $\mathcal{M}_{T}$. In magnetized plasmas, cyclotron-like resonances, in addition to the Langmuir-like resonance, contribute to the scattering of the \textit{T} waves. When scattered from the Langmuir-like resonance, both the wave energy coefficient and the scattering strength are finite. Therefore in this case, the angular dependence of $\mathcal{M}_{T}$ comes mostly from $k_3$, which reaches maximum for backscattering. In contrast, for scattering from cyclotron-like resonances, both the scattering strength and the wave energy coefficient can blow up. Their ratio, $\mathcal{M}_{T}$, goes to zero when the scattering angles are such that the \textit{L} wave frequency approaches either zero or the cyclotron frequencies. In addition, $\mathcal{M}_{T}$ can also become zeros at special angles where scattering from electrons and ions exactly cancel. Away from these special angles, scattering from cyclotron-like resonances, which increases with decreasing magnetic field, typically have growth rates that are comparable to scattering from Langmuir-like resonances. When the plasma parameters are known, we can determine the angular dependence of $\mathcal{M}_{T}$ using formula Eq.~(\ref{eq:MT}). This knowledge can be used to choose injection angles of two lasers such that their scattering is either enhanced or suppressed. Conversely, by measuring angular dependence of $\mathcal{M}_{T}$ in laser scattering experiments, one may be able to fit plasma parameters to match Eq.~(\ref{eq:MT}). This provide a diagnose method from which the magnetic field, as well as the plasma density and composition can be measured.

\subsection{$L\rightleftharpoons L+L$ scattering}\label{sec:LLL}
In this subsection, we consider the other scenario where the three-wave scattering happens between three resonant quasi-longitudinal waves. This happens, for example, when we launch an electrostatic wave into the plasma by some antenna arrays. When the wave power is strong enough to overcome damping, it may subsequently decay to two other waves if the resonance conditions can be satisfied. The decay waves are not necessarily electrostatic, but for the purpose of illustrating the general results in Sec.~\ref{sec:3waves}, we will only give examples where the two decay waves are also electrostatic. 

The coupling strength between three \textit{L} waves can be simplified using the approximation that the waves are quasi-longitudinal. Substituting $\mathbf{e}_i\simeq\hat{\mathbf{k}}_i$ into formula (\ref{eq:Thetaijl}) and using the frequency resonance condition (\ref{eq:resonantW}), the normalized scattering strength for \textit{LLL} scattering can be written as
\begin{eqnarray}
\label{eq:THETAL3}
\nonumber
\Theta^s\simeq&-&\frac{ck_1\omega_1}{\omega_2\omega_3}(\hat{\mathbf{k}}_1 \cdot\mathbb{F}_{s,2}^*\hat{\mathbf{k}}_2)(\hat{\mathbf{k}}_1\cdot\mathbb{F}_{s,3}^*\hat{\mathbf{k}}_3)\\
&+&\frac{ck_2\omega_2}{\omega_3\omega_1}(\hat{\mathbf{k}}_2 \cdot\mathbb{F}_{s,1}\hat{\mathbf{k}}_1)(\hat{\mathbf{k}}_2\cdot\mathbb{F}_{s,3}^*\hat{\mathbf{k}}_3)\\
\nonumber
&+&\frac{ck_3\omega_3}{\omega_1\omega_2}(\hat{\mathbf{k}}_3 \cdot\mathbb{F}_{s,1}\hat{\mathbf{k}}_1)(\hat{\mathbf{k}}_3\cdot\mathbb{F}_{s,2}^*\hat{\mathbf{k}}_2),
\end{eqnarray}
where $k_i:=|\mathbf{k}_i|$ is the magnitude of the wave vector, and $\hat{\mathbf{k}}_i$ is the unit vector along $\mathbf{k}_i$ direction. It is easy to recognize that $\hat{\mathbf{k}}_i\cdot(\mathbb{F}_{s,j}/\omega_j)\hat{\mathbf{k}}_j$ is the projection of quiver velocity $\hat{\mathbf{v}}_j$ in $\hat{\mathbf{k}}_i$ direction. Therefore, the couplings between three \textit{L} waves may also be interpreted as density beating. The first term in $\Theta^s$ is proportional to the rate of creating wave 1 by annihilating waves 2 and 3, the second term is proportional to the rate of annihilating waves 3 and $\bar{1}$ to create wave $\bar{2}$, and the last term can be interpreted similarly. The interference between these processes determines the overall scattering strength.

Having obtained expressions for the normalized scattering strength (\ref{eq:THETAL3}) and wave energy (\ref{eq:TTL_U3}), we can immediately evaluate the coupling coefficient (\ref{eq:coupling}), and find expressions for experimental observables. For example, the linear growth rate $\gamma_0$ [Eq.~(\ref{eq:GrowthRate})] of the parametric decay instability can be written as
\begin{equation}
\label{eq:GrowthRateL3}
\gamma_0=\gamma_L|\mathcal{M}_L|,
\end{equation}
where $\gamma_L$ is purely determined by the pump wave
\begin{equation}
\gamma_L=\frac{1}{2}ck_1|a_1|.
\end{equation}
The normalized growth rate for \textit{LLL} scattering
\begin{equation}
\label{eq:ML}
\mathcal{M}_L=\frac{\omega_p}{2ck_1}\bigg(\frac{\omega_p^2}{\omega_2\omega_3}\bigg)^{1/2}\mu_L,
\end{equation}
is the product of a kinematic factor with the coupling coefficient $\Gamma=\omega_p^2\mu/4$. In the \textit{LLL} approximation, the normalized coupling coefficient
\begin{equation}
\label{eq:uL}
\mu_L\simeq\sum_s\frac{Z_s}{M_s}\frac{\omega_{ps}^2}{\omega_p^2}\frac{\Theta^s_r}{(u_1u_2u_3)^{1/2}},
\end{equation}
where $\Theta^s_r$ is the real part of Eq.~(\ref{eq:THETAL3}). Again, notice when density of species $s$ goes to zero, its contribution to $\mu_L$ also goes to zero as expected.

To evaluate the normalized growth rate $\mathcal{M}_L$, we can use the following procedure to mimic what happens in an actual experiment. Suppose we know the species density and magnetic field, then we know what resonances are there in the plasma. We can then launch a pump wave at resonance frequency $\omega_1$ using some antenna array. The antenna array not only inject a wave at the given frequency, but also selects the wave vector $k_1$ and the wave direction $\theta_1$. 
To observe the decay waves, we can place a probe at some angle $\theta_2$ with respect to the magnetic field, and some azimuthal angle $\phi_2$ in a spherical coordinate. The probe can measure fluctuations of the plasma potential and therefore inform us about the wave frequency $\omega_2$. Then we immediately know $\omega_3=\omega_1-\omega_2$ from the three-wave resonance condition. Moreover, since the third wave is a magnetic resonance, the frequency $\omega_3$ constrains the angle $\theta_3$ at which the third wave can propagate. However, a simple probe cannot measure the wave vector, so we will have to solve $k_2$ and $k_3$ from the resonance condition (\ref{eq:resonantK}), which can be written in components as
\begin{eqnarray}
\label{eq:k32}
&&k_3^2=k_1^2+k_2^2-2k_1k_2\cos\alpha_2,\\
\label{eq:kpara}
&&k_3\cos\theta_3=k_1\cos\theta_1-k_2\cos\theta_2.
\end{eqnarray}
Here $\alpha_2=\alpha_2(\theta_1,\theta_2,\phi_2)$ is the angle between $\mathbf{k}_1$ and $\mathbf{k}_2$. The above system of quadratic equations have two solutions in general. This degeneracy comes from the symmetry $2\leftrightarrow3$, because we cannot distinguish whether the probe is measuring wave 2 or wave 3, both of which are electrostatic resonances. If the solutions $k_2$ and $k_3$ are both real and positive, then the three-wave resonance conditions can be satisfied, and three-wave decay will happen once the pump amplitude $a_1$ exceeds the damping threshold. In another word, we control $\omega_1$ and $\mathbf{k}_1$ by the antenna array, measure $\omega_2$ using probes, and infer $\omega_3$, $\mathbf{k}_2$, and $\mathbf{k}_3$ by solving resonance conditions. With these information, the analytical formula of the normalized growth rate $\mathcal{M}_L$ can be readily evaluated numerically.

\subsubsection{Parallel pump}\label{sec:LLLPara}
To demonstrate how to evaluate the normalized growth rate $\mathcal{M}_L$, consider the example where the pump wave is launched along the magnetic field ($\theta_1=0$). In an electron-ion plasma, this geometry allows the antenna to launch three electrostatic waves: the Langmuir wave, the electron cyclotron wave, or the ion cyclotron wave. In the regime where $\omega_p\sim|\Omega_e|\sim|\omega_p-\Omega_e|\gg\Omega_i$, four decay modes are allowed by the resonance conditions: $u\rightarrow l+l$, $l\rightarrow l+l$, $l\rightarrow l+b$, and $b\rightarrow b+b$, where we have labeled waves by the resonance branch they belong to, and $u$, $l$, and $b$ denote the upper, lower and bottom resonances. 

\begin{figure}[t]
	\includegraphics[angle=0,width=8cm]{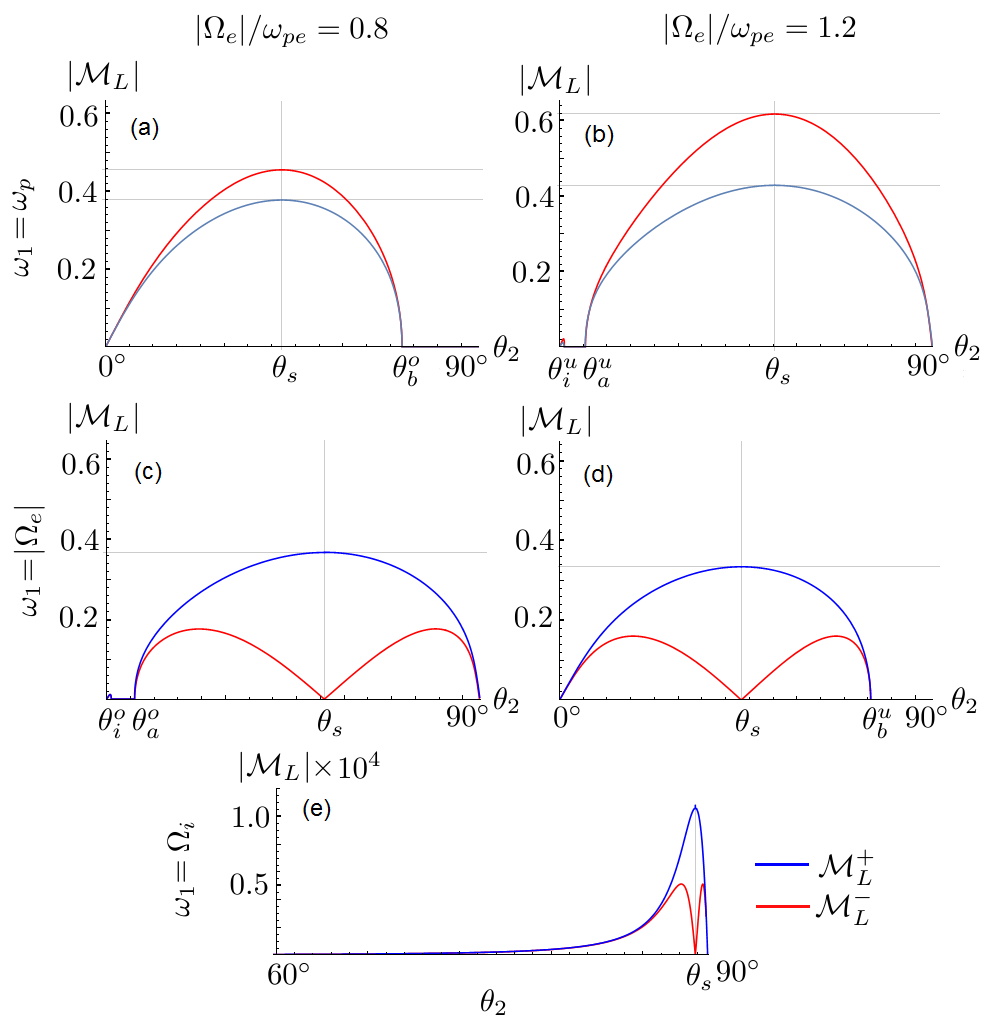}
	\caption{Scattering of a parallel electrostatic pump wave in uniform hydrogen plasmas, when observed at angle $\theta_2$ with respect to $\mathbf{k}_1\!\parallel\!\mathbf{B}_0$.  At each $\theta_2$, due to the degeneracy $2\leftrightarrow3$, the wave vector has two possible values $k_2^{\pm}$, corresponding to $\mathcal{M}_{L}^+$ (blue, $\theta_3>90^\circ$) and $\mathcal{M}_{L}^-$ (red, $\theta_3<90^\circ$). 
	The pump wave can be the Langmuir wave (a,b); the electron cyclotron wave (c,d); and the ion cyclotron wave (e). The normalized growth rate attains local extrema for symmetric scattering, where the two decay waves have the same frequency $\omega_r(\theta_s)=\omega_1/2$. In over-dense plasma [eg. (a),(c)], $u\rightarrow l,l$ happens for $\theta_2<\theta_b^o$, where $\omega_l(\theta_b^o)=\omega_p-|\Omega_e|$; $l\rightarrow l,l$ happens for $\theta_2>\theta_a^o$, where $\omega_l(\theta_a^o)=|\Omega_e|-\omega_{LH}$; and $l\rightarrow l_2,b_3$ happens for $\theta_2<\theta_i^o$, where $\omega_l(\theta_i^o)=|\Omega_e|-\Omega_i$. In under-dense plasma [eg. (b),(d)], $u\rightarrow l,l$ happens for $\theta_2<\theta_b^u$, where $\omega_l(\theta_b^u)=|\Omega_e|-\omega_p$; $l\rightarrow l,l$ happens for $\theta_2>\theta_a^u$, where $\omega_l(\theta_a^u)=\omega_p-\omega_{LH}$; and $l\rightarrow l_2,b_3$ happens for $\theta_2<\theta_i^u$, where $\omega_l(\theta_i^u)=\omega_p-\Omega_i$. Regardless of plasma density [(e)], $b\rightarrow b,b$ can always happen, for which the growth rate peaks near $\theta_s\sim88^\circ$, where the decay is symmetrical. The gray lines indicate the symmetric angles and the asymptotic maxima obtained in the text.}
	\label{fig:LLL_Para}
\end{figure}

First, let us consider the case where the pump wave is the Langmuir wave (Fig.~\ref{fig:LLL_Para}a, \ref{fig:LLL_Para}b). In this case, the magnetization factor $\gamma_1$ is finite, the wave energy coefficient $u_1=1$, and $\mathbb{F}_{s,1}\hat{\mathbf{k}}_1=\hat{\mathbf{k}}_1$. The normalized scattering strength (\ref{eq:THETAL3}) contains the following four simple inner products:  $(\hat{\mathbf{k}}_1\cdot\mathbb{F}_{s,2}^*\hat{\mathbf{k}}_2) =(\hat{\mathbf{k}}_2\cdot\mathbb{F}_{s,1}\hat{\mathbf{k}}_1)=\cos\theta_2$; $(\hat{\mathbf{k}}_1\cdot\mathbb{F}_{s,3}^*\hat{\mathbf{k}}_3)= (\hat{\mathbf{k}}_3\cdot\mathbb{F}_{s,1}\hat{\mathbf{k}}_1)=\cos\theta_3$, as well as two other inner products $(\hat{\mathbf{k}}_2\cdot\mathbb{F}_{s,3}^*\hat{\mathbf{k}}_3)=\cos\theta_2\cos\theta_3-\gamma_{s,3}^2\sin\theta_2\sin\theta_3$; and $(\hat{\mathbf{k}}_3\cdot\mathbb{F}_{s,2}^*\hat{\mathbf{k}}_2)=\cos\theta_3\cos\theta_2-\gamma_{s,2}^2\sin\theta_3\sin\theta_2$. Substituting these inner products into Eq.~(\ref{eq:THETAL3}), and using the resonance condition (\ref{eq:kpara}), the normalized scattering strength can be immediately found. 
In the above expressions, $\theta_2$ is the independent variable, and $\omega_2$ is measured. Then we can determine $\theta_3$ from $\omega_3(\theta_3)=\omega_1-\omega_2$ using Eq.~(\ref{eq:resonance}), and solve for $k_2$ and $k_3$ from Eqs.~(\ref{eq:k32}) and (\ref{eq:kpara}). Finally, with the above information, the normalized matrix element $\mathcal{M}_L$ can be readily evaluated.

When pumped at the Langmuir frequency ($\omega_1=\omega_p$), the resonance conditions constrain the plasma parameters and angles at which the three-wave decay can happen. In over-dense plasma (eg. Fig.~\ref{fig:LLL_Para}a), the Langmuir wave is in the upper resonance, so the resonance condition can be satisfied only if $\omega_p<2|\Omega_e|$. Having satisfied this condition, the $u\rightarrow l+l$ decay can happen if $\theta_2<\theta_b^o$, where $\theta_b^o$ is the angle such that $\omega_l(\theta_b^o)=\omega_p-|\Omega_e|$. In comparison, in under-dense plasma (eg. Fig.~\ref{fig:LLL_Para}b), the Langmuir wave is in the lower resonance, and therefore can always decay. One decay mode is $l\rightarrow l+l$, which can happen for $\theta_2>\theta_a^u$, where $\omega_l(\theta_a^u)=\omega_p-\omega_{LH}$. Another decay mode is $l\rightarrow l+b$. When $\omega_2=\omega_l$, this decay mode happens for $0<\theta_2<\theta_i^u$, where $\omega_l(\theta_i^u)=\omega_p-\Omega_i$; whereas when $\omega_2=\omega_b$, this decay mode can happen at any $\theta_2$. Finally, using the symmetry $2\leftrightarrow3$, the constrains on $\theta_3$ can be readily deduced.

For Langmuir wave pump, the normalized growth rate reaches maximum for symmetric decay, where $\omega_2=\omega_3=\omega_p/2$. Let us find the asymptotic expression of $\mathcal{M}_L$ in the symmetric case, so as to get a sense of how the normalized growth rate scales with plasma parameters. The symmetric angle $\theta_s$ can be solved from Eq.~(\ref{eq:resonance}). Using $\omega_p\sim|\Omega_e|\gg \Omega_i$, we find $\sin^2\theta_s\simeq3[1-\omega_p^2/(4\Omega_e^2)]/4$. Then the wave energy coefficient $u_{2}=u_3\simeq1+3\omega_p^2/(4\Omega_e^2-\omega_p^2)$, where the sub-dominant ion contribution in Eq.~(\ref{eq:TTL_U3}) has been dropped. To solve for the degenerate wave vectors in the symmetric case, it is more convenient to consider the two limits:  $\theta_2=\theta_s-\phi, \theta_3=\theta_s+\phi$, and $\theta_2=\theta_s-\phi, \theta_3=\pi-\theta_s-\phi$, and then let $\phi\rightarrow0$. Solving Eqs.~(\ref{eq:k32}) and (\ref{eq:kpara}) for the wave vectors, the two solutions are $k_2^-/k_1\simeq1/(2\cos\theta_s)$ and $k_2^+/k_1\simeq\sin\theta_s/(2\sin\phi)$. For the $k_2^-$ solution, all terms are finite, and the normalized scattering strength is dominated by electron contribution $\Theta_e^-\simeq-3ck_1[1+\omega_p^2/(2\Omega_e^2)]/(4\omega_p)$. Consequently, the normalized growth rate
for symmetric $k^-$ scattering
\begin{equation}
\label{eq:MLp-}
\mathcal{M}_L^-\Big(\omega_p\rightarrow\frac{\omega_p}{2},\frac{\omega_p}{2}\Big)\simeq\frac{3}{4}\Big(1-\frac{\omega_p^2}{4\Omega_e^2}\Big).
\end{equation}
Notice that this decay mode can happen only if $|\Omega_e|\ge\omega_p/2$. To see what happens to the $k_2^+$ solution, we need to keep the dominant terms, and expand $\omega_2\simeq\omega_p/2-\omega_s'\phi$ and $\omega_3\simeq\omega_p/2+\omega_s'\phi$, where the angular derivative 
of lower resonance $\omega_l(\theta)$ can be evaluated at the symmetric angle using Eq.~(\ref{eq:resonance}) to be $\omega_s'/\omega_p\simeq-2\Omega_e^2\sin(2\theta_s)/(2\Omega_e^2+\omega_p^2)$. Since ion terms does not contain singularity, the normalized scattering strength is again dominated  by electrons $\Theta_e^+\simeq3ck_1[1+5\omega_p^2/(4\Omega_e^2)]/(8\omega_p)$. Consequently, the normalized growth rate
for symmetric $k^+$ scattering is
\begin{equation}
\label{eq:MLp+}
\mathcal{M}_L^+\Big(\omega_p\rightarrow\frac{\omega_p}{2},\frac{\omega_p}{2}\Big)\simeq -\frac{\mathcal{M}_L^-}{2}\Big(1+\frac{3\omega_p^2/2}{\omega_p^2+2\Omega_e^2}\Big),
\end{equation}
where $\mathcal{M}_L^-$ is given by Eq.~(\ref{eq:MLp-}). Since $\omega_p\le2|\Omega_e|$, it is easy to see that $|\mathcal{M}_L^+|$ is always smaller than $|\mathcal{M}_L^-|$. Moreover, wave damping tends to be smaller for the $k_2^-$ solution. Therefore, the dominate decay mode in experiments will be the $k^-$ mode, where the two decay waves propagate symmetrically at angle $\theta_s$ with respect to the parallel pump wave.

Second, let us consider the case where the pump wave is the electron cyclotron wave (Fig.~\ref{fig:LLL_Para}c, \ref{fig:LLL_Para}d). In this case, $\beta_{e,1}\sim1$ and the magnetization factor $\gamma_{e,1}^2\simeq(\Omega_e^2/\omega_p^2-1)/\sin^2\theta_1$ approaches infinity, so the dominate contribution comes from electrons. Keeping track of dominate terms as $\theta_1\rightarrow0$ and using small angle expansion Eq.~(\ref{eq:wu_para}), the inner products 
$(\hat{\mathbf{k}}_2\cdot\mathbb{F}_{e,1}\hat{\mathbf{k}}_1 )\simeq\mp\gamma_{e,1}^2\sin\theta_1\sin\theta_2$, and  $(\hat{\mathbf{k}}_3\cdot\mathbb{F}_{e,1}\hat{\mathbf{k}}_1) \simeq\pm\gamma_{e,1}^2\sin\theta_1\sin\theta_3$.
The other four inner products that enters Eq.~(\ref{eq:THETAL3}) are the same as before. Keeping terms $\propto1/\sin\theta_1$, the leading term of the normalized scattering strength 
can be readily found.
Although the normalized scattering strength is divergent as $\theta_1\rightarrow0$, the normalized growth rate remains finite. This is because the divergence in $\Theta_e$ cancels the divergence in the wave energy coefficient $u_1\simeq(\omega_p^2-\Omega_e^2)^2/(\omega_p^2\Omega_e^2\sin^2\theta_1)$, which enters the denominator of $\mathcal{M}_L$. Follow procedure in the first example, the normalized growth rate can be readily obtained.

When intense electron cyclotron pump ($\omega_1=|\Omega_e|$) exceed the damping threshold, a number of decay modes are possible. In over-dense plasma (eg. Fig.~\ref{fig:LLL_Para}c), the electron cyclotron wave is in the lower resonance, and three-wave decay is always possible. One decay mode is $l\rightarrow l+l$, which can happen for $\theta_2>\theta_a^o$, where $\omega_l(\theta_a^o)=|\Omega_e|-\omega_{LH}$. Another decay mode is $l\rightarrow l+b$, which can happen for any $\theta_2$ if $\omega_2=\omega_b$, and can happen for $0<\theta_2<\theta_i^o$ if $\omega_2=\omega_l$, where $\omega_l(\theta_i^o)=|\Omega_e|-\Omega_i$. In comparison, in under-dense plasma (eg. Fig.~\ref{fig:LLL_Para}d), the electron cyclotron wave is in the upper resonance. The resonance condition can be satisfied if $|\Omega_e|<2\omega_p$, and $u\rightarrow l+l$ decay can happen if $\theta_2<\theta_b^u$, where $\omega_l(\theta_b^u)=|\Omega_e|-\omega_p$. We see the angular constrains for electron cyclotron pump decay is in reciprocal to that of the Langmuir pump.

For electron cyclotron pump, the normalized growth rate crosses zero and therefore vanish for symmetric $k^-$ decay, while reaching maximum for symmetric $k^+$ decay. Let us find the asymptotic expression for $\mathcal{M}_L^+$ to get a sense of how the normalized growth rate scales with plasma parameters. Again, we can find the symmetric angle $\theta_s$ from Eq.~(\ref{eq:resonance}), which gives $\sin^2\theta_s\simeq3[1-\Omega_e^2/(4\omega_p^2)]/4$. Then the wave energy coefficients $u_{2}=u_3\simeq2(1+2\omega_p^2/\Omega_e^2)/3$. To find the leading behavior of the scattering strength, consider the limit $\theta_2=\theta_s-\phi, \theta_3=\pi-\theta_s-\phi$, and let $\phi\rightarrow0$. In this limit, the wave vector $k_2^+/k_1\simeq\sin\theta_s/(2\sin\phi)\rightarrow\infty$, and the frequencies can be expanded by $\omega_2\simeq\omega_p/2-\omega_s'\phi$ and $\omega_3\simeq\omega_p/2+\omega_s'\phi$, where the angular derivative $\omega_s'$ can again be solved from Eq.~(\ref{eq:resonance}) to be $\omega_s'/\Omega_e\simeq2\omega_p^2\sin(2\theta_s)/(\Omega_e^2+2\omega_p^2)$. Keeping the dominate terms as $\phi\rightarrow0$, the normalized scattering strength $|\Theta_e^+|\simeq ck_1\sin(2\theta_s)(1-r^2)(1-r^2/4)/(\sin\theta_1\Omega_e)$, where $r:=|\Omega_e|/\omega_p$. Since the ion contributions are subdominate, the normalized growth rate for symmetric $k+$ scattering is
\begin{equation}
\label{eq:MLe+}
\Big|\mathcal{M}_L^+\Big(\!\Omega_e\!\rightarrow\!\frac{\Omega_e}{2},\!\frac{\Omega_e}{2}\!\Big)\Big|\!\simeq\! \frac{r}{4}\frac{\sqrt{(3\!-\!3r^2/4)^{3}(1\!+\!3r^2/4)}}{2+r^2}.
\end{equation}
We see $\mathcal{M}_L^+$ is nonzero for $0<r<2$, and reaches a maximum of $\sim0.38$ when $r\sim0.92$. The normalized growth rate can be related to the decay rate in experiments, once wave damping is taken into account. 

Finally, let us consider the case where the electrostatic pump wave is at ion cyclotron frequency (Fig.~\ref{fig:LLL_Para}e). Since $\Omega_i$ is much smaller than any other characteristic wave frequencies, the only possible decay mode is $b\rightarrow b+b$. Such decay can happen for any angle $\theta_2$, because the resonance conditions can always be satisfied. Similar to what happens in the previous example, the normalized growth rate $\mathcal{M}_L$ changes sign and therefore vanish for symmetric $k^-$ decay, while reaching maximum for symmetric $k^+$ decay. Now let us give an estimate of the maximum value of $\mathcal{M}_L^+$. Since the magnetization factor $\gamma_{1,i}^2\simeq\zeta/\tan^2\theta_1\rightarrow\infty$, where $\zeta:=M_i/Z_i\gg1$, the ion terms dominant. The divergent inner products are $(\hat{\mathbf{k}}_2\cdot\mathbb{F}_{i,1}\hat{\mathbf{k}}_1)\simeq\mp\gamma_{i,1}^2\sin\theta_1\sin\theta_2$ and $(\hat{\mathbf{k}}_3\cdot\mathbb{F}_{i,1}\hat{\mathbf{k}}_1)\simeq\pm\gamma_{i,1}^2\sin\theta_1\sin\theta_3$. The other four inner products are finite and similar to what we have before. Using these inner products and keep the leading terms, the normalized scattering $|\Theta_i^+|\simeq ck_1\Omega_e^2\cos\theta_s/(2\Omega_i^3\sin\theta_1)$, where we have expanded near the symmetric angle as before, with $\omega_s'\simeq9\Omega_e\sin(2\theta_s)/16$. The symmetric angle, very close to $\pi/2$, can be estimated from Eq.~(\ref{eq:wb_perp}) to be $\cos^2\theta_s\simeq\Omega_i/(3|\Omega_e|)$. The wave energy coefficients $u_1\simeq\omega_p^2|\Omega_e|/(\Omega_i^3\sin^2\theta_1)$, and $u_2=u_3\simeq16\omega_p^2/(9\Omega_i|\Omega_e|)$. Substituting these results into formula (\ref{eq:uL}), the normalized growth rate for symmetric $k^+$ decay is
\begin{equation}
\label{eq:MLi+}
\Big|\mathcal{M}_L^+\Big(\Omega_i\!\rightarrow\!\frac{\Omega_i}{2},\!\frac{\Omega_i}{2}\Big)\Big|\simeq \frac{3\sqrt{3}}{32}\frac{\Omega_i}{\omega_p}.
\end{equation}
We see in a typical plasma where $\omega_p\gg\Omega_i$, the decay mode $b\rightarrow b+b$ is orders of magnitude weaker than the other decay modes. Nevertheless, when compared with the pump frequency $\omega_1=\Omega_i$, the growth rate of the three-wave decay instability is not necessarily small.

\subsubsection{Perpendicular pump}\label{sec:LLLPerp}
In this subsection, we use another set of examples to illustrate how to evaluate the normalized growth rate $\mathcal{M}_L$, by considering the cases where the pump wave propagates perpendicular to the magnetic field. In this geometry, the pump frequency can either be the upper-hybrid frequency $\omega_{UH}$, or the lower hybrid frequency $\omega_{LH}$, in an electron-ion plasma. For three-wave decay to happen, the frequency resonance condition (\ref{eq:resonantW}) must be satisfied. Since the lower hybrid frequency $\omega_{LH}\gg\Omega_i$, it is not possible to match the frequency resonance condition with a \textit{LH} pump wave in a uniform plasma. By similar consideration, for a \textit{UH} pump wave, the decay mode $u\rightarrow u+u$ is also forbidden. However, other decays modes of the \textit{UH} pump are possible. Using expression $\omega_{UH}^2\simeq\omega_p^2+\Omega_e^2$, we see that $u\rightarrow u+b$ is always possible; $u\rightarrow u+l$ is possible if $2/\sqrt{\zeta}\lesssim r\lesssim \sqrt{\zeta}/2$, where $\zeta=M_i/Z_i\gg1$ is the normalized charge-to-mass ratio for ions; and $u\rightarrow l+l$ is possible only if $1/\sqrt{3}\leq r\leq \sqrt{3}$. Here, $r=|\Omega_e|/\omega_p$ is the ratio of electron cyclotron frequency to the plasma frequency. In this section, we will consider $r$ in the range where all three decay modes are possible. 

In addition to the frequency condition, the wave vector resonance conditions (\ref{eq:resonantK}) must also be satisfied for three-wave decay to happen. To see when Eq.~(\ref{eq:resonantK}) can be satisfied in this perpendicular geometry, it is convenient to discuss in the spherical coordinate where the polar angle $\theta$ is measured from the magnetic field $\mathbf{b}$, and the azimuthal angle $\phi$ is measured from $\mathbf{k}_1$. In this spherical coordinate, the wave vectors $\mathbf{k}_2$ and $\mathbf{k}_3$ are constrained on the two cones spanning angles $\theta_2, \pi-\theta_2$ and $\theta_3, \pi-\theta_3$. 
Then $\mathbf{k}_2$ and $\mathbf{k}_3$ can reside along the lines generated by cutting the two cones with a plane passing through $\mathbf{k}_1$. When $|\cos\theta_2|>|\cos\theta_3|$, the plane starts to intercept both cones when 
$|\cos\phi_3|\ge|\cos\phi_c|$, where the critical angle $\sin\phi_c=\tan\theta_2/\tan\theta_3$. When the strict inequality holds, for each $\mathbf{k}_3$, there are two solutions to $\mathbf{k}_2$ such that the resonance conditions (\ref{eq:resonantK}) is satisfied. By the exchange symmetry $2\leftrightarrow3$, we immediately know what happens when $|\cos\theta_2|<|\cos\theta_3|$. 
The resonance condition (\ref{eq:resonantK})  constrains where in the $\theta_2$-$\phi_2$ plane can the normalized growth rate $\mathcal{M}_L$ take nonzero values.

\begin{figure}[t]
	\includegraphics[angle=0,width=8cm]{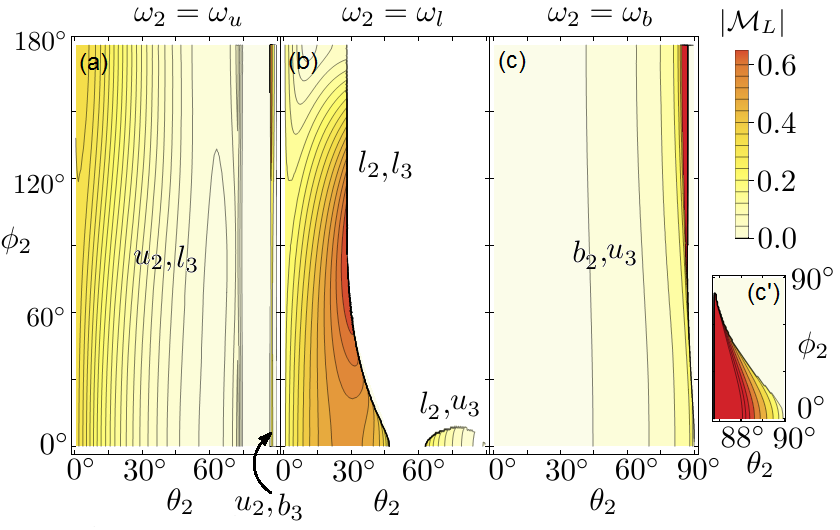}
	\caption{Normalized growth rate $|\mathcal{M}_L|$ when pumped by an upper-hybrid wave ($\mathbf{k}_1\!\perp\!\mathbf{B}_0$) in a uniform hydrogen plasma with $|\Omega_e|/\omega_p=1.2$. The growth rates are observed at polar angle $\theta_2$ with respect to $\mathbf{B}_0$ and azimuthal angle $\phi_2$ with respect to $\mathbf{k}_1$. When $\omega_2$ is on the upper resonance (a), the $u\rightarrow u_2,l_3$ decay can happen for $\theta_2<\theta_u^a$, where $\omega_u(\theta_u^a)=\omega_{UH}-\omega_{LH}$. In this case, an important decay channel has $\omega_2\sim|\Omega_e|$ propagating almost parallel to $\mathbf{B}_0$ in the backward direction, and $\omega_3\gg\omega_{LH}$ propagating almost perpendicular to $\mathbf{B}_0$ in the forward direction. This region corresponds to the $l_2,u_3$ region in (b), in which $\omega_2$ is on the lower resonance instead. The other decay mode is $u\rightarrow u_2,b_3$, which can happen in the narrow strip $\theta_2>\theta_u^b$ in (a), where $\omega_u(\theta_u^b)=\omega_{UH}-\Omega_i$. Equivalently, exchanging the labels to $b_2,u_3$, this decay mode can happen in the colored region in (c), in which $\omega_2$ is on the bottom resonance instead. For this decay mode, the dominant decay channel has $\omega_2\sim\omega_{UH}$ propagating almost perpendicular to $\mathbf{B}_0$ in the forward direction, and $\omega_3\sim\Omega_i$ propagating either in the forward or backward direction. The last decay mode is $u\rightarrow l_2,l_3$, which corresponds to the large colored region in (b). For this decay mode, the dominant decay channel is the symmetric decay, where $\omega_2\sim\omega_3\sim\omega_{UH}/2$ and both waves propagate at angles with $\mathbf{B}_0$ in the forward direction.}
	\label{fig:LLL_Perp}
\end{figure}

Having matched the resonance conditions, the normalize growth rate in the polar coordinate can be readily evaluated (Fig.~\ref{fig:LLL_Perp}). To understand the angular dependence of $\mathcal{M}_L$, it is useful to notice that due to the exchange symmetry $\mathcal{M}_L(2,3)=\mathcal{M}_L(3,2)$, the normalized growth rate $\mathcal{M}_L(\theta_2,\phi_2)$ in one region can be mapped to $\mathcal{M}_L(\theta_2',\phi_2')$ in anther region. To be more specific, when $\omega_2$ is on the upper resonance (Fig.~\ref{fig:LLL_Perp}a), the normalized growth rate $\mathcal{M}_L$ is nonzero in two regions. The first region is $\theta_2<\theta_u^a$, where $\omega_u(\theta_u^a)=\omega_{UH}-\omega_{LH}$. In this region, the decay mode $u_1\rightarrow u_2+l_3$ is allowed, where $\omega_3$ is on the lower resonance. By the exchange symmetry, this region can be mapped to the island on the bottom right corner of Fig.~\ref{fig:LLL_Perp}b, in which $\omega_2$ is on the lower resonance instead. The other region in Fig.~\ref{fig:LLL_Perp}a where $\mathcal{M}_L$ is nonzero is the narrow strip $\theta_2>\theta_u^b$, where $\omega_u(\theta_u^b)=\omega_{UH}-\Omega_i$. In this region, the decay mode $u_1\rightarrow u_2+b_3$ is allowed, where $\omega_3$ is on the bottom resonance. Exchanging $2\leftrightarrow3$, this region corresponds to the case where $\omega_2$ is on the bottom resonance instead (Fig.~\ref{fig:LLL_Perp}c). The remaining decay mode is $u_1\rightarrow l_2+l_3$, where both decay waves are on the lower resonance. This decay mode is allowed within the large region on the left of Fig.~\ref{fig:LLL_Perp}b. This region has a straight boundary at $\theta_2=\theta_l^m$, where $\omega_l(\theta_l^m)=\omega_{UH}/2$. To the left of this boundary, we have $\theta_2<\theta_3$, so there is only one solution for $k_2$. To the right of this boundary, we have $\theta_2>\theta_3$, so both  $k_2^-$ and $k_2^+$ solutions exist as long as $\sin\phi_2<\tan\theta_3/\tan\theta_2$. Whenever both solutions exist, Fig.~\ref{fig:LLL_Perp} shows the $k^-$ branch, which has weaker damping. In those degenerate cases, the $k^+$ branch is usually comparable with the $k^-$ branch. An exception is inserted in Fig.~\ref{fig:LLL_Perp}c', where the $k^+$ branch is dominant for $u_1\rightarrow b_2+u_3$ decay, corresponding to the forward scattering of the \textit{UH} pump with little frequency shift.

For the $u\rightarrow u_2+l_3$ decay (Fig.~\ref{fig:LLL_Perp}a), one important decay channel has $\omega_2\sim|\Omega_e|$ propagating almost parallel to $\mathbf{b}$ in the backward direction ($\phi_2=180^\circ$), and 
the other decay wave propagating almost perpendicular to $\mathbf{b}$ in the forward direction ($\phi_3=0^\circ$). To see how does $\mathcal{M}_L$ scales with plasma parameters, let us find its asymptotic expression when $\theta_2\rightarrow0$. In this limit $\omega_2\rightarrow|\Omega_e|$, so the magnetization factor $\gamma_{2,e}^2$ is divergent. Then the dominant terms of the coupling strength (\ref{eq:THETAL3}) comes from the $\mathbb{F}_{e,2}$ terms. The divergent inner products are $(\hat{\mathbf{k}_1}\cdot\mathbb{F}_{e,2}^*\hat{\mathbf{k}_2})\simeq-\gamma_{e,2}^2\sin\theta_2$ and $(\hat{\mathbf{k}_3}\cdot\mathbb{F}_{e,2}^*\hat{\mathbf{k}_2})\simeq-\gamma_{e,2}^2\sin\theta_2\sin\theta_3$, and we also need the finite inner products $(\hat{\mathbf{k}_1}\cdot\mathbb{F}_{e,3}^*\hat{\mathbf{k}_3})\simeq\gamma_{e,3}^2\sin\theta_3$ and $(\hat{\mathbf{k}_3}\cdot\mathbb{F}_{e,1}\hat{\mathbf{k}_1})\simeq\gamma_{e,1}^2\sin\theta_3$. Then the leading term of the normalized scattering strength is $\Theta_e\simeq ck_1\gamma_{e,1}^2\gamma_{e,2}^2\gamma_{e,3}^2 (\omega_1^2-\omega_3^2)\sin\theta_2\sin\theta_3/(\omega_1\omega_2\omega_3)$, where we have used the resonance condition $k_3\sin\theta_3=k_1$. The angle $\theta_3$ can be estimated from Eq.~(\ref{eq:resonance}) using $\omega_3\gg\Omega_i$, which gives $\sin^2\theta_3\simeq(\omega_3^2-\omega_p^2)(\omega_3^2-\Omega_e^2)/(\omega_p^2\Omega_e^2)$. Then the wave energy coefficient $u_3\simeq(2\omega_3^2-\omega_{UH}^2)(\omega_3^2-\Omega_e^2)$. As for the other two wave energy coefficients, using previous results, we know $u_1=\omega_{UH}^2/\omega_p^2$ and $u_2\simeq(\Omega_e^2-\omega_p^2)^2/(\Omega_e^2\omega_p^2\sin^2\theta_2)$. Substituting these into Eqs.~(\ref{eq:ML}) and (\ref{eq:uL}), we find the normalized growth rate 
\begin{equation}
\label{eq:MLul}
\Big|\mathcal{M}_L\big(\omega_{UH}\!\rightarrow\!|\Omega_e|,\omega_3\big)\Big|\!\simeq\! \frac{\omega_3(\omega_3+\omega_{UH})}{\omega_p\sqrt{2(\omega_{UH}^2-2\omega_3^2)}},
\end{equation}
where $\omega_3=\omega_{UH}-|\Omega_e|$ is the resonance frequency. From previous discussion, we know this decay mode can happen as long as $1/\sqrt{3}\leq r\lesssim\sqrt{\zeta}/2$. Within this parameter range, it is easy to see that Eq.~(\ref{eq:MLul}) decreases monotonically with increasing magnetic field. The maximum value $\mathcal{M}_L=\sqrt{3}/2$ is attained at $r=1/\sqrt{3}$, where $\omega_3=|\Omega_e|=\omega_{UH}/2$ such that the decay is symmetric.

For the $u\rightarrow l_2+l_3$ decay (Fig.~\ref{fig:LLL_Perp}b), the dominant decay channel is the symmetric decay, where $\omega_2=\omega_3=\omega_{1}/2$. In the symmetric decay geometry, we have $\theta_3=\pi-\theta_2$ and $\phi_3=-\phi_2$. Then the wave vector resonance condition becomes $k_2=k_3=k_1/(2\sin\theta_2\cos\phi_2)$. The symmetric decay angle $\theta_2=\theta_s$ can be estimated from Eq.~(\ref{eq:resonance}) using $\omega_2=\omega_{UH}/2\gg\Omega_i$, which gives $\cos^2\theta_s\simeq3\omega_{UH}^4/(16\omega_p^2\Omega_e^2)$. Since the frequencies are far away from cyclotron frequencies, all the magnetization factors are finite. Then the inner products $(\hat{\mathbf{k}_1}\cdot\mathbb{F}_{s,2}^*\hat{\mathbf{k}_2}) \simeq\gamma_{s,2}^2(\cos\phi_2+i\beta_{s,2}\sin\phi_2)\sin\theta_2$, $(\hat{\mathbf{k}_2}\cdot\mathbb{F}_{s,1}\hat{\mathbf{k}_1})\simeq\gamma_{s,1}^2(\cos\phi_2+i\beta_{s,1}\sin\phi_2)\sin\theta_2$,  $(\hat{\mathbf{k}_3}\cdot\mathbb{F}_{s,2}^*\hat{\mathbf{k}_2})\simeq-1+\gamma_{s,2}^2\sin^2\theta_2(2\cos^2\phi_2+i\beta_{s,2}\sin2\phi_2-\beta_{s,2}^2)$, and by exchanging $2\leftrightarrow 3$, we can easily find the other three inner products. Substituting these inner products into Eq.~(\ref{eq:THETAL3}), the normalized scattering strength becomes particularly simple when $\phi_2\rightarrow\pi/2$. In this limit $k_2,k_3\rightarrow\infty$, but the products $k_2\cos\phi_2=-k_3\cos\phi_3$ remains finite. Keeping nonzero terms as $\phi_2\rightarrow\pi/2$, the scattering strength simplifies to $\Theta_e^+\simeq- 2ck_1\omega_{UH}^3/[\omega_p^2(3\Omega_e^2-\omega_p^2)]$. The electron terms also dominate the wave energy coefficients $u_2=u_3\simeq2\omega_{UH}^2/(3\Omega_e^2-\omega_p^2)$. Gathering the above results, the normalized growth rate for symmetric $k^+$ scattering is
\begin{equation}
\label{eq:MLll}
\Big|\mathcal{M}_L^+\big(\omega_{UH}\!\rightarrow\!\frac{\omega_{UH}}{2},\frac{\omega_{UH}}{2}\big)\Big|\!\simeq\! \frac{\omega_p}{\omega_{UH}}.
\end{equation}
The above special value of $\mathcal{M}_L$ is approximately the maximum in Fig.~\ref{fig:LLL_Perp}b, where $\theta_2=\theta_s$ and $\phi_2=90^\circ$. Notice that this special case is singular in wave vector $k_2,k_3\rightarrow\infty$, and hence will be suppressed by wave damping. Therefore, the dominant decay channels observed in experiment will happen at smaller angle $\phi_2<90^\circ$ in the symmetric decay geometry.

Finally, for the $u\rightarrow b_2+u_3$ decay (Fig.~\ref{fig:LLL_Perp}c), the dominant decay channel has $\omega_2\sim\omega_{UH}$ propagating almost perpendicular to $\mathbf{b}$ in the forward direction, and $\omega_3\sim\Omega_i$ propagating either in the forward or backward direction. As an example, let us consider symmetric forward scattering where $\phi_2=\phi_3=0$ and $\theta_2=\pi-\theta_3=\theta_s$. In this geometry, $k_2^-=k_3^-=k_1/(2\sin\theta_s)$. Since $\theta_s\sim\pi/2$, we can estimate the symmetric angle using asymptotic expressions Eqs.~(\ref{eq:wu_perp}) and (\ref{eq:wb_perp}). Substituting these expressions into he frequency resonance condition (\ref{eq:resonantW}), we obtain $\cos^2\theta_s\simeq2\Omega_i\omega_{UH}^3/(\Omega_e^2\omega_p^2)\sim0$, where we have used that $\omega_p^2|\Omega_e|/(2\omega_{UH}^3)\lesssim0.2$ is always a small number. Then the wave energy $u_2\simeq u_1=\omega_{UH}^2/\omega_p^2$, and $u_3\simeq\omega_p^2[1+2\omega_{UH}^3/(\omega_p^2|\Omega_e|)]^2/(\Omega_i|\Omega_e|)$. Now that the magnetization factors are all finite, the inner products are simply $(\hat{\mathbf{k}_1}\cdot\mathbb{F}_{s,2}^*\hat{\mathbf{k}_2}) \simeq\gamma_{s,2}^2\sin\theta_2$, $(\hat{\mathbf{k}_2}\cdot\mathbb{F}_{s,1}\hat{\mathbf{k}_1})\simeq\gamma_{s,1}^2\sin\theta_2$,  $(\hat{\mathbf{k}_3}\cdot\mathbb{F}_{s,2}^*\hat{\mathbf{k}_2})\simeq\cos\theta_3\cos\theta_2+\gamma_{2,s}^2\sin\theta_3\sin\theta_2$, and the three other inner products can be obtained by exchanging $2\leftrightarrow 3$. Again, the scattering is mostly due to electrons, for which $\gamma_{e,1}^2\simeq\gamma_{e,2}^2\simeq\omega_{UH}^2/\omega_p^2$ and $\gamma_{e,3}^2\simeq-\omega_3^2/\Omega_e^2\ll\cos\theta^2_s$. Therefore, the dominant term comes from the second line of Eq.~(\ref{eq:THETAL3}), which gives the scattering strength $\Theta_e^-\simeq- ck_1\Omega_i\omega_{UH}^5/(\omega_3\Omega_e^2\omega_p^4)$. Substituting these results into formula (\ref{eq:ML}) and (\ref{eq:uL}), we immediately see that the normalized growth rate for forward scattering is
\begin{equation}
\label{eq:MLub}
\Big|\mathcal{M}_L^-\big(\omega_{UH}\!\rightarrow\!\omega_{UH},\Omega_i\big)\Big|\!\simeq\! \frac{\omega_p}{4\sqrt{\omega_{UH}|\Omega_e|}}\bigg(\frac{\omega_3}{\Omega_i}\bigg)^{1/2},
\end{equation}
where $\omega_3=\omega_b(\theta_s)\sim\Omega_i$ can be obtained from Eq.~(\ref{eq:wb_perp}). Using the above result, we can also find the symmetric nearly backward scattering $\mathcal{M}_L^+$ by replacing the coefficient $1/4$ with $k_2^+/(2k_1)$. The symmetric nearly backward scattering channel has divergent $k_2^+$, and therefore can have very large growth rate in the absence damping.   

\section{Conclusion and discussion}\label{sec:discussion}
In summary, we solve the cold fluid-Maxwell system to second order in the multiscale perturbation series in the most general geometry (Sec.~\ref{sec:fluid}), where a discrete spectrum of waves interact in triplets through quadratic nonlinearities [Eq.~(\ref{eq:E2s})]. Due to nonlinear interactions, three-wave scatterings change the envelopes of ``on-shell" waves as they advect, as well as generate a spectrum of `off-shell" waves due to wave beating. The coupling of wave triplets are described by the scattering strength $\mathbf{S}_{\mathbf{q},\mathbf{q}'}$ [Eq.~(\ref{eq:S})], which includes the effects of the $\mathbf{v}_{s1}\times\mathbf{B}_1$ nonlinearity 
, the $\mathbf{v}_{s1}\cdot\nabla_{(0)}\mathbf{v}_{s1}$ nonlinearity 
, as well as the $\nabla_{(0)}\cdot(n_{s1}\mathbf{v}_{s1})$ nonlinearity
. By introducing the forcing operator [Eq.~(\ref{eq:F})], we manage to give a convenient formula [Eq.~(\ref{eq:R})] for the three-wave scattering strength in the most general geometry.  

When there are only three resonant ``on-shell" waves participating in the interaction (Sec.~\ref{sec:3waves}), the three scattering strengths [Eq.~(\ref{eq:S23})] are closely related to one another due to action conservation. The action conservation laws are manifested by the three-wave equations [Eqs.~(\ref{eq:3waves1})-(\ref{eq:3waves3})], which describe how the amplitudes of waves evolve, regardless of the changes in their phases and polarizations. The three-wave equations contain one essential parameter, the coupling coefficient [Eq.~(\ref{eq:coupling})], whose explicit formula is given in terms of the wave energy coefficient [Eq.~(\ref{eq:Ucoef})] and the normalized scattering strength [Eq.~(\ref{eq:Theta3})]. The coupling coefficient contains five degrees of freedom, and can be readily evaluated once the participating waves and their geometry are specified.

The general formula of the scattering strength becomes particularly transparent once we quantize the classical three-wave Lagrangian. Using the quantized Lagrangian [Eq.~(\ref{eq:Lagrangian})], all six terms of the scattering strength arise from a single cubic interaction $\propto P^i(\partial_iA_j)J^j$ as six permutations of the Feynman diagrams [Eq.~(\ref{eq:Feynman})]. 
We postulate that this form of the three-wave interaction is independent of the plasma model that one uses to calculate the linear response. In this paper, the linear response is calculated using the cold fluid model. More generally, the linear response may be calculated using the kinetic model or even quantum models. Then, using the relation between the \textit{S} matrix element and the three-wave scattering strength [Eq.~(\ref{eq:MTheta})], the three-wave coupling coefficient may be directly computed without going through the perturbative solution of the equations.

To demonstrate how to evaluate the cold fluid coupling coefficient, we give a set of examples where all three participating waves are either quasi-transverse (\textit{T}) or quasi-longitudinal (\textit{L}) (Sec.~\ref{sec:quasi}). As an experimental observable, we compute the growth rate of the three-wave decay instability [Eq.~(\ref{eq:GrowthRate})], which is proportional to the coupling coefficient when wave damping is ignored. For \textit{TTL} decay (Sec.~\ref{sec:quasi}.\ref{sec:TTL}), the scattering is due to density perturbation of the \textit{L} wave, and the normalized growth rate is given conveniently by formula Eqs.~(\ref{eq:MT}) and (\ref{eq:uT}). For \textit{LLL} decay (Sec.~\ref{sec:quasi}.\ref{sec:LLL}), the scattering is due to density beating of three \textit{L} waves, and the normalized growth rate is given by the explicit formula Eqs.~(\ref{eq:THETAL3}), (\ref{eq:ML}) and (\ref{eq:uL}). We evaluate these formulas numerically for the cases where the pump wave is either parallel or perpendicular to the magnetic field, while the decay waves propagate at arbitrary angles. To facilitate understanding of the angular dependences, we also find asymptotic expressions of the normalized growth rate in limiting cases.

The above examples elucidate the previously unknown angular dependence of three-wave scattering when strong magnetic field is present. In contrast to the unmagnetized case, backscattering is not necessarily the fastest growing instability in a magnetized plasma. For example, in the \textit{TTL} scattering (Fig.~\ref{fig:TTL_Para},\ref{fig:TTL_Perp},\ref{fig:TTL_Perp1D}), which happens when two lasers interact via a magnetic resonance, exact backscattering may be suppressed, while nearly perpendicular scattering may be enhanced. For another example, in the \textit{LLL} scattering (Fig.~\ref{fig:LLL_Para},\ref{fig:LLL_Perp}), which can happen when an electrostatic wave launched by antenna arrays decay to two other longitudinal waves, symmetric decays are usually favored whenever possible, but asymmetric decays can also be important at special angles. 

The above collisionless, cold, fluid results will need to be modified when kinetic or collisional effects become important. Besides wave damping [Eq.~(\ref{eq:GrowthRateDamped})], a major modification comes from the alternation of the linear eigenmode structure, which will be constituted of Bernstein waves instead of the hybrid waves. In addition, weak coupling results obtained in this paper will need to be modified when three-wave interactions becomes strong. This happens when wave amplitudes become nonperturbative, so that relativistic effects becomes non-negligible, and linear eigenmode structure becomes strongly distorted. 

Despite of the above caveats, the importance of this work is twofold. First, the formulation we develop in this paper preserves the general mathematical structure, thereby enables profound simplifications of the most general results, from which illuminating physical consequences can be extracted. Second, the uniform, collisionless, and cold fluid results we have obtained serve as the baseline for understanding angular dependence of three-wave scattering in magnetized plasmas, which is important both for magnetic confinement devices, as well as laser-plasma interactions in magnetized environment.

\begin{acknowledgments} 
This research is supported by NNSA Grant No. DE-NA0002948 and DOE Research Grant No. DE-AC02-09CH11466.
\end{acknowledgments}

\begin{appendices}
\appendix
\counterwithin{figure}{section}
\renewcommand{\appendixname}{APPENDIX}

\section{\label{app:Multiscale}\MakeUppercase{Multiscale Perturbative Solution of System of ODEs}}
In Sec.~\ref{sec:fluid}, we use a multiscale expansion to solve a system of nonlinear hyperbolic partial differential equations. To facilitate understanding of the multiscale expansion, here we demonstrate how it can be successfully applied to the following system of ordinary differential equations, which are hyperbolic in the absence of perturbations
\begin{eqnarray}
\label{eq:dx}
\dot{x}&=&\phantom{+}y+\epsilon f(x,y),\\
\label{eq:dy}
\dot{y}&=&-x+\epsilon g(x,y).
\end{eqnarray}
where $\dot{x}$ and $\dot{y}$ denotes the time derivatives of $x(t)$ and $y(t)$, respectively, $f$ and $g$ are some polynomials, and $\epsilon\ll1$ is a small parameter enabling us to find the perturbative solution.

The above system of equations may be solved perturbatively using the expansion
\begin{eqnarray}
\label{eq:x}
x(t)&=&x_0(t)+\epsilon x_1(t)+\epsilon^2x_2(t)+\dots,\\
\label{eq:y}
y(t)&=&y_0(t)+\epsilon y_1(t)+\epsilon^2y_2(t)+\dots.
\end{eqnarray}
However, naive perturbative solution using only the above expansions will fail due to nonlinearity, by which the notorious secular terms will arise, which increase monotonically in time, and will quickly render the perturbative solutions invalid. To remove the secular terms, we need to also expand the time scales
\begin{eqnarray}
\label{eq:t}
t&=&t_0+\frac{1}{\epsilon} t_1+\frac{1}{\epsilon^2}t_2+\dots,\\
\label{eq:dt}
\partial_t&=&\partial_{0}+\epsilon\partial_{1}+\epsilon^2\partial_{2}+\dots,
\end{eqnarray}
where one unit of the slow time scale $t_n$ worth $1/\epsilon^n$ units of the fastest time scale $t_0$. By regarding different time scales as independent variables, the total time derivative is expressed in terms of the summation of derivative on each time scale $\partial_n:=\partial/\partial t_n$ using the chain rule. Substituting expansions (\ref{eq:x})-(\ref{eq:dt}) into Eqs.~(\ref{eq:dx}) and (\ref{eq:dy}) and collect terms according to their order in $\epsilon$, we can obtain a series of equations. 

The $\epsilon^0$-order equations are simply the equations for a simple harmonic oscillator
\begin{eqnarray}
\partial_0x_0-y_0&=&0,\\
\partial_0y_0+x_0&=&0.
\end{eqnarray}
For real valued $x$ and $y$, the general solution is well-known
\begin{eqnarray}
\label{eq:x0}
x_0&=&a_0e^{it_0}+c.c.,\\
\label{eq:y0}
y_0&=&ia_0e^{it_0}+c.c.,
\end{eqnarray}
where $c.c.$ stands for complex conjugate, and the complex amplitude $a_0=a_0(t_1,t_2,\dots)$ can be a function of slow variables. If we truncate the solution on this order, then $x$ and $y$ oscillate harmonically with constant amplitude. On the other hand, if we move on to the next order, perturbations $\epsilon f(x,y)$ and $\epsilon g(x,y)$ will in general cause the amplitude $a_0$ to vary on slow time scales, which will be described by higher order equations. 

The $\epsilon^1$-order equations start to couple perturbations on different time scales
\begin{eqnarray}
\partial_1x_0+\partial_0x_1-y_1-f_0&=&0,\\
\partial_1y_0+\partial_0y_1+x_1-g_0&=&0,
\end{eqnarray}
where $f_0:=f(x_0,y_0)$ and $g_0:=g(x_0,y_0)$, in which $x_0$ and $y_0$ are given by Eqs.~(\ref{eq:x0}) and (\ref{eq:y0}). The above two equations contain three unknowns $x_1,y_1$, and $\partial_1a_0$. Therefore, we can use the extra degree of freedom to remove secular terms. To do that, let us first separate variables $x_1$ and $y_1$ and rewrite the $\epsilon^1$-order equations as
\begin{eqnarray}
\label{eq:2x1}
\partial_0^2x_1+x_1+2\partial_1y_0&=&u_1,\\
\label{eq:2y1}
\partial_0^2y_1+y_1-2\partial_1x_0&=&v_1,
\end{eqnarray}
where the source terms are
\begin{eqnarray}
u_1[a_0]&:=&\partial_0f_0+g_0,\\
v_1[a_0]&:=&\partial_0g_0-f_0.
\end{eqnarray}
Substituting $x_0$ and $y_0$ into polynomials $f$ and $g$, we can write $f_0=\sum_nf_{0n}e^{int_0}+c.c.$, and $g_0=\sum_ng_{0n}e^{int_0}+c.c$, where $f_{0n}$ and $g_{0n}$ are some functionals of $a_0$. Then the source terms can be written similarly as $u_1=\sum_n u_{1n}e^{int_0}+c.c.$ and $v_1=\sum_n v_{1n}e^{int_0}+c.c.$, where $u_{1n}=g_{0n}+inf_{0n}$ and $v_{1n}=-f_{0n}+ing_{0n}$. 

To solve the $\epsilon^1$-order equations (\ref{eq:2x1}) and (\ref{eq:2y1}), we can match coefficients of Fourier exponents and split the equations into two sets . The first set of equations govern how the amplitude $a_0$ evolves on the slow time scale $t_1$, which can be written as $\partial_1x_0=-\frac{1}{2}(v_{11}e^{it_0}+c.c.)$ and $\partial_1y_0=\frac{1}{2}(u_{11}e^{it_0}+c.c.)$. These two equations are redundant, as can be seen from the relations between $x_0$ and $y_0$, as well as the definitions of $u_{11}$ and $v_{11}$. Both of these equations results in the same equation for $a_0$, which absorbs the secular term
\begin{equation}
\label{eq:d1a0}
\partial_1a_0=\frac{1}{2}(f_{01}-ig_{01}),
\end{equation}
where the right-hand-side is some functional of $a_0$. This first order ODE of $a_0$ can usually be integrated, from which $a_0$ will be a known function of $t_1$. The other sets of equations governs $x_1$ and $y_1$
\begin{eqnarray}
\label{eq:x1s}
\partial_0^2x_1+x_1&=&\sum_{n\ne 1}u_{1n}e^{int}+c.c.,\\
\label{eq:y1s}
\partial_0^2y_1+y_1&=&\sum_{n\ne 1}v_{1n}e^{int}+c.c.
\end{eqnarray}
Having removed the secular terms, the above equations are now secular-free, and can be readily solved by
\begin{eqnarray}
\label{eq:x1}
x_1=a_1e^{it_0}+\sum_{n\ne 1}\frac{u_{1n}}{1-n^2}e^{int_0}+c.c.,\\
\label{eq:y1}
y_1=b_1e^{it_0}+\sum_{n\ne 1}\frac{v_{1n}}{1-n^2}e^{int_0}+c.c..
\end{eqnarray}
The amplitudes $a_1$ and $b_1$ are clearly related by the $\epsilon^1$-order equations, which give
\begin{equation}
b_1=ia_1-\frac{1}{2}(f_{01}+ig_{01}).
\end{equation}
Notice that in the perturbation series Eq.~(\ref{eq:x}), we can always redefine $a_0+\epsilon a_1\rightarrow a_0'$. Hence it is sufficient to set the amplitude $a_1=0$. In this way, we will obtain a $x$-majored solution, where the amplitude of $e^{it_0}$ for $x$ is completely given by $a_0$, whereas amplitude $e^{it_0}$ for $y$ is given by the summation $b_0+\epsilon b_1+\dots$. Alternatively, by setting $b_1=0$, we can of course also obtain a $y$-majored solution, which we will not pursue here. For three-wave scattering studied in this paper, it is sufficient to truncate at this order. Then the solution is constituted of oscillations with slowly varying amplitudes.

To show the general structure of the multiscale expansion, here, it is instructive to carry out the solution to the next order. The $\epsilon^2$-order equations are
\begin{eqnarray}
\partial_2x_0+\partial_1x_1+\partial_0x_2-y_2-f_1&=&0,\\
\partial_2y_0+\partial_1y_1+\partial_0y_2+x_2-g_1&=&0,
\end{eqnarray}
where $f_1:=x_1\partial_xf_0+y_1\partial_yf_0$ and $g_1=x_1\partial_xg_0+y_1\partial_yg_0$. In the above two equations, there are three unknowns $x_2$, $y_2$ and $\partial_2a_0$. So again, we can use the extra degree of freedom to remove the secular terms. Separating variables $x_2$ and $y_2$, we can rewrite the equations as
\begin{eqnarray}
\label{eq:2x2}
\partial_0^2x_2+x_2+2\partial_2y_0&=&u_2\\
\label{eq:2y2}
\partial_0^2y_2+y_2-2\partial_2x_0&=&v_2.
\end{eqnarray}
Since we set $a_1=0$ for the $x$-majored solution, the source terms are functionals of $a_0$ only
\begin{eqnarray}
u_2[a_0]:&=&\partial_0f_1+g_1+\partial_1^2x_0-2\partial_1y_1-\partial_1 f_0,\\
v_2[a_0]:&=&\partial_0g_1-f_1+\partial_1^2y_0+2\partial_1x_1-\partial_1 g_0.
\end{eqnarray}
Again, since $f$ and $g$ are polynomials, we can write $f_1=\sum_nf_{1n}e^{int_0}+c.c.$, and $g_1=\sum_ng_{1n}e^{int_0}+c.c$. Then the source terms can be written similarly as $u_2=\sum_n u_{2n}e^{int_0}+c.c.$ and $v_2=\sum_n v_{2n}e^{int_0}+c.c.$, where $v_{21}=iu_{21}=i\partial_1^2a_0+ig_{11}-\partial_1g_{01}-f_{11}$, and for $n\ge2$, we have $u_{2n}=inf_{1n}-\partial_1f_{0n}+g_{1n}-2\partial_1v_{1n}/(1-n^2)$ and  $v_{2n}=ing_{1n}-\partial_1g_{0n}+f_{1n}+2\partial_1u_{1n}/(1-n^2)$.

To solve the $\epsilon^2$-order equations (\ref{eq:2x2}) and (\ref{eq:2y2}), we can use similar procedure to split the equations into two sets. The first set of equations are again redundant, and can be written as a single equation governing how the amplitude $a_0$ evolve on the slow time scale $t_2$
\begin{equation}
\label{eq:d2a0}
\partial_2a_0=\frac{1}{2}(f_{11}-ig_{11})-\frac{i}{4}\partial_1(f_{01}+ig_{01}).
\end{equation}
Regarding $t_1$ as a parameter, the above equation is a first order ODE for $a_0(t_2)$, which can usually be integrated. The second sets of equations are similar to Eqs.~(\ref{eq:x1s}) and (\ref{eq:y1s}), with $u_{1n}$ and $v_{1n}$ replaced by $u_{2n}$ and $v_{2n}$, respectively. The solutions to these secular-free equations are similar to  Eqs.~(\ref{eq:x1}) and (\ref{eq:y1}) with the order index ``1" replaced by the order index ``2", in which the second order amplitudes $a_2$ and $b_2$ are again related by the $\epsilon^2$-order equations
\begin{equation}
b_2=ia_2-\frac{1}{2}(f_{11}+ig_{11})-\frac{i}{4}\partial_1(f_{01}+ig_{01}).
\end{equation}
To obtain the $x$-majored solution, we again set $a_2$ to zero. By the obvious analogy between the $\epsilon^1$- and $\epsilon^2$-order equations, the above procedures can be readily extended to higher order in the perturbation series.

In summary, using the multiscale expansion Eqs.~(\ref{eq:x})-(\ref{eq:dt}), we convert a system of ODEs (\ref{eq:dx})-(\ref{eq:dy}) to a series of equations. The general solution for a system of hyperbolic ODEs is rapid oscillations with slowly varying amplitudes. The first order amplitude equation Eq.~(\ref{eq:d1a0}) governs how the amplitude varies on $t_1$ time scale, and the higher order amplitude equations, such as Eq.~(\ref{eq:d2a0}) governs how the amplitude evolves on even slower time scales. By summing up solutions on each order, which may include not only oscillations with fundamental frequency, but also higher harmonics such as Eqs.~(\ref{eq:x1}) and (\ref{eq:y1}), we can obtain a perturbative solution to the system of ODEs, majored in any one of its variables.

To see how the multiscale expansion work in practice, interested readers are encouraged to test it on the following two examples. The first is a linear example, where $f(x,y)=-x$ and $g(x,y)=0$. The exact solution to this linear case can be easily obtained. The second is a nonlinear example, where $f(x,y)=0$ and $g(x,y)=-x+2x^3$. The exact solutions to this nonlinear case are the Jacobi elliptic functions. One can expand the exact solutions in $\epsilon$, and check order by order that it matches the perturbative solution obtained using the multiscale expansion.

\section{\label{app:Linear}\MakeUppercase{Linear waves in cold magnetized plasmas}}
In Sec.~\ref{sec:fluid}.\ref{sec:first}, we obtain the first order electric field equation (\ref{eq:E1Fourier}) in the momentum space. The solutions to this matrix equation give the linear eigenmodes of the cold fluid-Maxwell system. In this appendix, we review properties of the linear waves, in order to facilitate understanding of their scatterings discussed in this paper.  

To discuss properties of the linear waves, it is convenient to choose the coordinate system where the uniform magnetic field is in the $z$-direction. In this coordinate, the forcing operator Eq.~(\ref{eq:F}) has matrix representation
\begin{eqnarray}
\mathbb{F}_{s,\mathbf{k}} =
\left( \begin{array}{ccc}
\gamma^2_{s,\mathbf{k}} & i\beta_{s,\mathbf{k}}\gamma^2_{s,\mathbf{k}} & 0 \\
-i\beta_{s,\mathbf{k}}\gamma^2_{s,\mathbf{k}} & \gamma^2_{s,\mathbf{k}} & 0 \\
0 & 0 & 1
\end{array} \right).
\end{eqnarray}
Having fixed the $z$-axis, we can rotate the coordinate system, such that the wave vector $\mathbf{k}=(k_\perp,0,k_\parallel)=k(\sin\theta,0,\cos\theta)$, where $\theta$ is the angle between $\mathbf{k}$ and $\mathbf{b}$. In this coordinate system, the matrix representation of the dispersion tensor (\ref{eq:Dk}) can be easily found. 
Then the first order electric field equation $\mathbb{D}_{\mathbf{k}}\mathbf{\mathcal{E}}^{(1)}_\mathbf{k}/\omega_{\mathbf{k}}^2=0$ can be written as
\begin{eqnarray}
\label{eq:E}
\left( \begin{array}{ccc}
S-n_\parallel^2 & -iD & n_\perp n_\parallel \\
iD & S-n^2 & 0 \\
n_\perp n_\parallel & 0 & P-n_\perp^2
\end{array} \right)
\left(\begin{array}{c} \mathcal{E}_x^{(1)} \\ \mathcal{E}_y^{(1)} \\\mathcal{E}_z^{(1)} \end{array}\right)=0,
\end{eqnarray}
where $n=ck/\omega$ is the refractive index, $n_\perp=n\sin\theta$, and $n_\parallel=n\cos\theta$ are projections in the perpendicular and parallel directions. Following Stix's notations \cite{Stix92}, the components of the dielectric tensor are
\begin{eqnarray}
S&=&1-\sum_s\frac{\omega_{ps}^2}{\omega^2-\Omega_s^2},\\
D&=&\sum_s\frac{\Omega_s}{\omega}\frac{\omega_{ps}^2}{\omega^2-\Omega_s^2},\\
P&=&1-\sum_s\frac{\omega_{ps}^2}{\omega^2}.
\end{eqnarray}
In the above expressions, we have omitted the $\mathbf{k}$-subscripts for both $\omega$ and $\mathcal{E}^{(1)}$. The expressions for $S$ and $D$ can be simplified, using identities in quasi-neutral electron-ion plasma, in which $n_e=Z_in_i$, so $\Omega_i\omega_{pe}^2+\Omega_e\omega_{pi}^2=0$ and $\Omega_i^2\omega_{pe}^2+\Omega_e^2\omega_{pi}^2+\omega_p^2\Omega_e\Omega_i=0$, where $\omega_p^2=\sum_s\omega_{ps}^2$ is the plasma frequency squared. 

\begin{figure}[t]
	\includegraphics[angle=0,width=8.5cm]{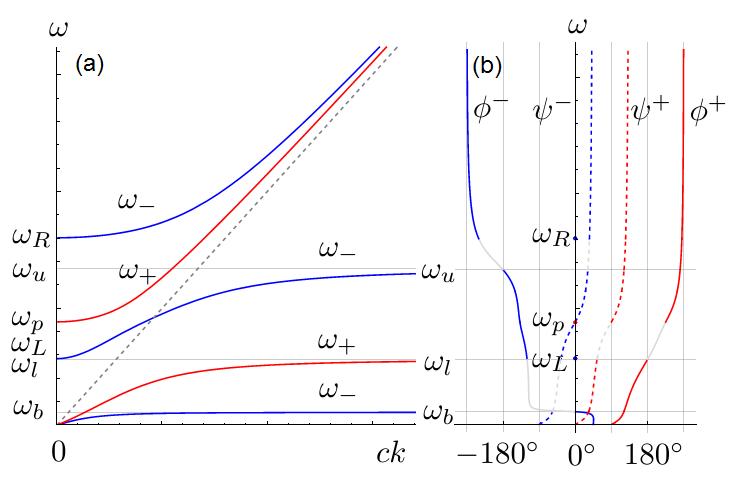}
	\caption{Linear wave dispersion relations (a) and polarization angles (b) in a cold electron-ion plasma with $m_i/m_e=10$ and $|\Omega_e|/\omega_{pe}=1.2$, when $\theta=45^\circ$. Both the $n^2_+$ (red) and the $n^2_-$ (blue) solutions contain an electromagnetic-like branch, and electrostatic-like branches. The electromagnetic-like branches asymptotes to vacuum light wave $\omega\rightarrow ck$ as $k\rightarrow\infty$, where the waves become transverse ($\phi\rightarrow90^\circ$, mod $180^\circ$). The electrostatic-like branches asymptotes to resonances $\omega\rightarrow \omega_r$ as $k\rightarrow\infty$, where the waves become longitudinal ($\phi\rightarrow 0^\circ$, mod $180^\circ$). The waves are in general elliptically polarized ($\psi\ne0^\circ$, mod $90^\circ$), except at special angles, . }
	\label{fig:Linear}
\end{figure}

The electric field equation (\ref{eq:E}) has nontrivial solution if and only if the dispersion tensor is degenerate. This is equivalent to requiring the determinant of the dispersion tensor to be zero, which gives a constraint between $\omega$ and $\mathbf{k}$, called the dispersion relation. In the above coordinate system, using Stix's notation, the dispersion relation can be written as
\begin{equation}
\label{eq:disp}
An^4-Bn^2+C=0,
\end{equation}
where the coefficients of the quadratic equation of $n^2$ are
\begin{eqnarray}
A&=&S\sin^2\theta+P\cos^2\theta,\\
B&=&RL\sin^2\theta+PS(1+\cos^2\theta),\\
C&=&PRL,
\end{eqnarray}
which are functions of $\omega$ only, independent of the wave vector. In the above expressions, $R=S+D$ and $L=S-D$ are the right- and left-handed components of the dielectric tensor. The quadratic dispersion relation (\ref{eq:disp}) has two solutions $n_{\pm}^2=(B\pm F)/(2A)$, where $F^2=B^2-4AC=(RL-PS)^2\sin^4\theta+4P^2D^2\cos^2\theta$. Since $F^2\ge0$, we see the two solutions $n_{\pm}^2$ are both real. However, $n_{\pm}^2$ is not always positive, so each solution may contain many branches, emanating from cutoff frequencies $\omega_c$, at which $C(\omega_c)=0$ so that $n^2=0$. For example, in electron-ion plasma (Fig.~\ref{fig:Linear}a), the cutoff frequencies are at $\omega_R$, $\omega_p$, and $\omega_L$, and the dispersion relation contains two electromagnetic-like branches, for which $\omega\rightarrow ck$ as $k\rightarrow\infty$, as well as three electrostatic-like branches, for which $\omega\rightarrow \omega_r$ as $k\rightarrow\infty$, where $\omega_r$ is some resonance frequencies.

\begin{figure}[b]
	\includegraphics[angle=0,width=8.5cm]{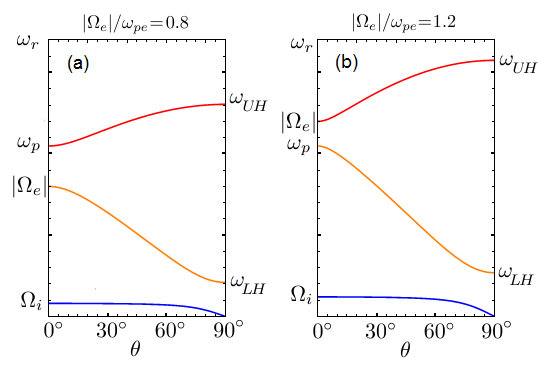}
	\caption{Resonance frequencies in electron-ion plasma with $m_i/m_e=10$. In over-dense plasma (a), as $\theta$ increases from $0^\circ$ to $90^\circ$, the upper resonance (red) increases from $\omega_p$ to $\omega_{UH}$; the lower resonance (orange) decreases from $|\Omega_e|$ to $\omega_{LH}$; and the bottom resonance (blue) decreases from $\Omega_i$ to zero. In under-dense plasma (b), as $\theta$ increases from $0^\circ$ to $90^\circ$, the upper resonance (red) increases from $|\Omega_e|$ to $\omega_{UH}$; the lower resonance (orange) decreases from $\omega_p$ to $\omega_{LH}$; and the bottom resonance (blue) decreases from $\Omega_i$ to zero. This figure can be used to read out the frequency shift $\Delta\omega$, once the scattering angle of the longitudinal wave is known.}
	\label{fig:Resonance}
\end{figure}

The resonance frequencies are asymptotic values of $\omega$ on electrostatic branches when $k\rightarrow\infty$. As the frequency approaches the resonance frequencies from the below, the refractive index $n^2_{\pm}\rightarrow\infty$, so we can find $\omega_r$ by solving $A(\omega_r)=0$. In electron-ion plasma, this equation for resonance frequencies can be explicitly written as
\begin{eqnarray}
\label{eq:resonance}
0&=&\omega_r^6-\omega_r^4(\omega_p^2+\Omega_e^2+\Omega_i^2)-\omega_p^2\Omega_e^2\Omega_i^2\cos^2\theta\\
\nonumber
&+&\omega_r^2[\omega_p^2(\Omega_e^2+\Omega_i^2)\cos^2\theta-\omega_p^2\Omega_e\Omega_i\sin^2\theta+\Omega_e^2\Omega_i^2].
\end{eqnarray}
The above cubic equation for $\omega_r^2$ has three solutions (Fig.~\ref{fig:Resonance}), which can be ordered from large to small as the upper $(\omega_u)$, lower $(\omega_l)$, and bottom $(\omega_b)$ resonance. When $\theta\sim 0$ or $\pi$, the resonance frequencies approaches $\omega_p, |\Omega_e|$, and $\Omega_i$. Keeping the next order angular dependence, the three resonance frequencies, when $\sin\theta\sim0$, can be approximated by
\begin{eqnarray}
\label{eq:wu_para}
\frac{\omega_r^2}{\omega_p^2}&\simeq&1-\frac{\Omega_e^2\sin^2\theta}{\Omega_e^2(2-\cos^2\theta)-\omega_p^2},\\
\label{eq:wl_para}
\frac{\omega_r^2}{\Omega_e^2}&\simeq&1-\frac{\omega_p^2\sin^2\theta}{\omega_p^2(2-\cos^2\theta)-\Omega_e^2},\\
\label{eq:wb_para}
\frac{\omega_r^2}{\Omega_i^2}&\simeq&1-\frac{\Omega_i}{|\Omega_e|}\tan^2\theta.
\end{eqnarray}
In the other limit, when $\theta\sim \pi/2$, the resonance frequencies approaches the upper-hybrid frequency $\omega_{UH}$, the lower hybrid frequency $\omega_{LH}$, and $0$. Keeping the next order angular dependence, the upper, lower, and bottom resonance frequencies, when $\cos\theta\sim 0$, can be approximated by 
\begin{eqnarray}
\label{eq:wu_perp}
\frac{\omega_u^2}{\omega_{UH}^2}&\simeq&1- \frac{\omega_p^2\Omega_e^2\cos^2\theta}{(\omega_p^2+\Omega_e^2)^2+\omega_p^2\Omega_e^2\cos^2\theta},\\
\label{eq:wl_perp}
\frac{\omega_l^2}{\omega_{LH}^2}&\simeq&1+\frac{\Omega_e^2\cos^2\theta}{\Omega_e^2\cos^2\theta+|\Omega_e|\Omega_i(1+\cos^2\theta)},\\
\label{eq:wb_perp}
\frac{\omega_b^2}{\Omega_i^2}
&\simeq&\frac{|\Omega_e|\cos^2\theta}{\Omega_i+|\Omega_e|\cos^2\theta}.
\end{eqnarray}
The above asymptotic expressions for resonance frequency $\omega_r$ are extremely useful when we approximate the scattering strength and wave energy coefficients.

When frequencies approaches resonances, the waves becomes longitudinal. On the other hand, the wave becomes transverse when frequencies approaches infinity. For intermediate frequencies, we can find the wave polarization by solving for eigenmodes of the electric field equation (\ref{eq:E}). 
In the wave coordinate $\hat{\mathbf{k}}, \hat{\mathbf{y}}$, and $\hat{\mathbf{k}}\times \hat{\mathbf{y}}$, we can write $\mathcal{E}_k=\mathcal{E}\cos\phi$, $\mathcal{E}_y=-i\mathcal{E}\sin\phi\cos\psi$, and $\mathcal{E}_\times=\mathcal{E}\sin\phi\sin\psi$, where we have omitted the superscript of $\mathcal{E}^{(1)}$. Then the polarization angles
\begin{eqnarray}
\tan\psi&=&\frac{Sn^2-RL}{n^2D\cos\theta}, \\
\tan\phi&=&\frac{P\cos\theta}{(n^2-P)\sin\theta\sin\psi}.
\end{eqnarray}
Notice that $\mathcal{E}_\times/\mathcal{E}_y=i\tan\psi$ is imaginary. Therefore, the wave is elliptically polarized in general. Also notice that the polarization ray $\hat{\mathcal{E}}$ is invariant under transformations $(\phi,\psi)\rightarrow(\phi\pm180^\circ,\psi)$ and  $(\phi,\psi)\rightarrow(-\phi,\psi\pm180^\circ)$. Therefore, the polarization angles (Fig.~\ref{fig:Linear}b) can be interpreted up to these identity transformations. Finally, notice that $\psi_\pm$ for the $n^2_\pm$ solutions satisfies the identity $\tan\psi_+\tan\psi_-=-1$. Hence, polarizations of these two frequency-degenerate eigenmodes are always orthogonal in the transverse plane.

\end{appendices}

%

\end{document}